\documentclass[prd,aps,showpacs,preprintnumbers,amssymb,nofootinbib]{revtex4}
\usepackage{graphicx}% Include figure files

\usepackage{epsf}
\usepackage{float}
\usepackage{array,booktabs}
\usepackage{epstopdf}
\usepackage{footnote}
\usepackage{threeparttable}

\def\e3p{$\eta \rightarrow 3 \pi$}
\newcolumntype{P}[1]{>{\centering\arraybackslash}p{#1}}

\begin{document}
\pdfoutput=1
\title{%
\hfill{\normalsize\vbox{%
\hbox{\rm }
 }}\\
{Chiral Nonet Mixing in
$\eta' \rightarrow \eta \pi\pi$ Decay}}

\author{Amir H. Fariborz,
$^{\it \bf a}$~\footnote[1]{Email:
 fariboa@sunyit.edu}}

\author{Joseph Schechter
 $^{\it \bf b}$~\footnote[2]{Email:
 schechte@phy.syr.edu}}

\author{Soodeh Zarepour
$^{\it \bf c}$~\footnote[3]{Email:
    soodehzarepour@shirazu.ac.ir}}

\author{Mohammad Zebarjad
$^{\it \bf c}$~\footnote[4]{Email:
  zebarjad@physics.susc.ac.ir}}

\affiliation{$^ {\bf \it a}$ Department of
Engineering, Science and Mathematics,
State University of New York, Institute of Technology,Utica, NY13504-3050, USA,}

\affiliation{$^ {\bf \it b}$ Department of Physics,
 Syracuse University, Syracuse, NY 13244-1130, USA,}

\affiliation{$^ {\bf \it c}$ Department of Physics,
 Shiraz University, Shiraz 71454, Iran}

\date{\today}

\newpage

\begin{abstract}
 Underlying mixing of scalar mesons is studied in $\eta'\rightarrow \eta\pi\pi$ decay  within a generalized linear sigma model of low-energy QCD which contains two nonets of scalar mesons and two nonets of pseudoscalar mesons (a quark-antiquark nonet and a four quark nonet).    The model has been previously employed in various investigations of the underlying mixings among scalar mesons below and above 1 GeV  (as well as those of their pseudoscalar chiral partners) and has provided a coherent global picture for the physical properties and quark substructure of these states.    The potential of the model is defined in terms of two- and four-quark chiral nonets,  and based on the number of underlying quark and antiquark lines in each term in the potential, a criterion for limiting the number of terms at each order of calculation (and systematically further improving the results thereafter).  At the leading order,  which corresponds to neglecting terms in the potential with higher than eight quark and antiquark lines,  the free parameters of the model have been previously fixed in detailed global fits to scalar and pseudoscalar experimental mass spectra below and above 1 GeV together with several low-energy parameters.  In the present work,  the same order of potential with fixed parameters is used to further explore the underlying mixings among scalar mesons in the $\eta'\rightarrow \eta\pi\pi$ decay.     It is found that the linear sigma model with only a single lowest-lying nonet is not accurate in predicting the decay width,  but inclusion of the mixing of this nonet with the next-to-lowest lying nonet,  together with the effect of final state interaction of pions,  significantly improves this prediction and agrees with experiment up to about 1\%.    It is also shown that while the prediction of the leading order of the generalized model for the Dalitz parameters is not close to the experiment, the model is able to give a reasonable prediction of the energy dependencies of the normalized decay amplitude squared and that this is expected to improve with further refinement of the complicated underlying mixings.  Overall this investigation provides further support for the global picture of scalar mesons: those below 1 GeV are predominantly  four-quark states and significantly mix with those above 1 GeV which are closer to the conventional p-wave quark-antiquark states.
\end{abstract}


\maketitle
\section{introduction}
The scalar mesons continue to attract the attention of many investigators for their important roles in low-energy QCD \cite{PDG}.   Although not all their properties have been fully uncovered, nevertheless a great deal of progress has been made over the past couple of decades
\cite{Weinberg_13}-\cite{vanBev}.
Now there seems to be an emerging agreement about their quark substructure.    Historically,   the light scalar mesons (below 1 GeV) with their low mass and inverted mass spectrum (isosinglet lighter than the isodoublet,  lighter than the heavier isosinglet which is nearly degenerate in mass with isovector) found a natural template in an ideally mixed four-quark MIT bag model \cite{Jaf}.      An ideally mixed pure four-quark  picture, while gives a perfect description of the mass spectra of the scalars below 1 GeV,   seems to need some distortions to be able to describe some of the decay channels of these states.      On the other hand,  the scalars above 1 GeV while seem to be close to the conventional p-wave quark-antiquark states, some of their properties deviate from such an idealized picture.    In short, the scalars below 1 GeV appear to be close to four-quark states with some distortions and those above 1 GeV appear to be close to quark-antiquark states with some distortions.    The natural question would be whether such distortions on the quark substructure of both of these sets of states is due to a mixing among these states.   The idea of mixing is intuitively understandable since some of the scalars below and above 1 GeV are very broad (such as, for example, $f_0(500)$ and $f_0(1370)$, or $K_0^*(800)$) and there is no reason that they should not refrain from mixing with members having the same quantum numbers in a nearby nonet (see refs. \cite{06_F}-\cite{Mec}).   In \cite{Mec}, the idea of such mixings and their effects on the properties of isovectors and isodoublets was studied within a nonlinear chiral Lagrangian model and was shown that allowing a four-quark scalar nonet below 1 GeV to slightly mix with a quark-antiquark scalar nonet above 1 GeV provides a natural explanation for certain aspects of the mass spectrum and decay properties of both nonets of scalars.    For example, it explains that when a pure four-quark nonet below 1 GeV mixes with a pure quark-antiquark nonet above 1 GeV,  due to level repulsion,  the scalar mesons below 1 GeV are pushed down in mass and hence become lighter than expected.    Also it shows that several unexpected mass and decay properties of the scalars above 1 GeV stem from this underlying mixing:  the fact that the experimental mass of $a_0(1450)$ is higher than that of $K_0^*(1430)$ (which is unexpected if these two states were to belong to the same pure quark-antiquark nonet) is due to a ``level-crossing'' that takes place in this mixing which also naturally explains several  unexpected decay properties of the states above 1 GeV \cite{Mec}.    In refs. \cite{global} (and refs. therein) such mixing patterns were further studied in a generalized linear sigma model.
The advantages of linear vs nonlinear model are: (a) the scalar and pseudoscalar states become chiral partners, form chiral nonets,  and the underlying chiral symmetry and its breakdown establishes connections and constrains on various parameters of the model (b) reliable experimental inputs on both scalar and pseudoscalar mesons can be used in determining the model parameters and (c) the status of some of the pseudoscalar states that are not quite established (such as $\eta(1405)$ which is stated to be a good ``non-${\bar q} q$'' candidate \cite{07_KZ}, or dynamically generated in $f_0(980)\eta$ channel \cite{10_AOR}) can be explored in this approach as well.   The main disadvantage of linear model vs nonlinear model is the fact that in scattering and decay processes one has to carefully deal with the individual contributions that are often large but tend to regulate each other in a very delicate manner (``local cancelations'').  This is a disadvantage compared to, for example,  chiral perturbation theory \cite{ChPT} where corrections are systematically controlled at different orders.    Nevertheless,  for the present objective of studying  the global picture for the family relations and mixings among various scalar states below 2 GeV,  the generalized linear model in which all such states are explicitly kept in the Lagrangian,  instead of being integrated out,  seems to be an efficient framework.   Although the description of $\eta'\rightarrow\eta\pi\pi$ seems to be beyond the immediate effectiveness of chiral perturbation theory \cite{bijnens11}, nevertheless,  this decay has been studied in some variations of this framework \cite{RCPT}.

\begin{figure}[H]

\begin{center}
\vskip .75cm
%-----------------------------------------
\epsfxsize = 2.7cm
 \includegraphics[height=3cm]{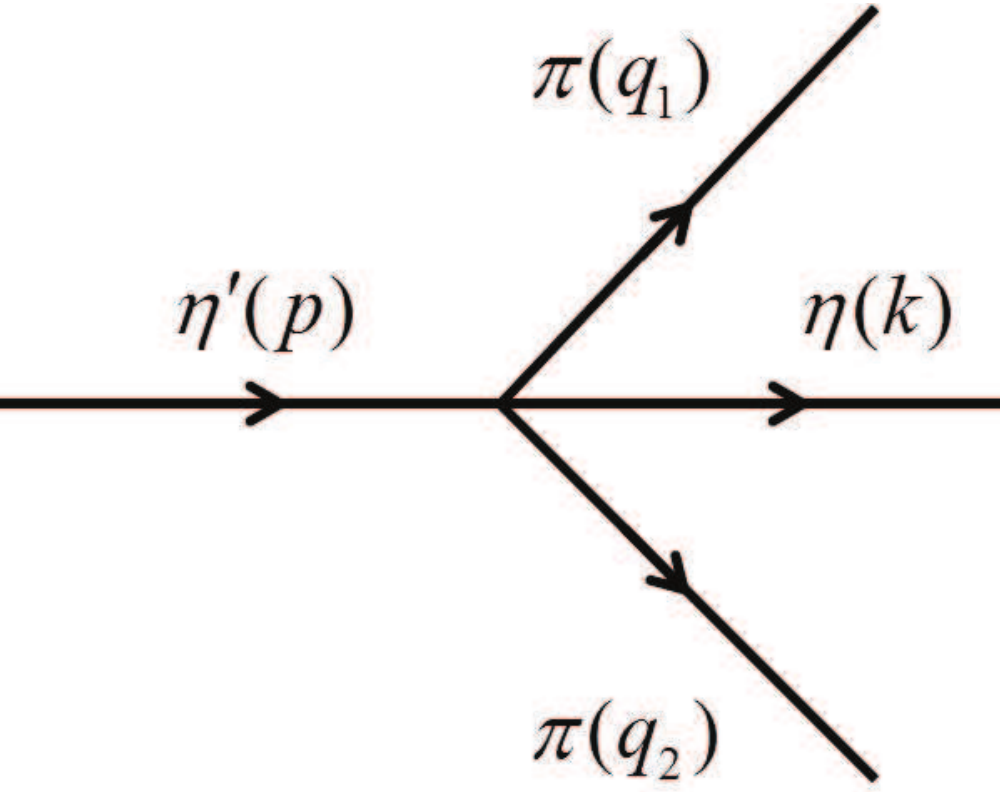}
\hskip 1cm
\epsfxsize = 3.2cm
 \includegraphics[height=3cm]{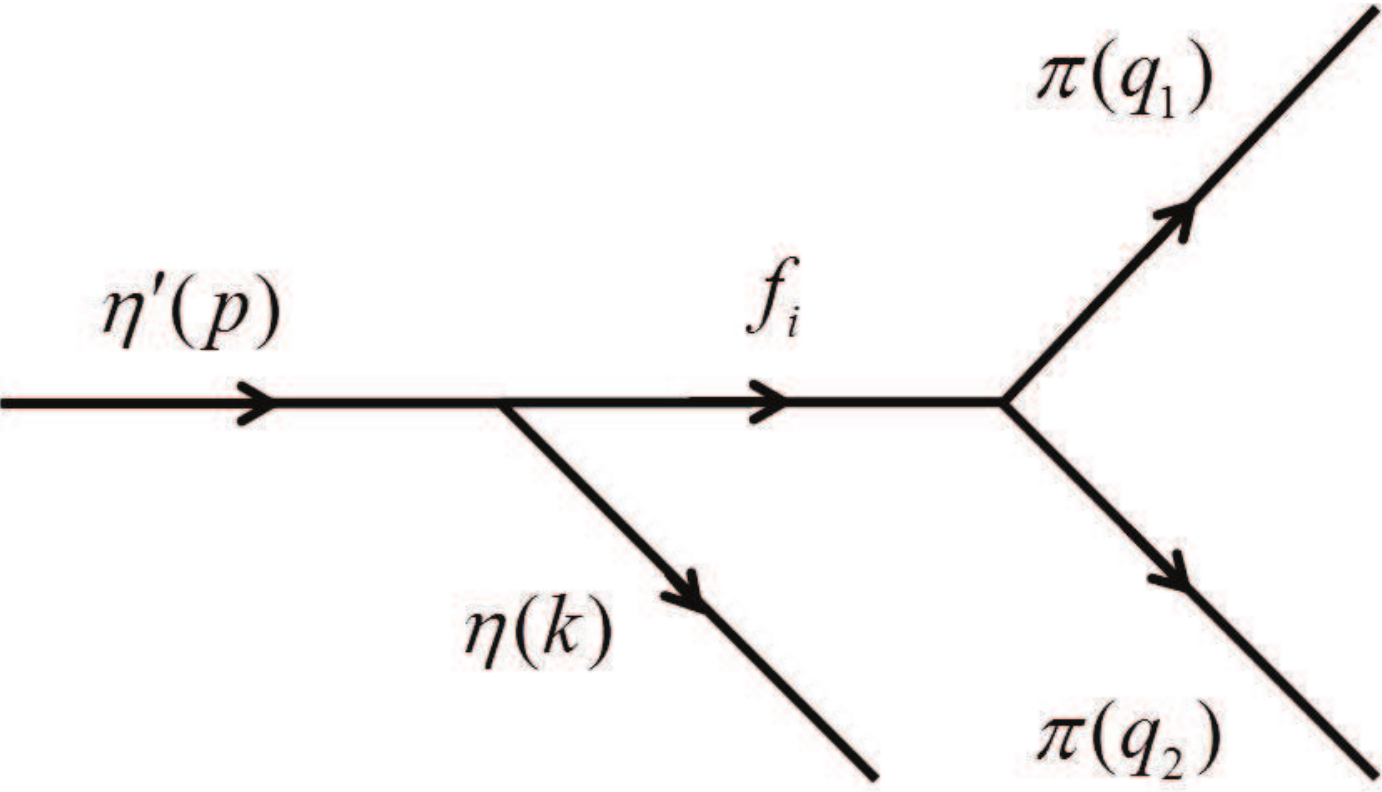}
\hskip .75cm
\epsfxsize = 3.2cm
 \includegraphics[height=3cm]{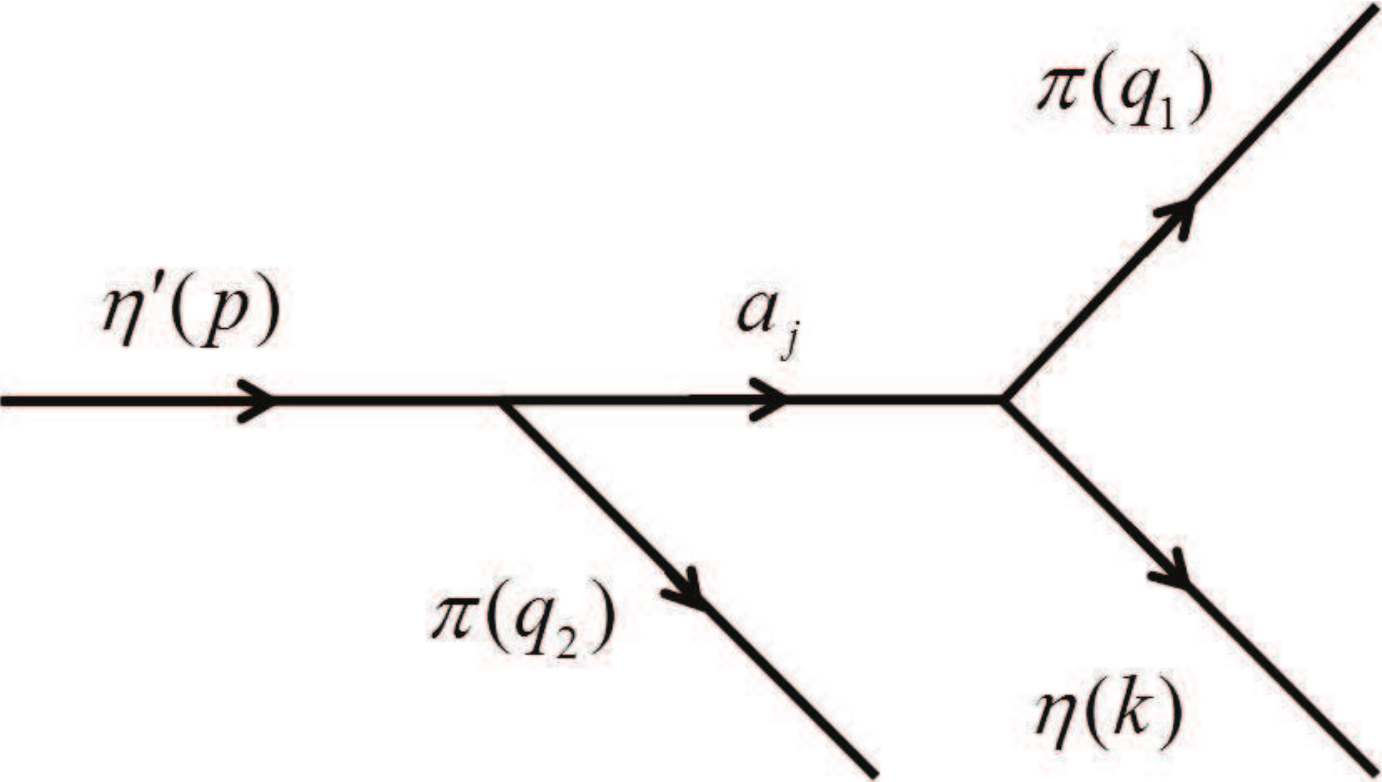}
\caption{Feynman diagrams representing the decay $\eta'\rightarrow\eta\pi\pi$:  Contact term (left), contribution of isosinglet scalars (middle) and contribution of isovectors (right). }
\label{F_FD}
\end{center}
\end{figure}

The tree-level Feynman diagrams representing the $\eta'\rightarrow \eta\pi\pi$ decay are shown in Fig. \ref{F_FD}.  These include a four-point interaction diagram (contact diagram) together with diagrams representing the contributions of isovector and isosinglet  scalar mesons.   This is a suitable decay channel for studying the role of scalar mesons and their underlying mixing patterns.     To probe the effect of such underlying mixings, we use both a single-nonet SU(3) linear sigma model, as well as a generalized version that contains two nonets of scalar mesons (a two-quark nonet and a four-quark nonet).    In either case, the computation of the partial decay width,   and the energy dependencies of the normalized decay amplitude, are the points of contact with experiment.
The individual amplitudes are
\begin{eqnarray}
M_{4p} &=& - \gamma^{(4)},\nonumber\\
M_{f_i}&=& \sqrt{2}\gamma_{f_i\eta\eta^{'}}\gamma_{f_i\pi\pi}\frac{1}{m_{f_i}^2+(p-k)^2}\nonumber\\
&=& \sqrt{2}\gamma_{f_i\eta\eta^{'}}\gamma_{f_i\pi\pi}\frac{1}
{m_{f_i}^2+ \left[ m_{\eta^{'}}^2-m_{\eta}^2-2m_{\eta^{'}}(w_1+w_2)\right]},
\nonumber\\
 M_{a_j} &=&  \gamma_{a_j\pi\eta^{'}} \gamma_{a_j\pi\eta}\left[  \frac{1}{m_{a_j}^2+(p-q_2)^2}+\frac{1}{m_{a_j}^2+(p-q_1)^2}\right]\nonumber\\
 &=&  \gamma_{a_j\pi\eta^{'}} \gamma_{a_j\pi\eta}\left[  \frac{1}
 {m_{a_j}^2+(-m_{\eta^{'}}^2-m_{\pi}^2+2m_{\eta^{'}}w_2)}+\frac{1}{m_{a_j}^2+(-m_{\eta^{'}}^2-m_{\pi}^2+2m_{\eta^{'}}w_1)}\right],
 \nonumber\\
\label{M_indv_temp}
\end{eqnarray}
where the subscripts $i$ and $j$ run over the number of isosingle and isovector intermediate states, respectively,  $\omega_1$ and $\omega_2$ are the pion energies, and the coupling constants are defined as
\begin{eqnarray}
-{\cal L} &=&
\frac{1}{2}\gamma^{(4)}\eta \eta' \mbox{\boldmath ${\pi}$} \cdot
{\mbox{\boldmath ${\pi}$}} +
 \frac{\gamma_{f_i \pi \pi}}{\sqrt 2}
f_i \mbox{\boldmath ${\pi}$} \cdot
{\mbox{\boldmath ${\pi}$}}
+ \gamma_{f_i \eta \eta} f_i  \eta  \eta
+  \gamma_{f_i \eta \eta'} f_i  \eta  \eta'
+ \gamma_{a_j \pi\eta} {\bf a_j} \cdot  \mbox{\boldmath
${\pi}$}  \eta +
 \gamma_{a_j \pi\eta'} {\bf a_j} \cdot  \mbox{\boldmath
${\pi}$}  \eta'
+ \cdots.
\label{gamma_temp}
\end{eqnarray}

Following the standard calculation,   the partial decay width is then
obtained from
\begin{equation}
\Gamma_{\eta^{'}\rightarrow \eta\pi\pi} = \frac{1}{64\pi^3 m_{\eta'}}\int dw_1 dw_2 |M|^2,
\label{DW_temp}
\end{equation}
with the total amplitude
\begin{equation}
M=M_{4p} + \sum_{i} M_{f_i} + \sum_{j} M_{a_j}.
\label{M_tot_temp}
\end{equation}
Equations (\ref{M_indv_temp}), (\ref{gamma_temp}), (\ref{DW_temp}) and (\ref{M_tot_temp}) serve as our ``templates'' for various investigations in this work.   The experimental data for decay width \cite{PDG} is given in Table \ref{T_DW_exp}.

\begin{table}[H]

\centering

\caption{Experimental decay width of $\eta'\rightarrow \eta \pi^+ \pi^-$  (first column), $\eta'\rightarrow \eta \pi^0 \pi^0$ (second column) and
$\eta'\rightarrow \eta \pi \pi$ in the isospin invariant limit (last column).  }
\renewcommand{\tabcolsep}{1pc} % enlarge column spacing
\renewcommand{\arraystretch}{1.2} % enlarge line spacing
\smallskip
 \begin{threeparttable}

\begin{tabular}{lccP{2.0cm}P{2.0cm}P{4.0cm}}

\hline
\centering
     & Exp. [$\eta'\rightarrow \eta \pi^+ \pi^-$]   & Exp. [$\eta'\rightarrow \eta \pi^0 \pi^0$] & Exp. (averaged \tnote{a} )  \\ \hline
%\centering
$\Gamma$ (MeV) & $0.086 \pm 0.004$ & $0.0430 \pm 0.0022$ & $0.086 \pm 0.003$ \\  \\\hline

\end{tabular}

\begin{tablenotes}
\item[a] For the average value ${\bar x} +\delta{\bar x}$ of measurements  $x_i+\delta x_i$, we use ${\bar x}=\sum_i x_i w_i/\sum_i w_i ; \delta {\bar x}=(\sum_i w_i)^{-1/2}$ with the weight $w_i=1/(\delta x_i)^2$, and $\delta x_{\rm{total}}= \sqrt{\delta x^2_{\rm{syst.}}+\delta x^2_{\rm{stat.}}}$

\end{tablenotes}
\label{T_DW_exp}
\end{threeparttable}
\end{table}

In addition to the partial decay width,  the energy dependence of the normalized decay amplitude squared can be compared with experiment.     For this comparison, it is common to use Dalitz variables
\begin{eqnarray}
X&=&{\sqrt{3}\over Q} \left( \omega_1 - \omega_2 \right), \nonumber \\
Y&=& - {{2+m_\eta/m_\pi}\over Q} \left( \omega_1 + \omega_2 \right) - 1 +
{{2+m_\eta/m_\pi}\over Q} \left( m_{\eta'} - m_\eta\right),
\end{eqnarray}
where $Q=m_{\eta'}-m_\eta - 2 m_\pi$.    Then the normalized decay amplitude squared can be expanded in powers of $X$ and $Y$.   In the generalized parametrization \cite{PDG}
\begin{equation}
{\cal M}^2  =  { {M(X,Y)^2} \over {M(0,0)^2 }} = 1 + a\, Y + b\, Y^2 + c \, X + d\, X^2 + \cdots,
\label{M_ren}
\end{equation}
where  $a$, $b$, $c$, and $d$ are real-valued parameters and $c=0$ in the isospin invariant limit.  The experimental data \cite{PDG} for $a$, $b$ and $d$ are given in Table \ref{T_abd_exp}.    See also \cite{Moskov,07_BJ}.

\begin{table}[H]
\centering
\label{T_sn_DP}
\caption{Experimental Dalitz slope parameters for $\eta'\rightarrow \eta \pi^+ \pi^-$ (first column), $\eta'\rightarrow \eta \pi^0 \pi^0$
 (second column) and $\eta'\rightarrow \eta \pi \pi$ in iso-spin invariant limit (third column). }
\renewcommand{\tabcolsep}{0.4pc} % enlarge column spacing
\renewcommand{\arraystretch}{2.5} % enlarge line spacing
\begin{tabular}{lccccccP{1.0cm}P{1.0cm}P{1.0cm}P{1.0cm}P{1.0cm}P{1.0cm}P{1.0cm}}
\hline
 Parameter       & Exp. [$\eta'\rightarrow \eta \pi^+ \pi^-$]   & Exp. [$\eta'\rightarrow \eta \pi^0 \pi^0$] & Exp. (averaged)  \\
        & VES \cite{VES}  & GAM4\cite{GAM4}& iso-spin invariant limit \\ \hline
a &  $-0.127 \pm 0.016 \pm 0.008$& $-0.066 \pm 0.016 \pm 0.003$ &  $-0.094 \pm 0.012$ &   \\
b& $-0.106 \pm 0.028 \pm 0.014$ & $-0.063 \pm 0.028 \pm 0.004$  &  $-0.082 \pm 0.021$\\
d & $-0.082 \pm 0.017 \pm0.008$ & $-0.067 \pm 0.020 \pm 0.003 $ &  $-0.075 \pm 0.014$  \\ \hline
\end{tabular}\\[2pt]
\label{T_abd_exp}
\end{table}

In Sec. II we present the predictions of single nonet SU(3) linear sigma model for the $\eta'\rightarrow\eta\pi\pi$ decay.   We then present a brief review of the double nonet generalized linear sigma model in Sec. III,  followed by its predictions for the relevant two-body decays in Sec. IV and  of the $\eta'\rightarrow\eta\pi\pi$ decay in Sec. V.    We give our approximation for the effect of final state interactions in Sec. VI and a summary and discussion of the results in Sec. VII.

%###################################################

\section{Single nonet approach}
The role of scalar mesons in $\pi\pi$, $\pi K$ and $\pi \eta$ scattering channels was extensively studied in a single nonet SU(3) linear sigma model in \cite{LsM}. It was shown that when the tree-level scattering amplitudes are unitarized with the simple K-matrix unitarization method,  the model is able to explain the experimental data on the $I$=$J$=0 $\pi\pi$ scattering amplitude up to around 1.2 GeV.    The first pole found in this unitarized amplitude clearly agrees with the properties of the light and broad sigma meson (with $m_\sigma=0.457$ GeV and $\Gamma_\sigma = 0.632$ GeV), and the second pole agrees with the properties of $f_0(980)$ (with $m_{f_0}$=0.993 GeV and $\Gamma_{f_0}=$0.051 MeV).     Within the same framework, a light and broad kappa meson (with $m_\kappa$=0.798-0.818 GeV and $\Gamma_\kappa$ = 0.257-0.614 GeV) was identified in the studies of $I=1/2$,$J=0$, $\pi K$ scattering amplitude.   Similarly,  a coherent picture was observed in the studies of $I=1$, $J=0$,  $\pi \eta$ scattering amplitude in which a scalar resonance with the properties of $a_0(980)$ is clearly detected (with $m_{a_0}$ = 0.890-1.013 GeV and $\Gamma_{a_0}$=0.109-0.241 GeV).      These investigations were carried out within a non-renormalizable linear sigma model in which the Lagrangian has the general structure
\begin{equation}
{\cal L} = - \frac{1}{2} {\rm Tr}
\left( \partial_\mu M \partial_\mu M^\dagger
\right)
- V_0 \left( M \right) - V_{SB},
\label{sn_LsM}
\end{equation}
where the chiral field $M$ is constructed out of scalar nonet $S$ and pseudoscalar nonet $\phi$,
\begin{equation}
M=S + i \phi,
\end{equation}
and transforms linearly under chiral transofrmation
\begin{equation}
M \rightarrow U_L M U_R^\dagger,
\end{equation}
and $V_0$ is an arbitrary function of the independent
SU(3)$_{\rm L}\times$SU(3)$_{\rm R}\times$U(1)$_{\rm V}$ invariants
\begin{eqnarray}
I_1 &=& {\rm Tr} \left( M M^\dagger \right),  \hskip 1cm
I_2 = {\rm Tr} \left( M M^\dagger M M^\dagger \right),\nonumber \\
I_3 &=& {\rm Tr} \left[\left( M M^\dagger \right)^3 \right], \hskip 1cm
I_4 = 6 \left( {\rm det} M + {\rm det} M^\dagger \right).
\end{eqnarray}
The symmetry breaker $V_{SB}$ has the minimal form
\begin{equation}
V_{SB} = - 2 {\rm Tr} (A S),
\label{V_SB}
\end{equation}
where $A=$ diag ($A_1, A_2, A_3$) are proportional to the three ``current'' type quark masses.   The vacuum values satisfy
\begin{equation}
\langle S_a^b \rangle = \alpha_a \delta_a^b.
\end{equation}
In the isospin invariant limit
\begin{equation}
A_1 = A_2 \ne A_3, \hskip 1cm \alpha_1 = \alpha_2 \ne \alpha_3.
\end{equation}
Using ``generating equations'' that express the chiral symmetry of $V_0$ together with the minimum equation
\begin{equation}
\left\langle
{
{\partial V}
\over
{\partial S_a^b}
}
\right\rangle
=0,
\end{equation}
masses of pseudoscalars are completely determined based on the underlying chiral symmetry together with the choice of symmetry breakers (both U(1)$_{\rm A}$ and SU(3)$_{\rm L} \times$ SU(3)$_{\rm R}$ $\rightarrow$ SU(2) isospin).    The scalar masses on the other hand are not all predicted;  in the most general case only the mass of isodoublet kappa meson is predicted, whereas
if the renormalizability is imposed the isovector mass and one of the isosinglet masses are determined.    It is found in \cite{LsM} that it is necessary not to  impose the renomalizability condition in order to be able to fit to the $\pi \pi$ and  $\pi K$ scatttering amplitudes and to get a reasonable description of $\pi\eta$ amplitude.    In the nonrenormalizable case, the ``bare'' scalar masses $m_{BARE}(\sigma)$, $m_{BARE}(f_0)$ and $m_{BARE}(a_0)$ (i.e. the Lagrangian masses which are different than the physical masses that are related to the poles of the appropriate unitarized scattering amplitudes) and the scalar mixing angle $\theta_s$  are found from fits to various low-energy data in \cite{LsM}.    Here we use the same set of parameters to study the $\eta'\rightarrow \eta\pi\pi$ decay.      In this case the required coupling constants in our ``template'' equations (\ref{M_indv_temp})-(\ref{M_tot_temp}) are computed from the ``generating equations'' that express the symmetry of the Lagrangian  (\ref{sn_LsM}) (a computational algorithm is presented in \cite{LsM_Maple}):
\begin{eqnarray}
\gamma^{(4)}&=&\sum_{a,b} \left\langle \frac{\partial^4 V}{\partial \phi_1^2 \partial\phi_2^1\partial\phi_a^a\partial\phi_b^b}\right\rangle_0 (R_{\phi})_{2}^a(R_{\phi})_{3}^b,\nonumber\\
\gamma_{a_0 \pi\eta}&=&\sum_a \left\langle \frac{\partial^3 V}{\partial S_1^2 \partial\phi_a^a\partial\phi_2^1}\right\rangle_0 (R_{\phi})_{2}^a,\nonumber\\
\gamma_{a_0 \pi\eta'}&=&\sum_a \left\langle \frac{\partial^3 V}{\partial S_1^2 \partial\phi_a^a\partial\phi_2^1}\right\rangle_0 (R_{\phi})_{3}^a,\nonumber\\
\gamma_{f_i \pi\pi}&=&{1\over \sqrt{2}}\,\sum_a \left\langle \frac{\partial^3 V}{\partial S_a^a \partial\phi_1^2 \partial\phi_2^1}\right\rangle_0 (R_s)_{i+1}^a, \nonumber\\
\gamma_{f_i \eta\eta'}&=& \sum_{a,b,c} \left\langle \frac{\partial^3 V}{\partial S_a^a \partial\phi_b^b\partial\phi_c^c}\right\rangle_0 (R_s)_{i+1}^a(R_{\phi})_{2}^b(R_{\phi})_{3}^c,\nonumber\\
\end{eqnarray}
where the ``bare'' couplings and the rotation matrices ($R_s$ and $R_\phi$) are given in Appendix A.    Here $f_1=\sigma$ and $f_2=f_0(980)$.   We find
\begin{equation}
\Gamma\left(\eta'\rightarrow\eta\pi\pi\right)=
0.61 \pm 0.01 \, {\rm MeV}  \hskip 2.0cm {\rm Single \hskip 0.15cm nonet \hskip 0.15cm (\textbf{bare}\: result).}
\end{equation}
Clearly, despite the success of the nonrenormalizable single nonet SU(3) linear sigma model in describing the low-energy scatterings discussed above,   it estimates this partial decay width about seven times larger than the experimental value displayed in Table \ref{T_DW_exp}.

\begin{figure}[H]
\begin{center}
\vskip 1cm
\epsfxsize = 7.5cm
 \includegraphics[width=8cm,height=5.8cm]{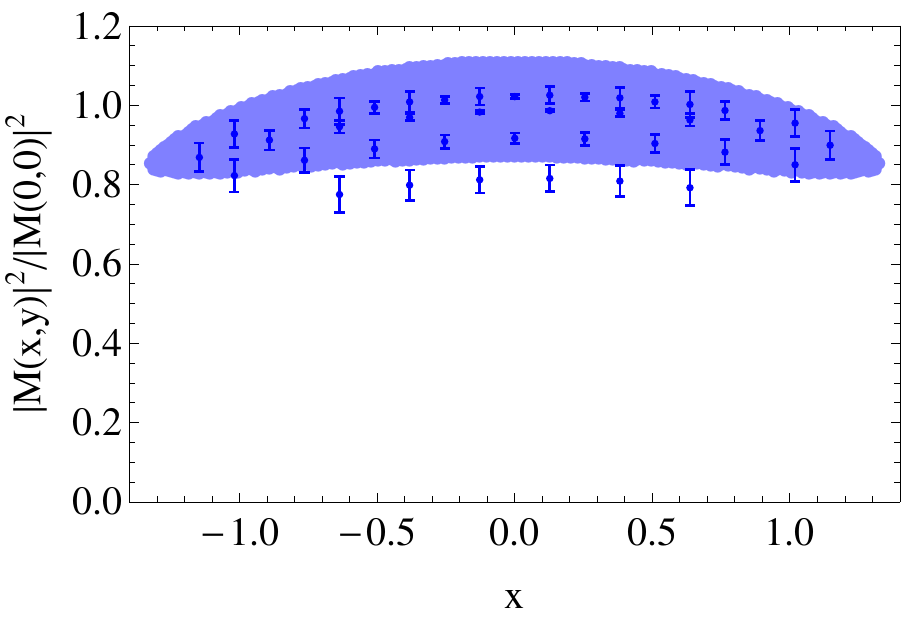}
\hskip 1cm
\epsfxsize = 7.5cm
 \includegraphics[width=8cm,height=6cm]{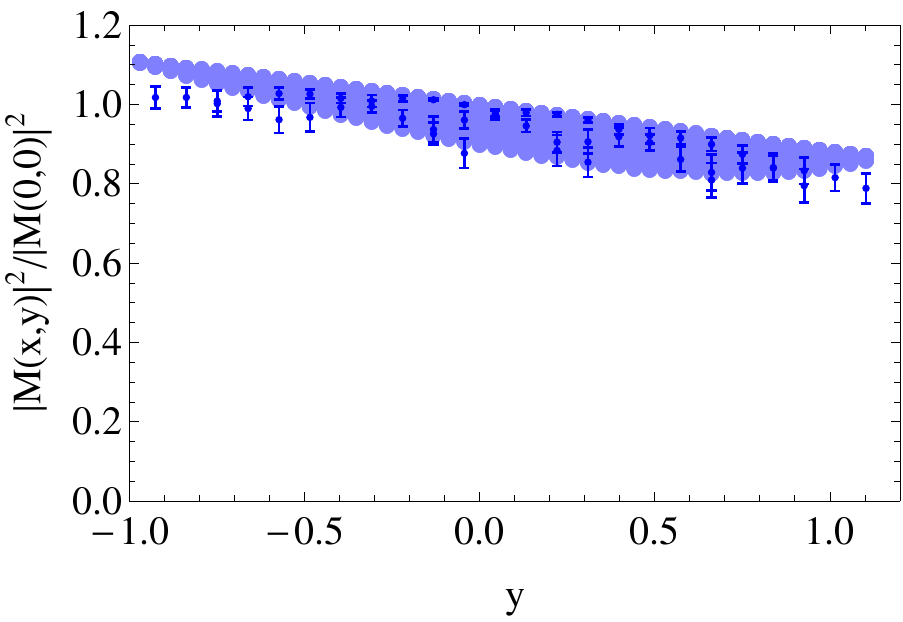}
 \caption{ Projections of $ |\hat{M}|^2=|M(x,y)|^2/|M(0,0)|^2$ onto the $ y - |\hat{M}|^2$ and
 $x - |\hat{M}|^2  $ planes (single nonet model).}
 \label{F_sn_ED}
\end{center}
\end{figure}

\begin{figure}[H]
\begin{center}
\vskip 1cm
\epsfxsize = 7.5cm
 \includegraphics[width=7cm,height=5cm]{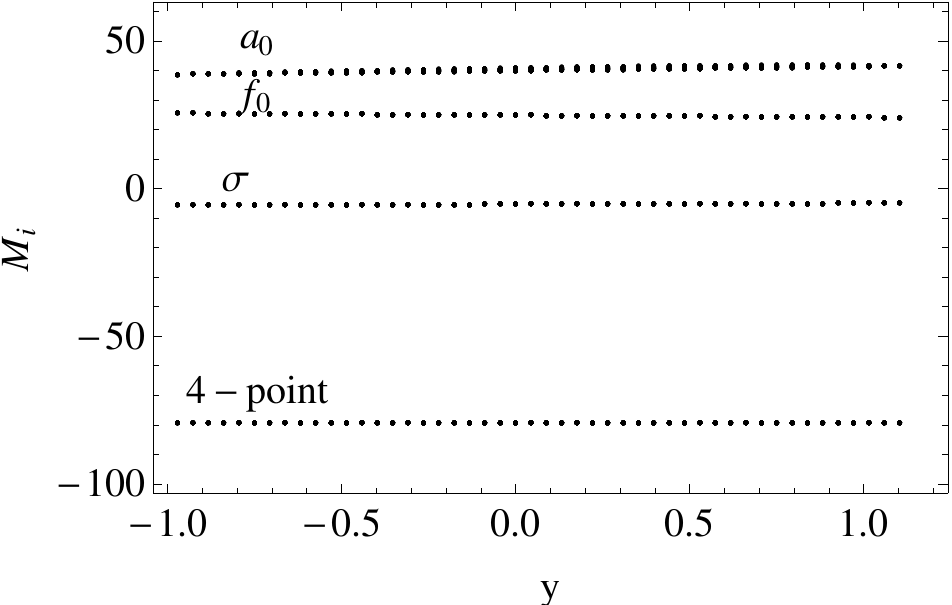}
 \caption{Individual contributions to the $\eta'\rightarrow\eta\pi\pi$ decay amplitude in single nonet model.  The large contribution of contact term $M_{4p}$ is balanced with the contributions of $f_0(980)$ and $a_0(980)$.}
 \label{F_sn_indv}
\end{center}
\end{figure}

The energy dependence of the normalized decay amplitude squared is compared with experiment in Fig. \ref{F_sn_ED} and the Dalitz parameters that characterize the energy expansion of this
amplitude squared are given in Table \ref{T_sn_ED}.   Comparing with the averaged experimental values of Table \ref{T_abd_exp}, we see that there is a qualitative order of magnitude agreement, at best.    This lack of accuracy of the single nonet approach raises the natural question of whether the underlying mixing among scalar mesons (which are clearly important players in this decay) has a noticeable effect on these estimates.    One of the important roles of the scalars is to balance the large contribution due to the contact term ($M_{4p}$) as can be seen in Fig. \ref{F_sn_indv}.    Moreover, the eta systems (both the two below 1 GeV as well as those above 1 GeV) can mix and have a nontrivial effect on this decay estimate.    The single nonet approach does not take these mixing effects among the scalars and among the pseudoscalars into account which can have important consequences for this partial decay width.   This motivates us to further study this decay within the generalized linear sigma model (that contains two scalar nonets and two pseudoscalar nonets)     in this investigation.

\begin{table}[H]
\centering
\caption{The predicted Dalitz
 parameters in single nonet linear sigma model of ref. \cite{LsM}.}
\renewcommand{\tabcolsep}{0.4pc} % enlarge column spacing
\renewcommand{\arraystretch}{2.5} % enlarge line spacing
\begin{tabular}{lcP{1.0cm}P{1.0cm}}
\hline
 Parameter       & single nonet model  \\  \hline
a & $-0.114\pm 0.001$   \\
b & $-0.001 \pm 0.001$  \\
d & $-0.063 \pm 0.001$  \\ \hline
\end{tabular}
\label{T_sn_ED}
\end{table}

%%%%%%%%%%%%%%%%%%%%%%%%%%%%%%%%%%%%%%%%%%%%%%%

\section{Brief Review of the generalized linear sigma model}

%%%%%%%%%%%%%%%%%%%%%%%%%%%%%%%%%%%%%%%%%%%%%%%

   The model employs the 3$\times$3 matrix
chiral nonet fields \cite{global}:
\begin{equation}
M = S +i\phi, \hskip 2cm
M^\prime = S^\prime +i\phi^\prime.
\label{sandphi}
\end{equation}
The matrices $M$ and $M'$ transform in the same way under
chiral SU(3) transformations but
may be distinguished by their different U(1)$_A$
transformation properties. $M$ describes the ``bare"
 quark-antiquark scalar and pseudoscalar nonet fields while
$M'$ describes ``bare" scalar and pseudoscalar fields
containing two quarks and two antiquarks. At the
symmetry level in which we are working,
it is unnecessary to
further specify the four quark field configuration.
The four quark field may, most generally,
 be imagined as some linear
combination of a diquark-antidiquark  and a
``molecule" made of two quark-antiquark ``atoms".

The general Lagrangian density which defines our model is
\begin{equation}
{\cal L} = - \frac{1}{2} {\rm Tr}
\left( \partial_\mu M \partial_\mu M^\dagger
\right) - \frac{1}{2} {\rm Tr}
\left( \partial_\mu M^\prime \partial_\mu M^{\prime \dagger} \right)
- V_0 \left( M, M^\prime \right) - V_{SB},
\label{mixingLsMLag}
\end{equation}
where $V_0(M,M^\prime) $ stands for a function made
from SU(3)$_{\rm L} \times$ SU(3)$_{\rm R}$
(but not necessarily U(1)$_{\rm A}$) invariants
formed out of
$M$ and $M^\prime$.

 As  previously discussed \cite{global}, the
 leading choice of terms
corresponding
to eight or fewer underlying quark plus antiquark lines
 at each effective vertex
reads:
\begin{eqnarray}
V_0 =&-&c_2 \, {\rm Tr} (MM^{\dagger}) +
c_4^a \, {\rm Tr} (MM^{\dagger}MM^{\dagger})
\nonumber \\
&+& d_2 \,
{\rm Tr} (M^{\prime}M^{\prime\dagger})
     + e_3^a(\epsilon_{abc}\epsilon^{def}M^a_dM^b_eM'^c_f + h.c.)
\nonumber \\
     &+&  c_3\left[ \gamma_1 {\rm ln} (\frac{{\rm det} M}{{\rm det}
M^{\dagger}})
+(1-\gamma_1){\rm ln}\frac{{\rm Tr}(MM'^\dagger)}{{\rm
Tr}(M'M^\dagger)}\right]^2.
\label{SpecLag}
\end{eqnarray}
     All the terms except the last two (which mock up the axial anomaly)
      have been chosen to also
possess the  U(1)$_{\rm A}$
invariance.   A possible term $\left[{\rm Tr} (M M^{\dagger})\right]^2$ is neglected for simplicity because it violates the OZI rule.
The symmetry breaking term which models the QCD mass term
takes the form given in Eq. ({\ref{V_SB}).   The model allows for two-quark condensates,
$\alpha_a=\langle S_a^a \rangle$ as well as
four-quark condensates
$\beta_a=\langle {S'}_a^a \rangle$.
Here we assume  isotopic spin
symmetry so A$_1$ =A$_2\ne$ A$_3$ and:
\begin{equation}
\alpha_1 = \alpha_2  \ne \alpha_3, \hskip 2cm
\beta_1 = \beta_2  \ne \beta_3.
\label{ispinvac}
\end{equation}

 We also need the ``minimum" conditions,
\begin{equation}
\left< \frac{\partial V_0}{\partial S}\right> + \left< \frac{\partial
V_{SB}}{\partial
S}\right>=0,
\quad \quad \left< \frac{\partial V_0}{\partial S'}\right>
=0.
\label{mincond}
\end{equation}

There are twelve parameters describing the Lagrangian and the
vacuum: Six coupling constants
 given in Eq.(\ref{SpecLag}), the two quark mass parameters,
($A_1=A_2,A_3$) and the four vacuum parameters ($\alpha_1
=\alpha_2,\alpha_3,\beta_1=\beta_2,\beta_3$). Ten of these parameters ($c_2$, $c_4^a$, $d_2$, $e_3^a$, $\alpha_1$, $\alpha_3$, $\beta_1$, $\beta_3$, $A_1$, $A_3$) are determined using the four minimum
equations  together with the following  six experimental inputs for the masses, pion decay constant and the ratio of strange to non-strange quark masses:
\begin{eqnarray}
 m[a_0(980)] &=& 984.7 \pm 1.2\, {\rm MeV},
\nonumber
\\ m[a_0(1450)] &=& 1474 \pm 19\, {\rm MeV},
\nonumber \\
 m[\pi(1300)] &=& 1300 \pm 100\, {\rm MeV},
\nonumber \\
 m_\pi &=& 137 \, {\rm MeV},
\nonumber \\
F_\pi &=& 131 \, {\rm MeV},
\nonumber \\
{A_3 \over A_1} &=& 20\rightarrow 30.
\label{inputs1}
\end{eqnarray}
Clearly, $m[\pi(1300)]$ and  $A_3/A_1$ have large uncertainties which in turn dominate the
uncertainty of predictions.

 The remaining two parameters ($c_3$ and $\gamma_1$) only affect the isosinglet pseudoscalars (whose properties also
 depend on the ten parameters discussed above).    However, there are several choices for determination of these two parameters depending on how the the  four isosinglet pseudoscalars predicted in this model are matched to many experimental candidates below 2 GeV.   The two lightest predicted by the model ($\eta_1$ and $\eta_2$)  are identified with $\eta(547)$ and $\eta'(958)$ with masses:
 \begin{eqnarray}
m^{\rm exp.}[\eta (547)] &=& 547.853 \pm
0.024\, {\rm
MeV},\nonumber \\
m^{\rm exp.}[\eta' (958)] &=& 957.78 \pm 0.06
\, {\rm
MeV}.
\end{eqnarray}
For the two heavier ones ($\eta_3$ and $\eta_4$),   there are six ways that they can be identified with the four experimental candidates above 1 GeV:  $\eta(1295)$,  $\eta(1405)$,  $\eta(1475)$, and $\eta(1760)$ with masses,
\begin{eqnarray}
m^{\rm exp.}[\eta (1295)] &=& 1294 \pm 4\, {\rm
MeV},\nonumber \\
m^{\rm exp.}[\eta (1405)] &=& 1409.8 \pm 2.4 \,
{\rm
MeV},
\nonumber \\
m^{\rm exp.}[\eta (1475)] &=& 1476 \pm 4\, {\rm
MeV},\nonumber \\
m^{\rm exp.}[\eta (1760)] &=& 1756 \pm 9 \,
{\rm
MeV}.
\end{eqnarray}
This led to six scenarios considered in detail in \cite{global}.
The two experimental inputs for determination of the two parameters $c_3$ and $\gamma_1$ are taken to be  Tr$M_\eta^2$ and det$M_\eta^2$, i.e.
\begin{eqnarray}
{\rm Tr}\, \left(  M^2_\eta  \right) &=&
{\rm Tr}\, \left(  {M^2_\eta}  \right)_{\rm exp},
\nonumber \\
{\rm det}\, \left( M^2_\eta \right) &=&
{\rm det}\, \left( {M^2_\eta} \right)_{\rm exp}.
\label{trace_det_eq}
\end{eqnarray}
Moreover,  for each of the six scenarios,  $\gamma_1$ is found from a quadratic equation, and as a result, there are altogether twelve possibilities for determination of $\gamma_1$ and $c_3$.    Since only Tr and det of experimental masses are imposed for each of these twelve possibilities, the resulting  $\gamma_1$ and $c_3$ do not necessarily recover the exact individual experimental masses,  therefore the best overall agreement between the predicted masses (for each of the twelve possibilities) were examined in \cite{global}.   Quantitatively,  the
goodness of each solution was measured by the smallness of
the following quantity:
\begin{equation}
\chi_{sl} =
\sum_{k=1}^4
{
 {\left| m^{\rm theo.}_{sl}(\eta_k)  -
  m^{\rm exp.}_{s}(\eta_k)\right|}
                 \over
  m^{\rm exp.}_{s}(\eta_k)
},
\label{E_chi_sl}
\end{equation}
in which $s$ corresponds to the scenario
(i.e. $s= 1 \cdots 6$) and
$l$ corresponds to the solution number
(i.e. $l=$ I, II).   The quantity $\chi_{sl}
\times 100$ gives the overall percent
discrepancy between our theoretical prediction
and experiment.   For the six scenarios and
the two solutions for each scenario,
$\chi_{sl}$ was analyzed  in ref. \cite{global}.    Some of these  scenarios, such as those  involving $\eta(1405)$ are clearly not favored.   This suggests that $\eta(1405)$ is of a more complicated quark substructure that can be probed by the present model, and this is consistent with the investigation of ref. \cite{10_AOR} in which it is shown that this state may be dynamically generated in $f_0(980)\eta$ interaction.
For the third scenario (corresponding to identification of $\eta_3$ and $\eta_4$ with experimental candidates $\eta(1295)$ and $\eta(1760)$) and  solution I the best agreement with the mass spectrum of the eta system was obtained (i.e. $\chi_{3\rm{I}}$ was the smallest).      For the present analysis too,  all six scenarios are examined and it is again found that the best overall result (both for the partial decay width of $\eta'\rightarrow \eta\pi\pi$ as well as the energy dependence of its squared decay amplitude) is obtained for scenario ``3I'' consistent with the analysis of ref. \cite{global}.
In this work,  we only present the result of ``3I'' scenario.    To reduce the model uncertainty for the analysis of $\eta'\rightarrow \eta\pi\pi$ decay,  we have further refined the numerical study of ref. \cite{global} for scenario ``3I'' and have  displayed the result in Fig. \ref{F_chi_31}, in which $\chi_{3\rm{I}}$ is plotted over the parameter space $m[\pi(1300)]$-$A_3/A_1$  that are two of the model inputs with largest experimental uncertainties.

\begin{figure}[!htb]
\begin{center}

\epsfxsize = 5cm
 \includegraphics[width=8cm,height=7cm]{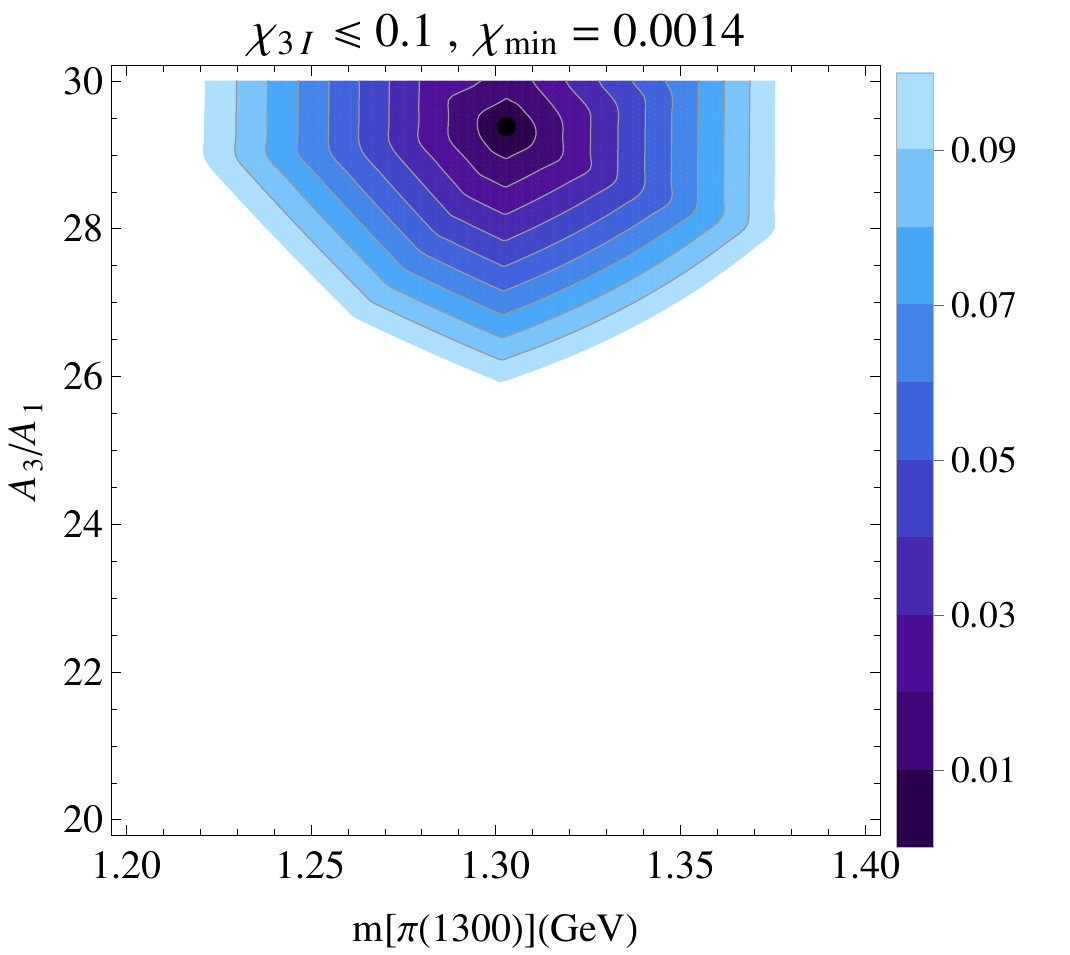}
\hskip 1cm
\caption{Contour plot of function $\chi_{3{\rm I}}$ [defined in Eq. (\ref{E_chi_sl})] over the $m[\pi(1300)]$-$A_3/A1$ plane for scenario ``3I'' in which the four isosinglet pseudoscalar states predicted by this model $\eta_1$, $\eta_2$,
$\eta_3$ and $\eta_4$ are identified with the four experimental candidates $\eta(547)$, $\eta'(958)$, $\eta(1295)$ and $\eta(1760)$, respectively.   The minimum of $\chi_{3{\rm I}}$ occurs at $m[\pi(1300)]=1.30$ GeV and $A_3/A_1$=29.40, at which it has a value of  $\chi_{3{\rm I}}^{\rm min} < 0.0015$, and shows an overall uncertainty of less than 0.15\% between the four isosinglet pseudoscalar masses predicted by the model and the central values of the four experimental masses.   (Note:   the total experimental uncertainty $\sum_i \Delta m^{\rm exp.}_i/m^{\rm exp.}_i \approx 0.0083$ where $m^{\rm exp.}_i \pm \Delta m^{\rm exp.}_i, i=1..4$ denote the four experimental masses.)   }
\label{F_chi_31}
\end{center}
\end{figure}

Consequently,  all twelve parameters of the model (at the present order of approximation) are evaluated by the method discussed above using four minimum equations and eight experimental inputs.   The uncertainties of the experimental inputs result in uncertainties on the twelve model parameters which in turn result in uncertainties on  physical quantities that are computed in this model.    In the work of ref. \cite{global} all rotation matrices describing the underlying mixing among two- and four-quark components for each spin and isospin states are computed.
For scalars:
\begin{equation}
\left[
\begin{array}{cc}
a_0^+(980)\\
a_0^+(1450)
\end{array}
\right]
=
L_a^{-1}
\left[
\begin{array}{cc}
S_1^2\\
{S'}_1^2
\end{array}
\right],
%---------------------------------------------------
\hskip 2cm
%---------------------------------------------------
\left[
\begin{array}{cc}
K_0(800)\\
K_0^*(1430)
\end{array}
\right]
=
L_\kappa^{-1}
\left[
\begin{array}{cc}
S_1^3\\
{S'}_1^3
\end{array}
\right],
%---------------------------------------------------
\hskip 2cm
%---------------------------------------------------
\left[
\begin{array}{cc}
f_1\\
f_2\\
f_3\\
f_4
\end{array}
\right]
=
L_0^{-1}
\left[
\begin{array}{cc}
f_a\\
f_b\\
f_c\\
f_d
\end{array}
\right],
\label{E_SRot}
\end{equation}
where $L_a^{-1}$, $L_\kappa^{-1}$ and  $L_0^{-1}$ are the rotation matrices for $I=1$, $I=1/2$ and $I=0$ respectively; $f_i, i=1..4$ are four of the physical isosinglet scalars below 2 GeV (in this model $f_1$ and $f_2$ are clearly identified with $f_0(500)$ and $f_0(980)$ and the two heavier states resemble two of the heavier isosinglet scalars above 1 GeV); and
\begin{eqnarray}
f_a&=&\frac{S^1_1+S^2_2}{\sqrt{2}} \hskip .7cm
n{\bar n},
\nonumber  \\
f_b&=&S^3_3 \hskip .7cm s{\bar s},
\nonumber    \\
f_c&=&  \frac{S'^1_1+S'^2_2}{\sqrt{2}}
\hskip .7cm ns{\bar n}{\bar s},
\nonumber   \\
f_d&=& S'^3_3
\hskip .7cm nn{\bar n}{\bar n}.
\label{f_basis}
\end{eqnarray}

For pseudoscalars:
\begin{equation}
\left[
\begin{array}{cc}
\pi^+(137)\\
\pi^+(1300)
\end{array}
\right]
=
R_\pi^{-1}
\left[
\begin{array}{cc}
\phi_1^2\\
{\phi'}_1^2
\end{array}
\right],
%---------------------------------------------------
\hskip 2cm
%---------------------------------------------------
\left[
\begin{array}{cc}
K^+(496)\\
{K'}^+(1460)
\end{array}
\right]
=
R_K^{-1}
\left[
\begin{array}{cc}
\phi_1^3\\
{\phi'}_1^3
\end{array}
\right],
%---------------------------------------------------
\hskip 2cm
%---------------------------------------------------
\left[
\begin{array}{cc}
\eta_1\\
\eta_2\\
\eta_3\\
\eta_4
\end{array}
\right]
=
R_0^{-1}
\left[
\begin{array}{cc}
\eta_a\\
\eta_b\\
\eta_c\\
\eta_d
\end{array}
\right],
\label{E_PRot}
\end{equation}
where $R_\pi^{-1}$, $R_K^{-1}$ and  $R_0^{-1}$ are the rotation matrices for $I=1$, $I=1/2$ and $I=0$ pseudoscalars respectively; $\eta_i, i=1..4$ are four of the physical isosinglet pseudoscalars below 2 GeV; and
\begin{eqnarray}
\eta_a&=&\frac{\phi^1_1+\phi^2_2}{\sqrt{2}} \hskip .7cm
n{\bar n},
\nonumber  \\
\eta_b&=&\phi^3_3 \hskip .7cm s{\bar s},
\nonumber    \\
\eta_c&=&  \frac{{\phi '}^1_1 + {\phi'}^2_2}{\sqrt{2}}
\hskip .7cm ns{\bar n}{\bar s},
\nonumber   \\
\eta_d&=& {\phi '}^3_3
\hskip .7cm nn{\bar n}{\bar n}.
\label{eta_basis}
\end{eqnarray}

In the present work,  we use the results obtained in \cite{global} to compute the decay properties of $\eta'\rightarrow\eta\pi\pi$ without introducing any new parameters and  find a reasonable agreement between the model prediction and experiment.    This provides further test of the underlying two and four-quark mixing
among scalar mesons below and above 1 GeV  and the appropriateness of the generalized linear sigma model developed in \cite{global} and reference therein.

%%%%%%%%%%%%%%%%%%%%%%%%%%%%%%%%%%%%%%%%%%%%%%%%

\section{Two body decays}

%%%%%%%%%%%%%%%%%%%%%%%%%%%%%%%%%%%%%%%%%%%%%%%%

Since the scalar-pseudoscalar-pseudoscalar coupling constants are essential in analyzing the $\eta' \rightarrow \eta \pi \pi$ decay, for orientation  we first calculate some of  these couplings that appear in the prediction of the model for the main two-body decays of the scalar mesons below 1 GeV (for states above 1 GeV additional components such as mixing with glueballs would have to be included and will be presented in future works).  The three decay widths that are particularly relevant for our analysis are,
  \begin{eqnarray}
\Gamma[f_i\longrightarrow \pi\pi]&=& 3\Big(\frac{   q\,\gamma_{f_i\pi\pi}^2 }{8 \pi m_{f_i}^2}\Big)\nonumber\\
\Gamma[a_{j}\longrightarrow \pi\eta] &=&  \frac{\, q\, \gamma_ {a_{j}\pi \eta}^2}{8 \pi m_{a_{j}}^2}\nonumber\\
\Gamma[K_0^*\longrightarrow \pi K]&=& 3\Big(\frac{\, q\, \gamma_ {\kappa\pi K}^2}{16 \pi m_{\kappa}^2}\Big)\nonumber\\
\end{eqnarray}
where q is the center of mass momentum of the final state mesons (for a generic two-body decay $A\longrightarrow BC$ by $q=\sqrt{[m_A^2-(m_B+m_C)^2][m_A^2-(m_B-m_C)^2]}/(2 m_A$)).
The coupling constants are related to the bare couplings:
\begin{eqnarray}
\gamma_{f_i\pi\pi} &=&
{1\over {\sqrt{2}}}
\left\langle
{{\partial^3 V}
\over
{\partial f_i \, \partial \pi^+ \, \partial \pi^-}}
\right\rangle
=
{1\over {\sqrt{2}}}
\sum_{I,A,B}
\left\langle
{{\partial^3 V}
\over
{
 \partial f_I \,
 \partial (\phi_1^2)_A \,
 \partial (\phi_2^1)_B
}}
\right\rangle
(L_0)_{I i} \,
(R_\pi)_{A1} \,
(R_\pi)_{B1},
% ............................................
\nonumber \\
\gamma_{a\pi\eta} &=&
\left\langle
{{\partial^3 V}
\over
{\partial a^- \, \partial \pi^+ \, \partial \eta}}
\right\rangle
=
\sum_{A,B,I}
\left\langle
{{\partial^3 V}
\over
{
 \partial (S^2_1)_A \,
 \partial (\phi_1^2)_B \,
 \partial \eta_I
 }}
\right\rangle
(L_a)_{A1} \,
(R_\pi)_{B1} \,
(R_0)_{I1},
% ..............................................
 \nonumber \\
\gamma_{\kappa K \pi} &=&
\left\langle
{{\partial^3 V}
\over
{\partial \kappa^0 \, \partial K^- \, \partial \pi^+}}
\right\rangle
=
\sum_{A,B,C}
\left\langle
{{\partial^3 V}
\over
{
 \partial (S_2^3)_A \,
 \partial  (\phi_3^1)_B \,
 \partial (\phi_1^2)_C
 }}
\right\rangle
(L_\kappa)_{A1} \,
(R_K)_{B1} \,
(R_\pi)_{C1},
\end{eqnarray}
where $A$, $B$  and $C$ can take values of 1 and 2 (with 1 referring to nonet $M$ and 2 referring to nonet $M'$) and $I$ is a placeholder for  {\it a},{\it b},{\it c} and {\it d} that
respectively represent the four bases in Eq.
(\ref{f_basis}) and (\ref{eta_basis})  . $L_0$, $R_\pi$, $L_a$, $R_0$, $L_\kappa$, $R_K$  are the rotation matrices defined in previous Sec. III.  The bare coupling constants are all given in Appendix A.
The kappa coupling is defined as:
$
-{\cal L} =
 \frac{\gamma_{\kappa K \pi}}{\sqrt 2} \left(
 {\bar K} \mbox{\boldmath ${\tau}$} \cdot
{\mbox{\boldmath ${\pi}$}}
\kappa + h.c. \right) + \cdots.
$

We begin with the decay width of $f_0(500)$ to two pions which is the benchmark test of any low-energy QCD model.   At the present level of approximation,  the main uncertainties in fixing the free parameters of the model are on experimental inputs for the ratio of strange to nonstrange quark masses ($A_3/A_2$) and on the mass of $\pi(1300)$ resonance.  Hence, the $m[\pi(1300)]$-$A_3/A_1$ plane is numerically scanned and the decay width is computed.   The result is displayed in Fig. \ref{F_Cplots_2body}   showing that for most parts of the parameter space the lightest isosinglet state $f_0(500)$ (or $\sigma$) is broad with the decay width comparable to the latest PDG result.   The decay width averaged over the entire parameter space is
\begin{equation}
\Gamma[f_0(500) \rightarrow \pi\pi] = 530 \pm 100 \, {\rm  MeV},
\end{equation}
where the uncertainty represents one standard deviation around the average.
This is consistent with the decay width predicted in this model from the pole of the  K-matrix unitatized $\pi\pi$ scattering amplitude.   Therefore,  the model clearly detects  a light
and broad isosinglet scalar meson.

Similarly,  the prediction of the model over the $m[\pi(1300)]$-$A_3/A_1$ plane for $\Gamma[f_0(980) \rightarrow \pi\pi]$, $\Gamma[a_0(980) \rightarrow \pi\eta]$ and $\Gamma[K_0^*(800) \rightarrow \pi K]$ are shown in Fig. \ref{F_Cplots_2body} with the averaged values:
\begin{eqnarray}
\Gamma[f_0(980) \rightarrow \pi\pi] &=& 35 \pm 27 \,\,{\rm MeV}, \nonumber\\
\Gamma[a_0(980) \rightarrow \pi\eta] &=& 57 \pm 44\,\, {\rm MeV}, \nonumber\\
\Gamma[K_0^*(800) \rightarrow \pi K] &=& 58 \pm 90\,\, {\rm MeV}. \nonumber\\
\end{eqnarray}

\begin{figure}[H]
\begin{center}
\vskip .75cm
%-----------------------------------------
\epsfxsize = 6cm
  \includegraphics[width=6cm,height=5cm]{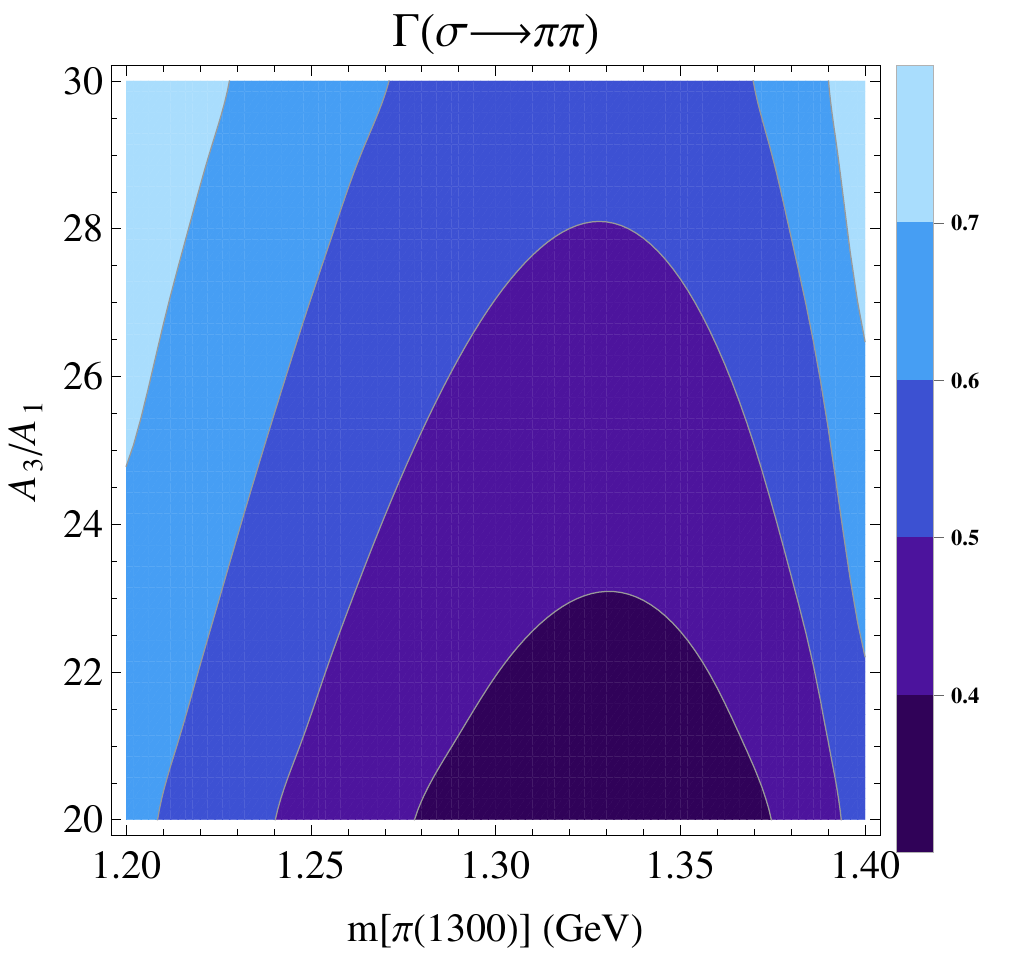}
\hskip 1cm
\epsfxsize = 6cm
  \includegraphics[width=6cm,height=5cm]{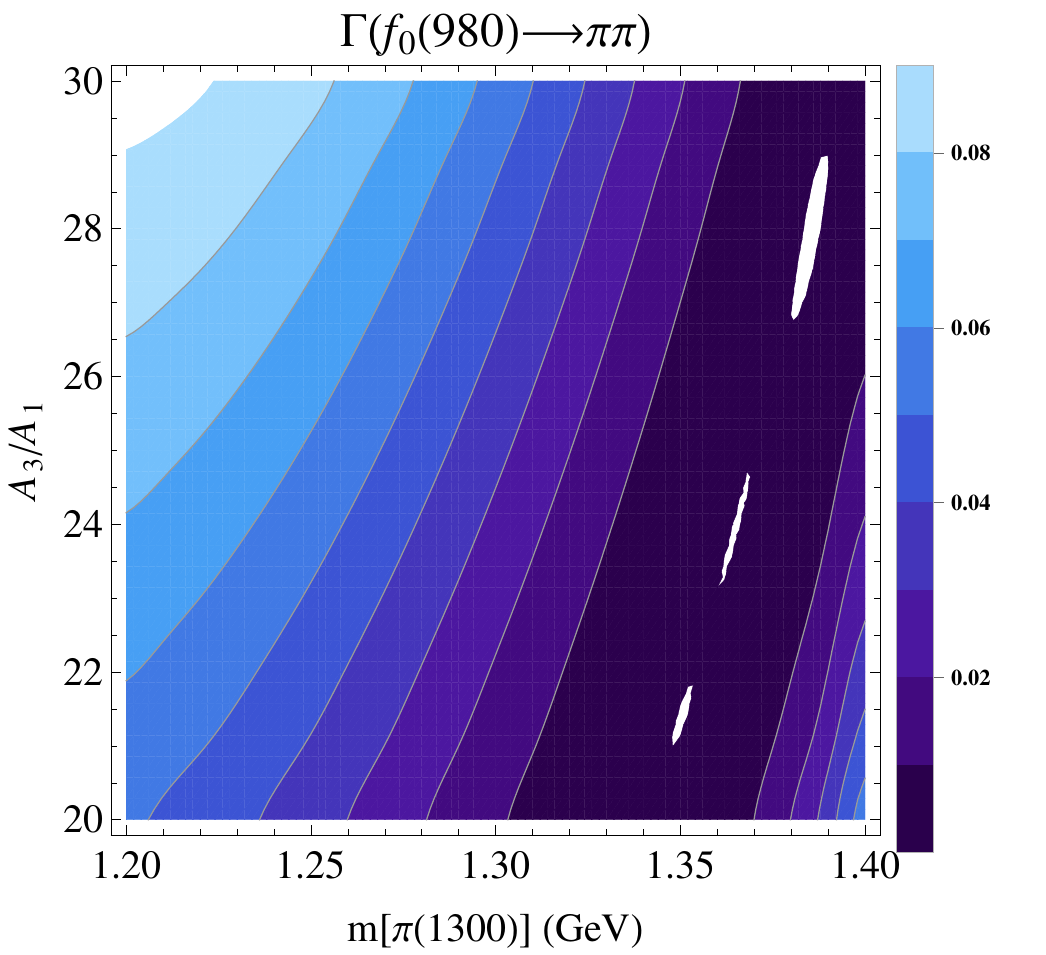}
\vskip .5cm
%-----------------------------------------
\epsfxsize = 6cm
  \includegraphics[width=6cm,height=5cm]{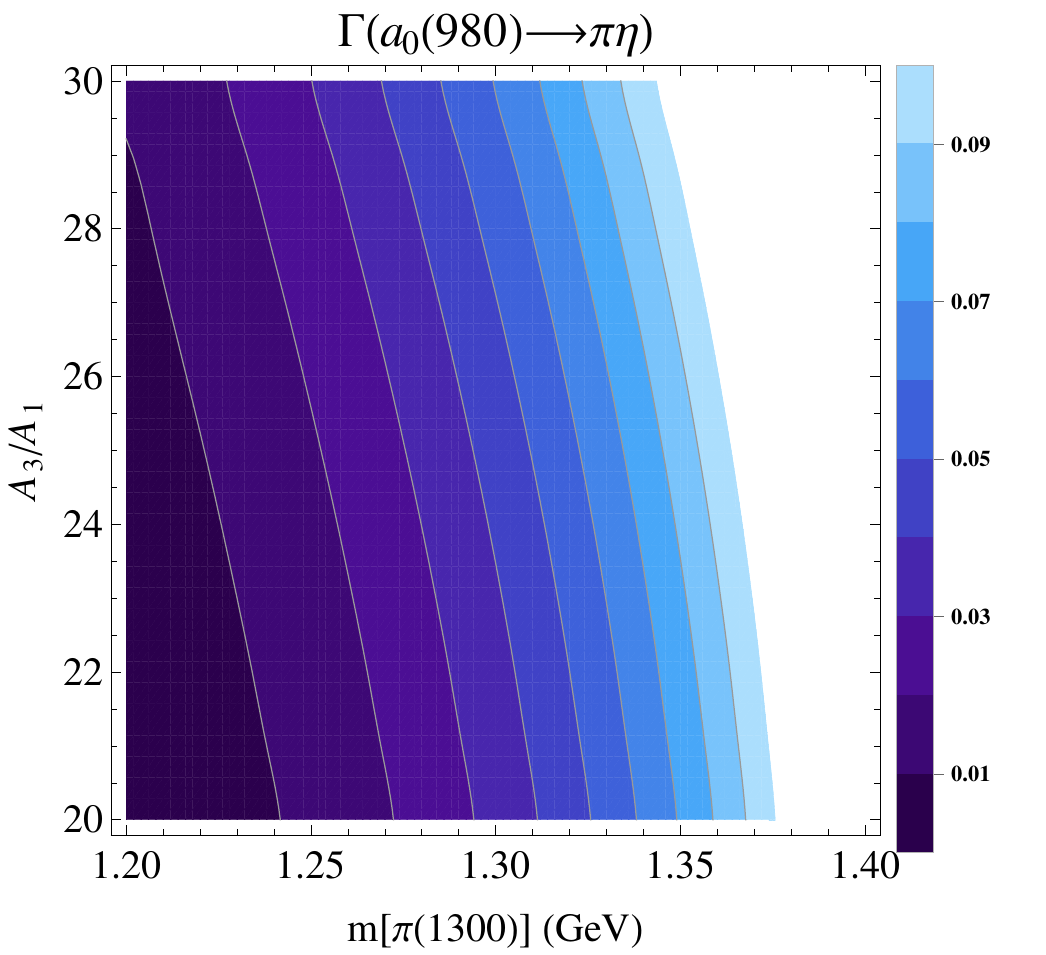}
\hskip 1cm
\epsfxsize = 6cm
  \includegraphics[width=6cm,height=5cm]{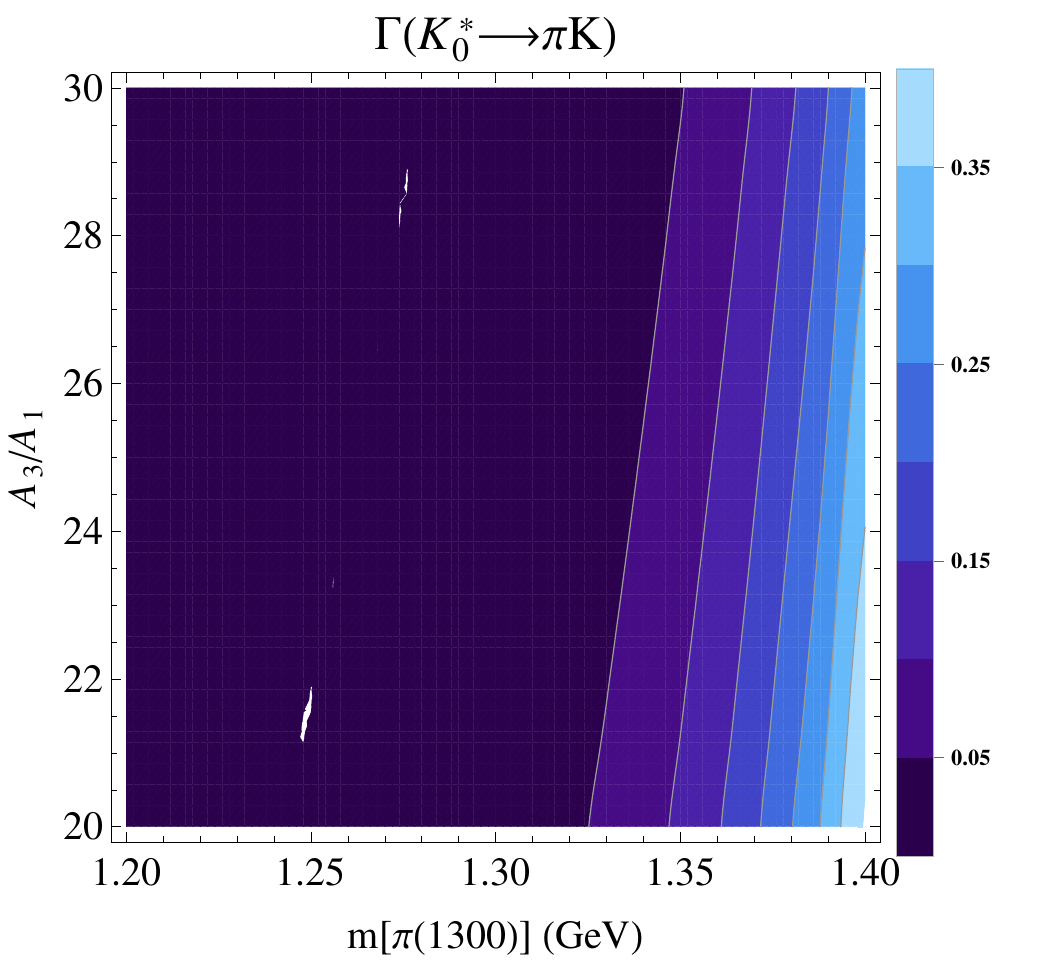}
\caption{Contour plots of the prediction of the model for the main two-body decay widths of light scalar mesons over the $m[\pi(1300)]$-$A_3/A_1$ plane:   $\Gamma[f_0(500)\rightarrow\pi\pi]$ (top left) is predicted to be very large; $\Gamma[f_0(980)\rightarrow\pi\pi]$ (top  right) and
$\Gamma[a_0(980)\rightarrow\pi\eta]$ (bottom left) are within the  expected experimental ranges;  $\Gamma[K_0^*(800)\rightarrow\pi K]$ (bottom right) near high $m[\pi(1300)]$ mass  is large,  and in addition receives unitarity corrections due to the $\pi K$ final-state interation.}
\label{F_Cplots_2body}
\end{center}
\end{figure}

The first three overlap with the expected experimental ranges \cite{PDG}. The averaged decay width of $K_0^*(800)$ is not as large as expected, even though  we see in Fig. \ref{F_Cplots_2body} that there is a region in the parameter space (toward high values of $m[\pi(1300)]$) where this decay width has the right order of magnitude.   However, in a separate work \cite{mixing_piK}, it is shown that the prediction of the model for the $I=1/2$, $J=0$,  $\pi K$ scattering amplitude describes the experimental data well  up to around 1 GeV.  It is also shown that the poles of the K-matrix unitarized  scattering amplitude (the $\kappa$ pole) results in a light and broad $K_0^*(800)$ with a  mass around 710-770 MeV   and decay width around 610-700 MeV.    We interpret the reduction in mass and the increase in the decay width to be the effect of the final state interactions of $\pi K$ which are estimated by the simple K-matrix method.

The main two-body decay channels of the light scalars presented in this section are in a reasonable agreement with the experiment.    This gives an initial test of some of the scalar-pseudoscalar-pseudoscalar coupling constants that will be incorporated in the study of $\eta'\rightarrow\pi\pi$ decay in the next section.

%%%%%%%%%%%%%%%%%%%%%%%%%%%%%%%%%%%%%%%%%%%%%%%%

\section{The ``bare'' prediction of the generalized linear sigma model for $\eta'\rightarrow\eta\pi\pi$ decay}

%%%%%%%%%%%%%%%%%%%%%%%%%%%%%%%%%%%%%%%%%%%%%%%%

In this section we present the ``bare'' prediction of the model (i.e. without  unitarity corrections due to the final state interaction of pions) for decay width and the energy dependencies of the normalized decay amplitude squared.    In next Sec.  we include the effect of these unitarity corrections.
The Feynman diagrams of Fig. \ref{F_FD} include the contact term interaction together with the contributions of the four isosinglet scalars ($f_1$, $\cdots$, $f_4$) as well as the two isovector scalars ($a_1$ and $a_2$).   Some of the scalar-pseudoscalar-pseudoscalar coupling constants were discussed in previous sections and the remaining ones are as follows:

\begin{eqnarray}
 \gamma^{(4)} &=& \sum_{I,J,A,B}\left\langle\frac{\partial^{4}V}{\partial\eta_{I}\partial\eta_{J}\partial  (\phi_{1}^{2})_{A}\partial(\phi_{2}^{1})_{B}}\right\rangle (R_{0})_{I1}(R_{0})_{J2}(R_{\pi})_{A1}(R_{\pi})_{B1},\nonumber \\
 \gamma_{f_{i}\eta\eta^{'}} &=& \left\langle\frac{\partial^{3}V}{\partial f_{i}\partial\eta\partial \eta^{'} }\right\rangle = \sum_{K,I,J}\left\langle\frac{\partial^{3}V}{\partial f_{K}\partial\eta_{I}\partial\eta_{J}}\right\rangle (L_{0})_{Ki}(R_{0})_{I1}(R_{0})_{J2},\nonumber \\
 \gamma_{a_j\pi\eta^{'}}&=&\left\langle\frac{\partial^{3}V}{\partial a_j^{+}\partial\pi^{-}\partial \eta^{'}}\right\rangle=\sum_{A,B,I}\left\langle\frac{\partial^{3}V}{\partial (S_{1}^{2})_{A}\partial(\phi_{2}^{1})_{B}\partial \eta_{I}}\right\rangle (L_{a})_{Aj}(R_\pi)_{B1}(R_0)_{I2},
\end{eqnarray}
where $K$, $I$,and $J$ run over the bases $a$, $b$, $c$ and $d$ defined in Eqs. (\ref{f_basis}) and (\ref{eta_basis}), and $A$ and $B$ can take values of 1, 2 (with 1 referring to nonet $M$ and 2 to nonet $M^{'}$) and the rotation matrices
are  all defined in Eqs. (\ref{E_SRot}) and (\ref{E_PRot}).   All ``bare'' coupling constants are calculated and presented in Appendix B.

We first note that the known ``current algebra'' result for this decay is recovered by decoupling the four-quark nonet $M'$ and imposing the large scalar mass limit (see Appendix C).   This illustrates how contributions of scalar mesons balance the large contribution of the four-point interaction and results in the known small ``current algebra'' result.

It is important to examine the ``bare'' predictions first in order to be able to then test different methods of unitarity corrections that in turn shed light on the important issue of final state interactions.
Using the physical coupling constants defined above (together with those discussed in previous section) we compute the partial decay width by incorporating these couplings into  our ``template'' equations (\ref{M_indv_temp})-(\ref{M_tot_temp}).
The ``bare'' predictions for scenario 3I (previously defined in Fig. \ref{F_chi_31}) are plotted in Fig. \ref{F_DW_bare} for the range of $m[\pi(1300)]$ and several values of $A_3/A_1$.  Although the model prediction is of comparable order of magnitude to the experiment and gets closer to the experimental bounds for low values of $m[\pi(1300)]$, overall it is larger than that of experiment.  The result is however closer to the experiment compared to that predicted by the single nonet approach.   To find the best agreement we search for  the values of $m[\pi(1300)]$ and $A_3/A_1$ that minimize function $\chi_\Gamma$ defined as
\begin{equation}
\chi_\Gamma (m[\pi(1300)], A_3/A_1) =  { {|\Gamma^{\rm theo}(m[\pi(1300)], A_3/A_1) - \Gamma^{\rm exp.}|} \over \Gamma^{\rm exp.}}.
\label{chi}
\end{equation}
We also use a $\chi^2$ fit for doublecheck.   The best predicted decay widths from $\chi$ and $\chi^2$-fit are found with $m[\pi(1300)]=1.22\pm0.01$ and $A_3/A_1= 30.00\pm0.25$:

\begin{figure}[H]
\begin{center}
\vskip 1cm
\epsfxsize = 10cm
 \includegraphics[width=10cm,height=7cm]{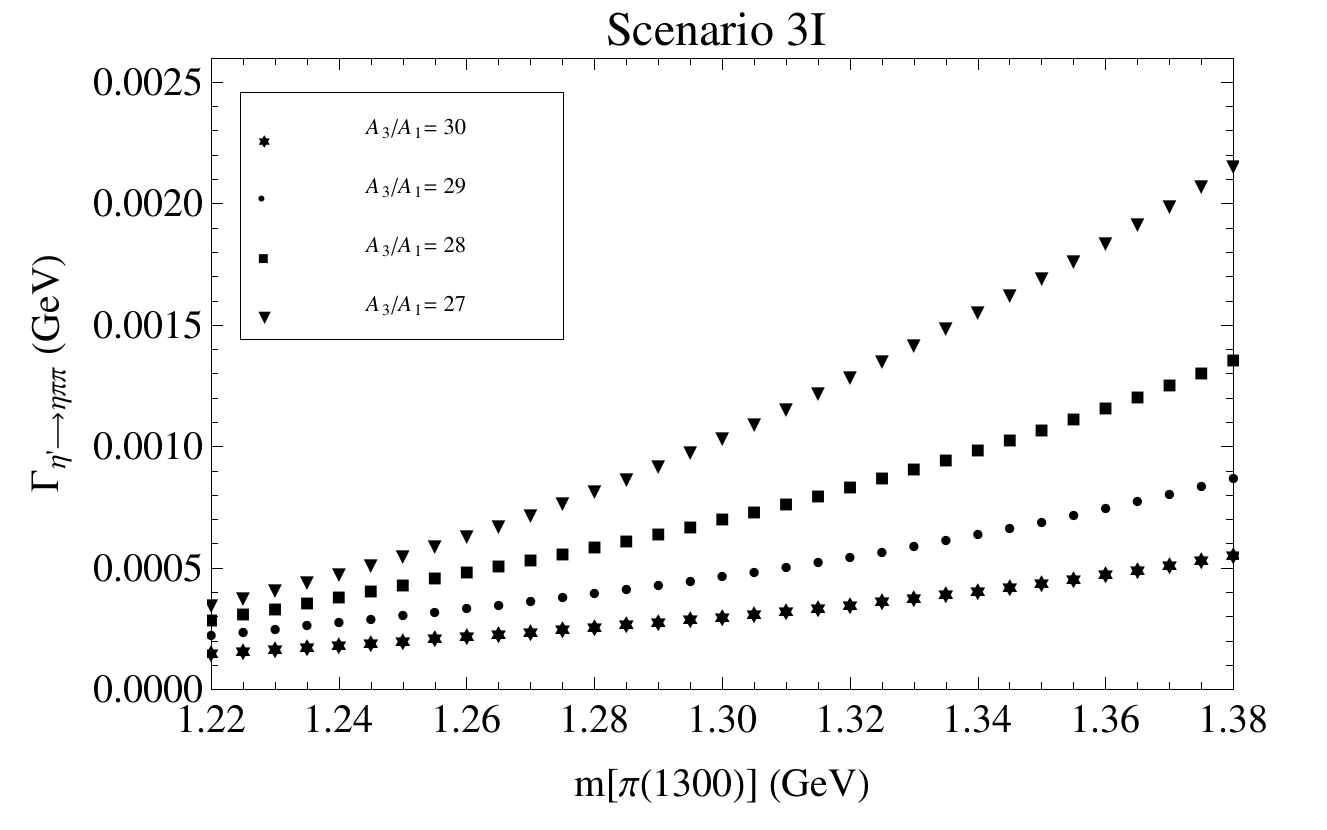}
 \caption{``Bare'' prediction (without unitarity corrections) of the generalized linear sigma model for partial decay width of $\eta'\rightarrow\eta\pi\pi$.}
\label{F_DW_bare}
\end{center}
\end{figure}

\begin{equation}
\Gamma\left(\eta'\rightarrow\eta\pi\pi\right)=
0.15 \pm 0.01 \, {\rm MeV} \hskip 2.0cm  {\rm Generalized \hskip 0.15cm linear\hskip 0.15cm sigma \hskip 0.15cm model \hskip 0.15cm (\textbf{bare} \: result).}
\label{DW_bare}
\end{equation}

The ``bare'' prediction for the energy dependence of the normalized decay amplitude squared is shown in Fig. \ref{F_mmp_bare_ED} and compared with the averaged experimental data of Table \ref{T_abd_exp}.  The best fits to the Dalitz parameters result in best values of $m[\pi(1300)] = 1.38$ GeV and $A_3/A_1=28.75$  which are within the parameter space of the model [Eq. (\ref{inputs1})] however do not coincide with the best  values of these parameters found in the partial decay width analysis in Eq. (\ref{DW_bare}).   This shows that although inclusion of mixing among scalar and among pseudoscalars clearly improves the model predictions, nevertheless,   it is necessary to account for the effect of final state interactions.    A general characteristic of the linear sigma model is the cancelation of large four-point contribution with those of scalar mesons which for the ``bare'' predictions is shown in Fig. \ref{F_mmp_bare_indv}.

\begin{figure}[H]
\begin{center}
\vskip 1cm
\epsfxsize = 10cm
 \includegraphics[width=8cm,height=6cm]{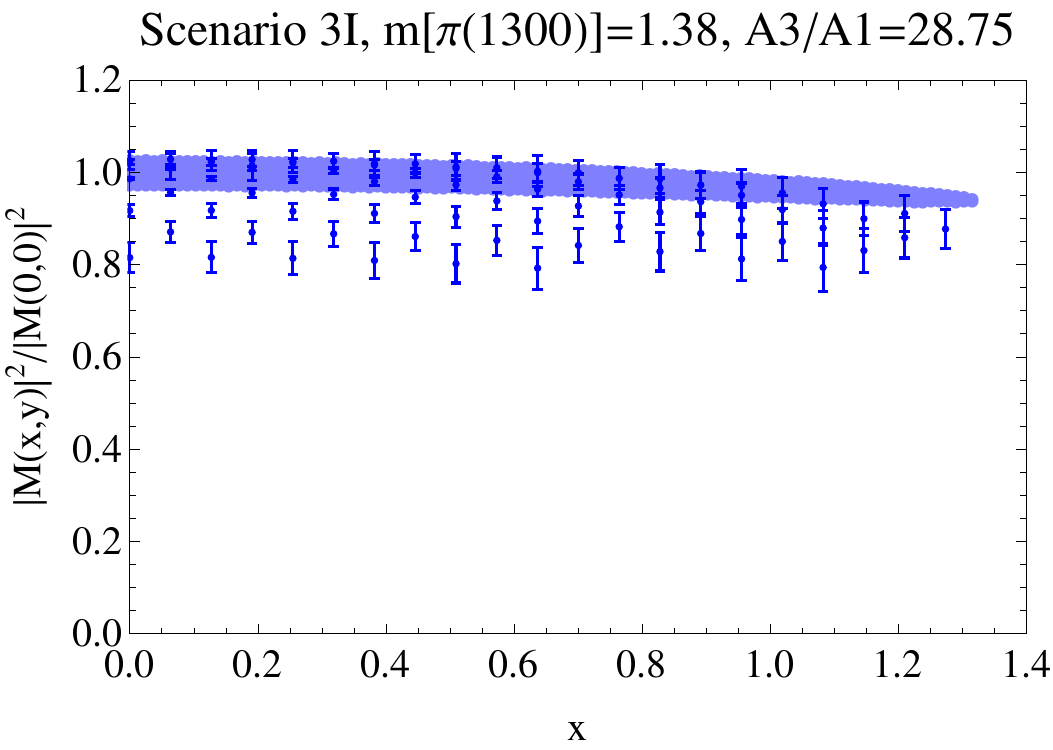}
 \hskip 1cm
\epsfxsize = 10cm
 \includegraphics[width=8cm,height=6cm]{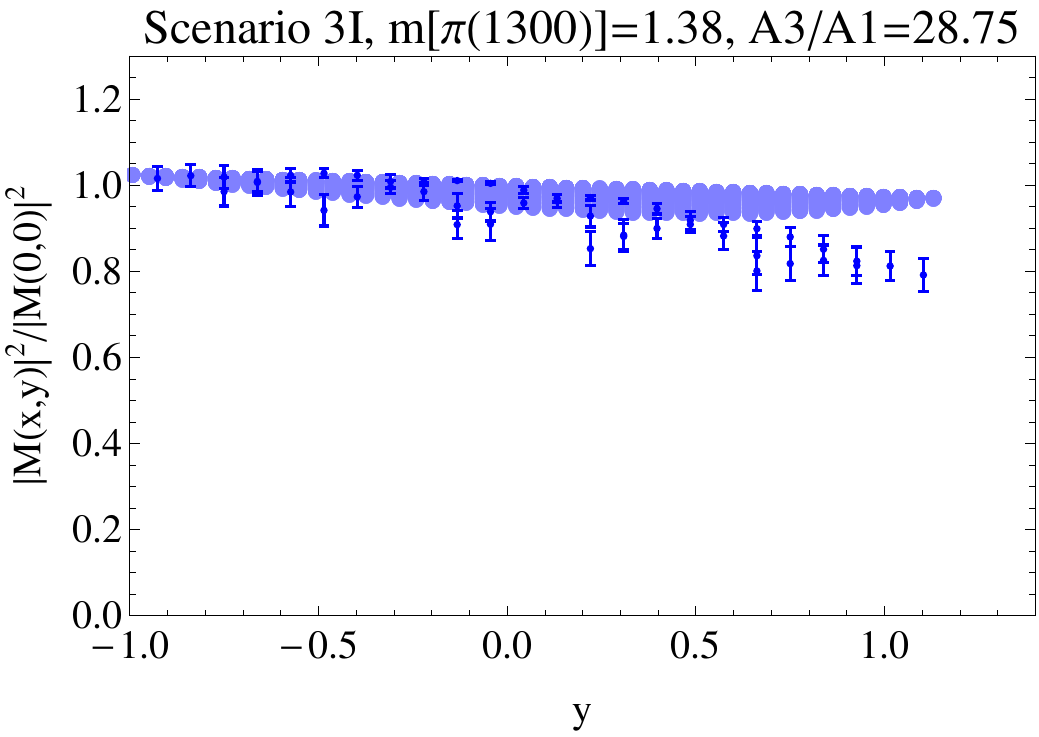}
 \caption{
 Projections of $ |\hat{M}|^2=|M(x,y)|^2/|M(0,0)|^2$ onto the $ y - |\hat{M}|^2$ and
 $x - |\hat{M}|^2  $ planes (``bare'' prediction of the generalized linear sigma model).}
 \label{F_mmp_bare_ED}
\end{center}
\end{figure}

\begin{table}[H]
\center
\caption{Dalitz  parameters obtained in fitting the ``bare'' generalized linear sigma model to experiment in a $\chi$-fit [best point at $m[\pi(1300)]=1.38 \pm 0.02$ and $A_3/A_1=28.75^{+1.25}_{-1.75}$] and a $\chi^2$-fit [best point at $m[\pi(1300)]=1.38\pm0.01$ and $A_3/A_1=27.25^{+1.50}_{-0.25}$].   }
\renewcommand{\tabcolsep}{0.4pc} % enlarge column spacing
\renewcommand{\arraystretch}{2.5} % enlarge line spacing
\begin{tabular}{lccccccP{0.5cm}P{1.0cm}P{1.0cm}P{1.0cm}P{1.0cm}P{1.0cm}}
\hline
  Parameter   & $\chi$-fit  & $\chi^2$-fit  \\ \hline
a & $-0.024^{+0.025}_{-0.017} $& $-0.039^{+0.015}_{-0.003}$   \\
b & $0.0001^{+0.0110}_{-0.0034} $ &$ 0.008^{+0.002}_{-0.008}$  \\
d & $-0.029^{+0.012}_{-0.001} $ & $-0.020 ^{+0.003}_{-0.009}$ \\ \hline

\end{tabular}\\[2pt]
\label{T_mmp_bare_DP}
\end{table}

\begin{figure}[H]
\begin{center}
\vskip 1cm
\epsfxsize = 10cm
 \includegraphics[width=8cm,height=6cm]{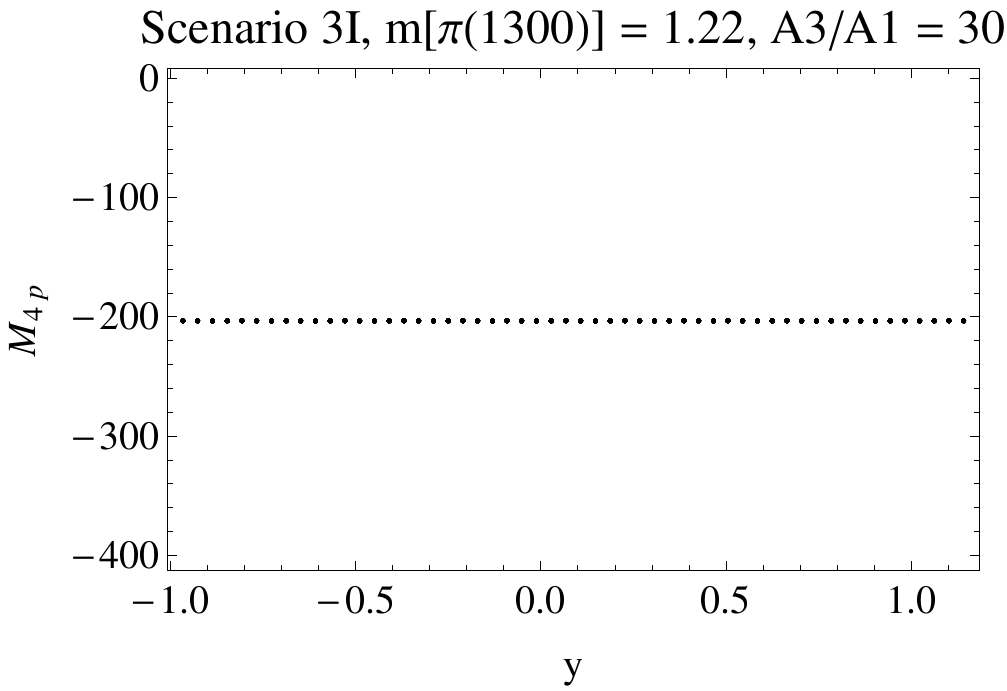}
 \hskip 1cm
\epsfxsize = 10cm
 \includegraphics[width=8cm,height=6cm]{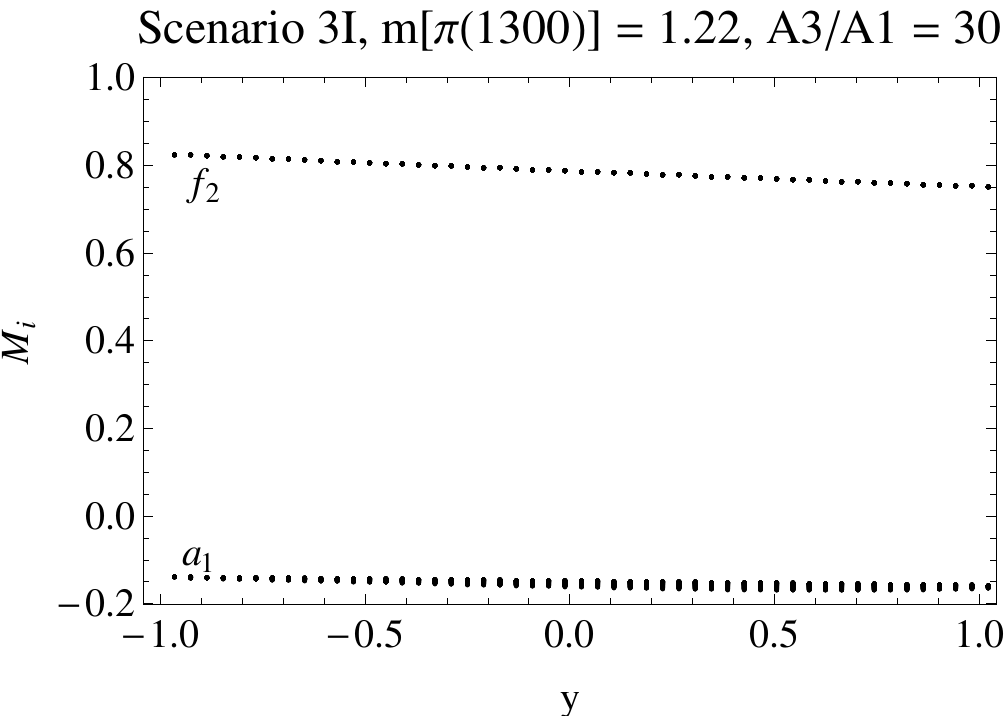}
 \vskip 1cm
\epsfxsize = 10cm
 \includegraphics[width=8cm,height=6cm]{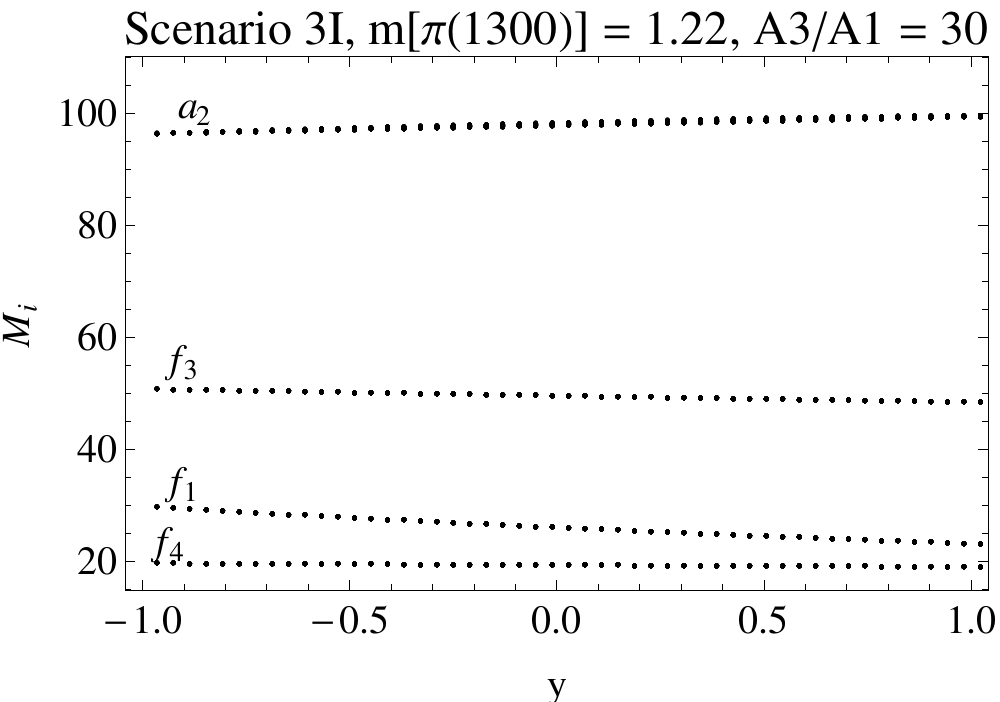}
 \caption{Individual contributions to the ``bare'' decay amplitude of $\eta'\rightarrow\eta\pi\pi$.   The large contribution of the contact terms is balanced with the large contributions of scalar mesons.}
 \label{F_mmp_bare_indv}
\end{center}
\end{figure}

\section{Unitarity corrections}

In principle there are corrections due to the final-state interactions of $\pi\pi$ and $\pi\eta$.    These effects have been studied within the present model in ref. \cite{mixing_pipi} in which the final-state interactions of pions were studied in unitarization of $\pi\pi$ scattering amplitude,  and recently in unitarization of $\pi K$  and $\pi\eta$ scattering amplitudes in \cite{mixing_piK,mixing_pieta}.   In the $\pi\pi$ analysis it is found that the effect of the final-state interactions on the properties of the  sigma meson is large and this manifests itself in the substantial  difference between the ``bare'' sigma mass (Lagrangian mass) and the physical sigma mass found from the  pole of the K-matrix unitarized $I=J=0$,  $\pi\pi$ scattering amplitude  (it is found \cite{mixing_pipi} that the physical mass of sigma is around 480 MeV and its decay width is  450-500 MeV).   On the contrary,  the properties of $a_0(980)$ probed in the $\pi\eta$ scattering analysis \cite{mixing_pieta} does not show a significant shift between the ``bare'' mass of $a_0(980)$ (Lagrangian mass) and that probed in the K-matrix unitarized $\pi\eta$ scattering amplitude.    Since we are investigating the $\eta'\rightarrow\eta\pi\pi$ decay within the same framework of refs. \cite{mixing_pipi,mixing_pieta},  we take the effect of $\pi\pi$ final state-interactions to be the dominant one.

Our main motivation in this work is to learn about the scalar meson mixing patterns, therefore, it is natural for us to approximate the unitarity corrections in a language that is explicitly expressed in terms of the shifts in the scalar mesons properties (from their ``bare'' Lagrangian values to their physical values).   For this purpose,    the K-matrix provides a reasonable tool to both account for unitarity corrections as well as to probe the underlying mixings.
The K-matrix has the advantage  of not introducing any new parameters into the analysis, hence,  allows establishing a direct connection between the ``bare'' Lagrangian properties of scalars and the physical properties of scalars probed in fits to appropriate experimental data.     We follow the prior work presented in \cite{mixing_pipi} in which a detailed analysis of $I=J=0$,  $\pi\pi$ scattering amplitude is given.  The K-matrix unitarized scattering amplitude is given by
\begin{equation}
T_0^0 = {  {T_0^0}^B \over { 1 - i\, {T_0^0}^B} },
\label{T00_unitary}
\end{equation}
where ${T_0^0}^B$ is the ``bare'' scattering amplitude calculated from the Lagrangian.   It is shown in \cite{mixing_pipi} that
\begin{equation}
{T_0^0}^B = T_\alpha + \sum_i  {  {T_\beta^i } \over
{m_{f_i}^2 - s}},
\label{T00B}
\end{equation}
with
\begin{eqnarray}
T_\alpha &=&
{1\over 64 \pi}
\sqrt{1 - {4 m_\pi^2\over s}}\,
\left[-5\, \gamma^{(4)}_{\pi\pi} +
  { 2 \over {p_\pi^2}}\,
   \sum_i \gamma_{f_i\pi\pi}^2\,  {\rm ln} \left(1 +  {{4
p_\pi^2}\over m_{f_i}^2} \right)
\right],
\nonumber \\
T_\beta^i &=&
{3\over 16 \pi}
\sqrt{1 - {4 m_\pi^2\over s}}\, \gamma_{f_i\pi\pi}^2,
\end{eqnarray}
where $p_\pi = \sqrt{s - 4 m_\pi^2} / 2$, the scalar-pseudoscalar-pseudoscalar couplings $\gamma_{f_i\pi\pi}$ are defined in Sec. I, and $\gamma^{(4)}_{\pi\pi}$ is  the pion four-point coupling constant.
It is shown in \cite{LsM} that the K-matrix unitarized amplitude (\ref{T00_unitary}) can be expressed as a constant background and a  sum over simple poles
\begin{equation}
{T_0^0} \approx {\tilde T}_\alpha +
\sum_i  {  {  {\tilde T}_\beta^i } \over {z_i - s}},
\label{T00_unitary_expand}
\end{equation}
where ${\tilde T}_\alpha$ is the constant (complex) background,  the simple poles  $z_i= {\tilde m}_i^2 - i {\tilde m} {\tilde \Gamma}_i$ with ${\tilde m}_i$ and ${\tilde \Gamma}_i$ being interpreted as the physical mass and decay width of the $i$-th isosinglet scalar meson, respectively,  and ${\tilde T}_\beta^i$ are the residues.   Moreover, it can be shown that
\begin{equation}
\left| {\tilde T}_\beta^i \right| \approx {\tilde m}_i {\tilde \Gamma}_i,
\label{residue}
\end{equation}
which resemble the corresponding numerators in ``bare'' amplitude (\ref{T00B}) where
\begin{equation}
\left. T_\beta^i \right|_{s=m_i^2} =  m_i \Gamma_i.
\label{bare_T_beta}
\end{equation}
Comparing (\ref{T00B}), (\ref{T00_unitary_expand}), (\ref{residue}) and (\ref{bare_T_beta}) we see that unitarity corrections effectively shift the isosinglet scalar masses and decay widths
\begin{eqnarray}
m_i &\rightarrow& {\tilde m}_i \nonumber \\
\Gamma_i &\rightarrow& {\tilde \Gamma}_i
\label{Ushift}
\end{eqnarray}

In the decay $\eta'\rightarrow\eta\pi\pi$ the unitarity corrections for the sigma meson are the most important ones.     We account for these corrections by shifting the  ``bare'' mass and decay width to two pions according to (\ref{Ushift}).      The second shift in (\ref{Ushift}) can also be expressed in terms of the shift in coupling constant, i.e.
\begin{eqnarray}
m_{\sigma} &\rightarrow& {\tilde m}_\sigma, \nonumber \\
\gamma_{\sigma\pi\pi} &\rightarrow& {\tilde \gamma}_{\sigma\pi\pi},
\label{mg_shift}
\end{eqnarray}
where ${\tilde m}_\sigma$ and ${\tilde \gamma}_{\sigma\pi\pi}$ are those found from the lowest pole
$z_1 = {\tilde m}_\sigma^2 - i {\tilde m}_\sigma \Gamma_\sigma$ of the scattering amplitude \cite{mixing_pipi}
and since $\Gamma_\sigma  \approx \Gamma [\sigma\rightarrow\pi\pi]$,
\begin{equation}
{\tilde \gamma}_{\sigma\pi\pi} =
\sqrt{
       {16 \pi {\tilde m}_\sigma^2 \Gamma_\sigma}
                         \over
                         {3\sqrt{{\tilde m}_{\sigma}^2 - 4 m_\pi^2}}
     }.
\end{equation}
Recalculating the partial decay width of $\eta'\rightarrow\eta\pi\pi$ (presented in the previous Sec.) with the new substitutions (\ref{mg_shift}) we find the  results displayed in Fig. \ref{F_uc_DW},  showing that the model predictions easily cross into the experimental range.   The same effect can be taken into account for the $f_0(980)$, but that has a negligible effect on the results presented.   On the two dimensional parameter space of the model ($m[\pi(1300)]$, $A_3/A_1$) the point that gives the best agreement with the experimental value of decay width is (1.29 GeV, 29.75) obtained by  minimizing $\chi$ defined in Eq.(\ref{chi}) (as well as  by minimizing the conventional $\chi^2$).   The decay width in this case is
\begin{equation}
\Gamma \left(\eta'\rightarrow\eta\pi\pi\right) =
0.085^{+0.003}_{-0.002}\, {\rm MeV} \hskip 2.0cm {\rm Generalized \hskip 0.15cm linear\hskip 0.15cm sigma \hskip 0.15cm model \hskip 0.15cm (\textbf{unitarized}\: result)}.
\end{equation}
This result is within 1.2\% of experimental data on the decay width.

  The energy dependencies of the normalized decay amplitude squared are plotted in Fig. \ref{F_uc_ED}, and fits to the Dalitz parameters are given in Table \ref{T_uc_DP}.   It is found that the point ($m[\pi(1300)]$, $A_3/A_1$) = (1.38 GeV, 29.75) gives the best agreement with the  experiment.  Although this point and the best point for the decay width (presented above)  are both within the parameter space of the model,  they do not coincide, showing the need for further improvement of this complicated decay and will be further discussed in next section.   The general feature of linear sigma model in which scalar mesons ``conspire'' to balance the large contribution of the contact term can be seen in Fig. \ref{F_uc_indv}.

\begin{figure}[H]
\begin{center}
\vskip 1cm
\epsfxsize = 5cm
 \includegraphics[width=8.5cm,height=5.7cm]{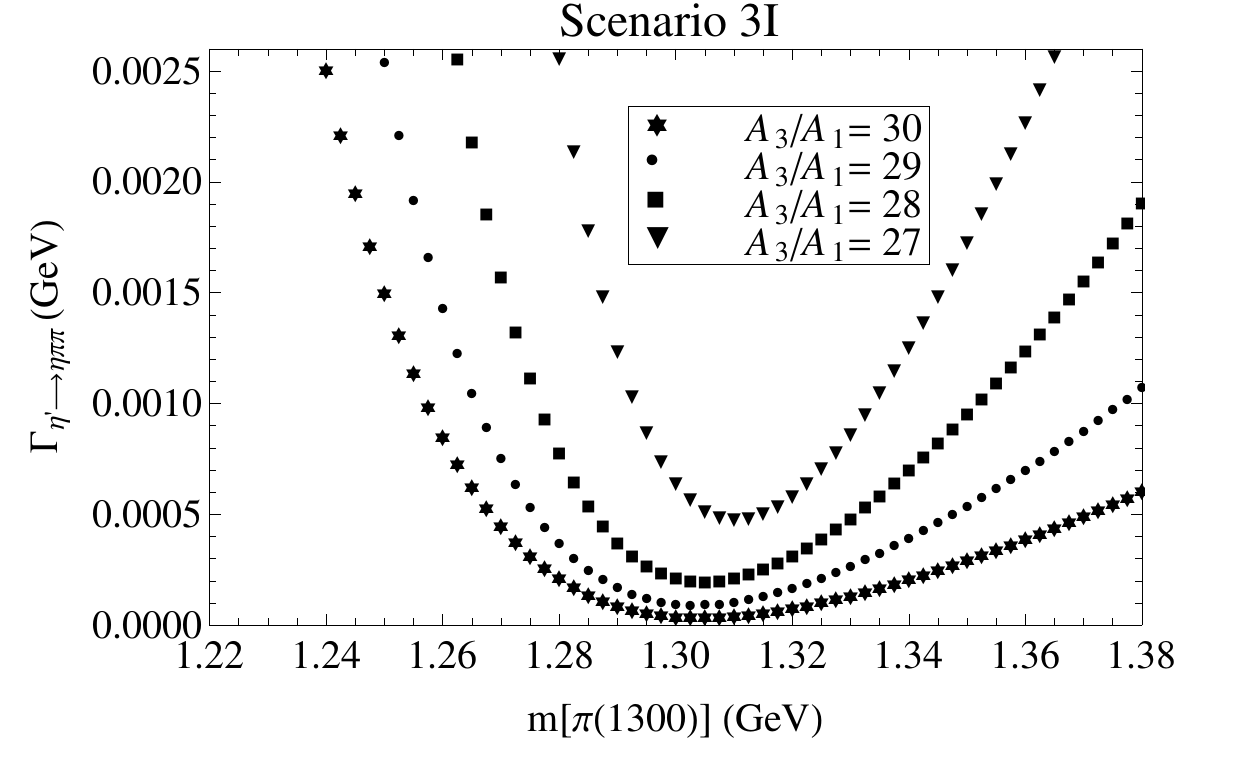}
 \hskip 0.1cm
\epsfxsize = 5cm
 \includegraphics[width=8.5cm,height=5.7cm]{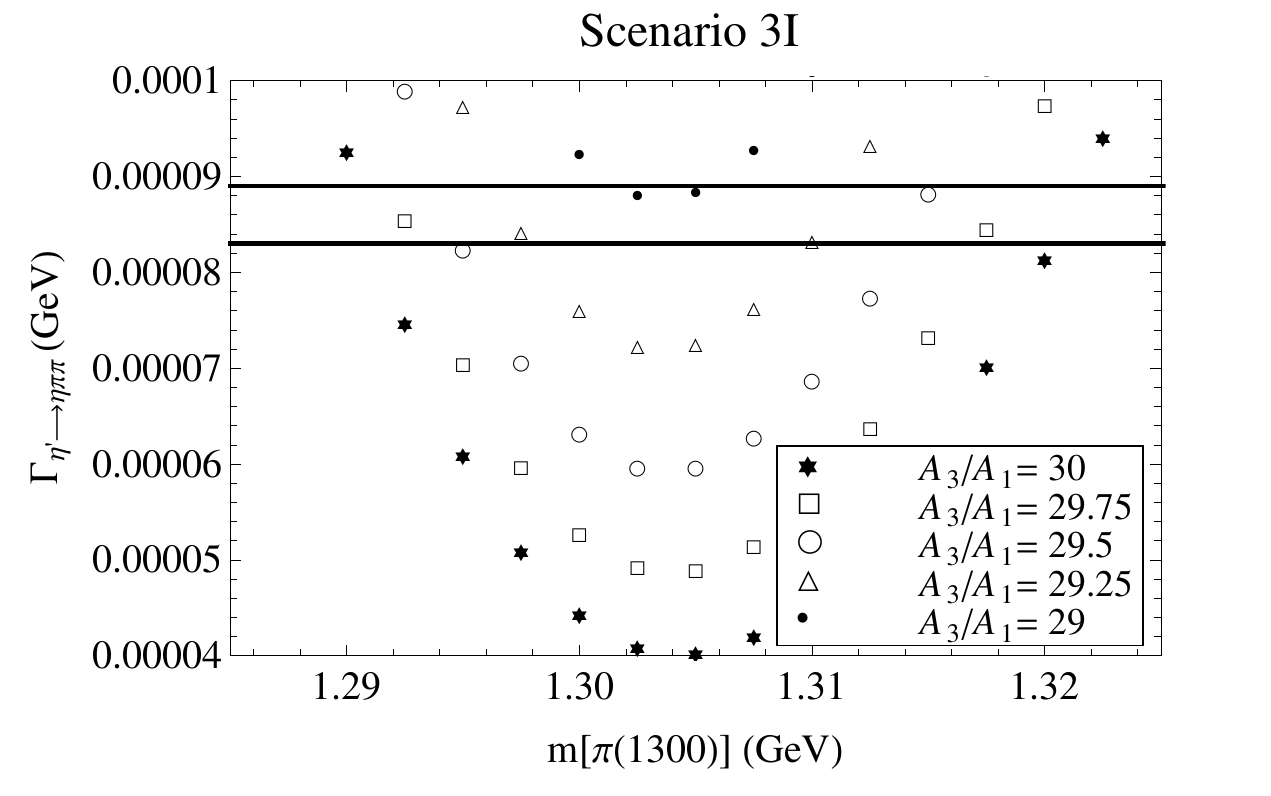}
 \caption{Prediction of the generalized linear sigma  model for the partial decay width of $\eta'\rightarrow\eta\pi\pi$.   The final-state interactions of pions are taken into account by shifting the mass and coupling constant of sigma meson according to  Eq. (\ref{mg_shift}).}
 \label{F_uc_DW}

\end{center}
\end{figure}

\begin{figure}[H]
\begin{center}
\vskip 1cm
\epsfxsize = 5cm
 \includegraphics[width=8.5cm,height=5.7cm]{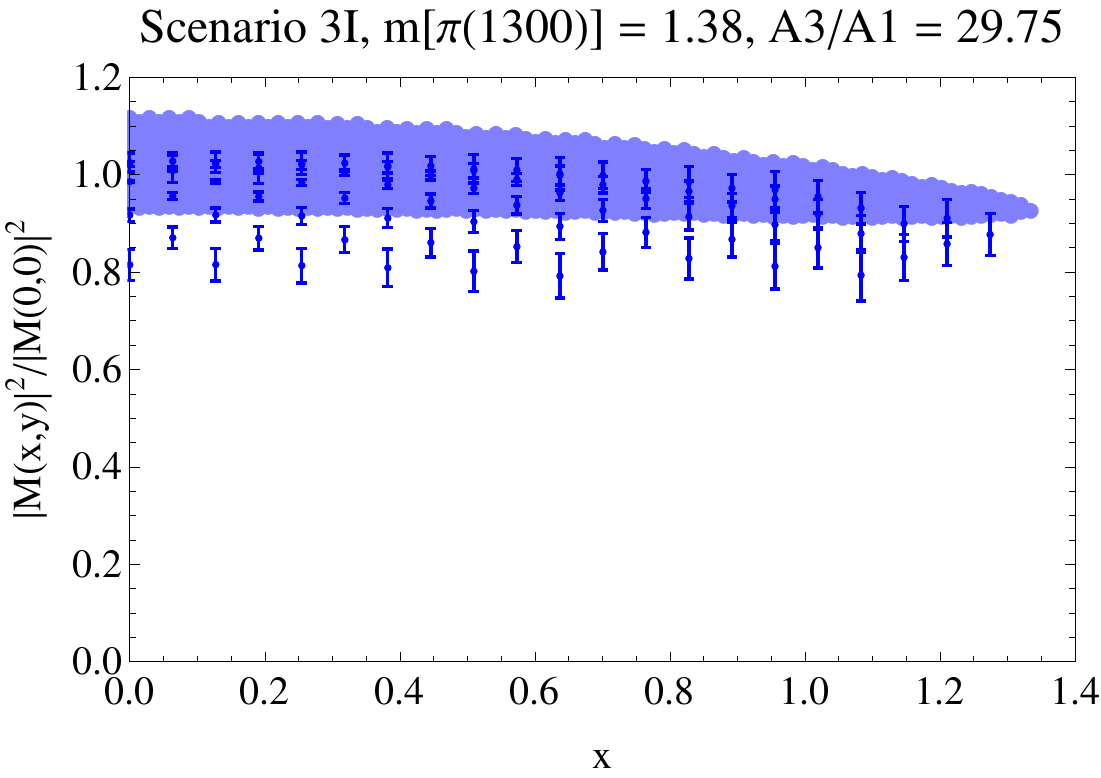}
 \hskip 0.1cm
\epsfxsize = 5cm
 \includegraphics[width=8.5cm,height=5.7cm]{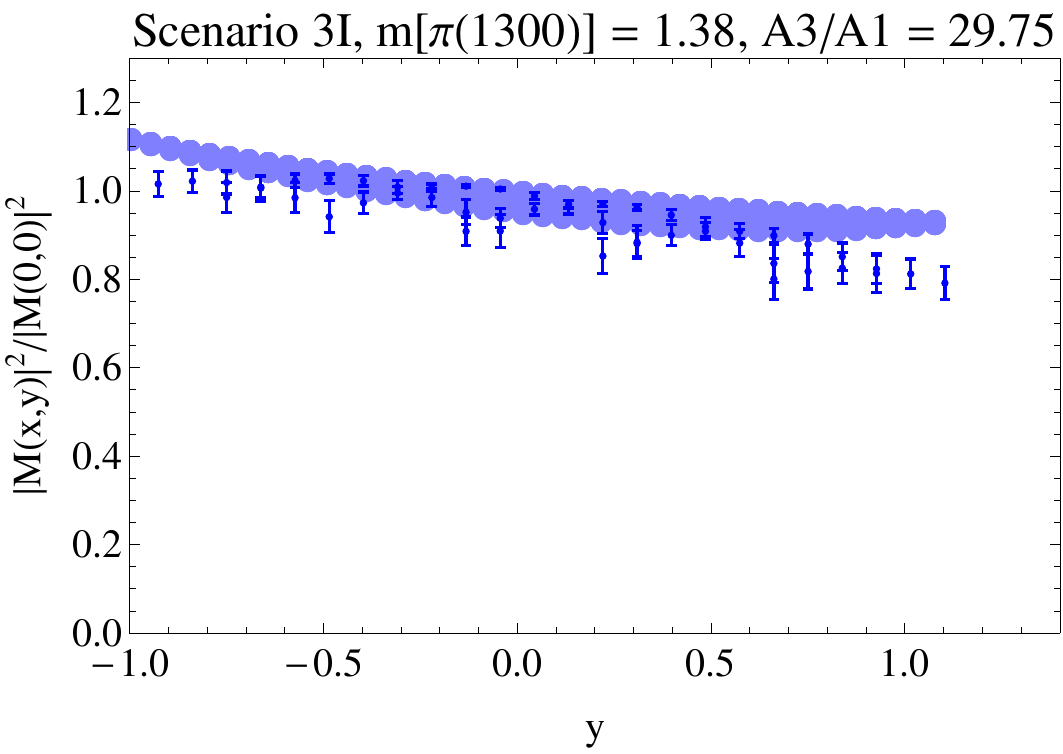}
 \caption{Projects of the normalized decay amplitude squared  onto planes containing $x$ and $y$ parameters (shaded regions) are compared with the experimental data (error bars).      The final-state interactions of pions are taken into account by shifting the mass and coupling constant of sigma meson according to  Eq. (\ref{mg_shift}).}
 \label{F_uc_ED}
\end{center}
\end{figure}

\begin{table}[H]
\center
\caption{
Dalitz parameters in unitarized generalized linear sigma model from fits (both $\chi$-fit as well as $\chi^2$ fit) to experiment. The presented results are the closest agreement with experiment that occur at point ($m[\pi(1300)]$,$A_3/A_1$) = (1.38$\pm0.01$ GeV, 29.75$\pm 0.25$).}
\renewcommand{\tabcolsep}{0.4pc} % enlarge column spacing
\renewcommand{\arraystretch}{2.5} % enlarge line spacing
\begin{tabular}{lccccccP{1.0cm}P{1.0cm}P{1.0cm}P{1.0cm}P{1.0cm}P{1.0cm}}
\hline
 Parameter & $\chi$-fit  & $\chi^2$-fit  \\ \hline
a &  $-0.079^{+0.019}_{-0.021}$ & $-0.079\pm0.019$  \\
b & $ 0.024^{+0.010}_{-0.009}$ & $0.024\pm0.009$ \\
d &  $-0.028\pm0.001$ & $-0.028\pm0.001$ \\ \hline

\end{tabular}\\[2pt]
\label{T_uc_DP}
\end{table}

\begin{figure}[H]
\begin{center}
\vskip 1cm
\epsfxsize = 5cm
 \includegraphics[width=8.5cm,height=5.7cm]{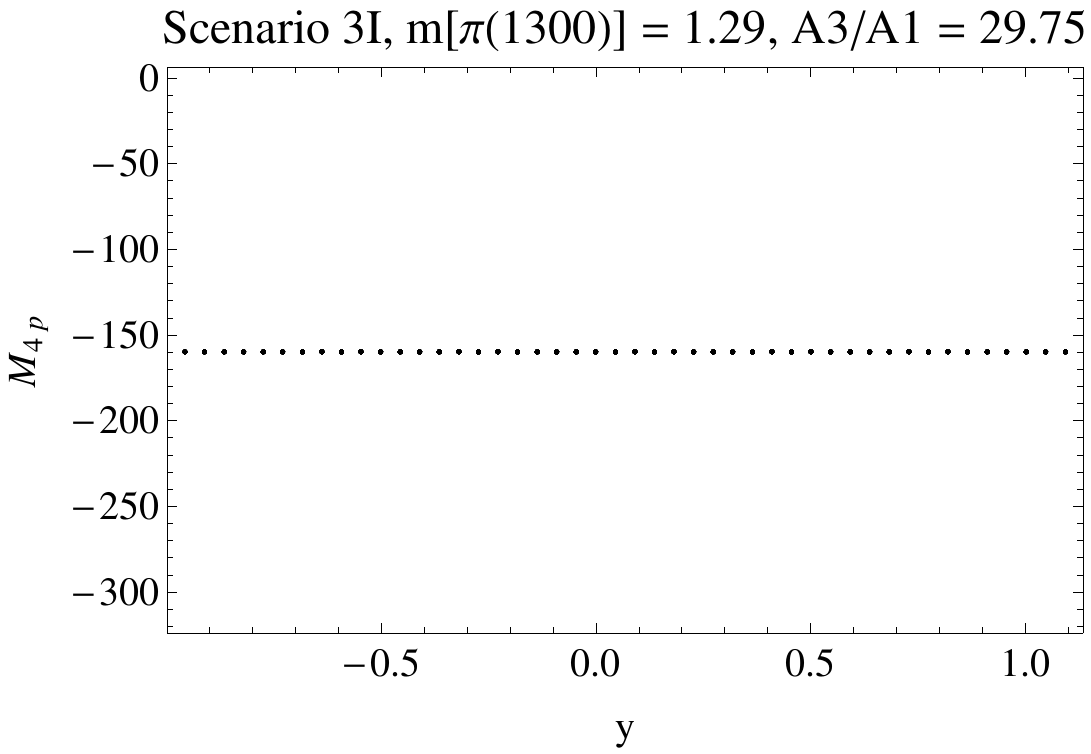}
 \hskip 0.1cm
\epsfxsize = 5cm
 \includegraphics[width=8.5cm,height=5.7cm]{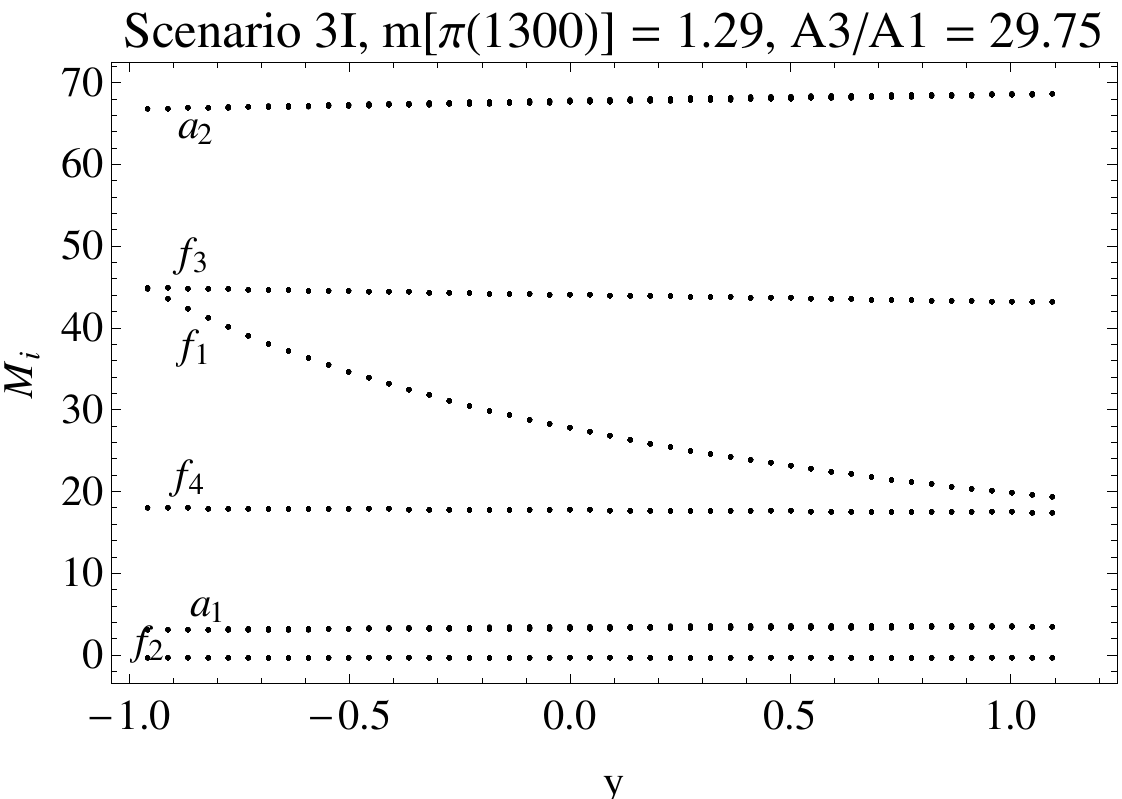}
 \caption{Individual contributions to the decay amplitude.  The final-state interactions of pions are taken into account by shifting the mass and coupling constant of sigma meson according to  Eq. (\ref{mg_shift}).}
\label{F_uc_indv}
\end{center}
\end{figure}

Similarly, we can estimate the final state interactions for the single nonet model of Sec. II.   We find that the decay width improves
\begin{equation}
\Gamma \left(\eta'\rightarrow\eta\pi\pi\right) = 0.35 \pm 0.01 \hskip 0.15cm {\rm MeV} \hskip 2.0cm {\rm Single\hskip 0.15cm nonet \hskip 0.15cm (\textbf{unitarized } \: result)}.
\end{equation}
However, the energy dependencies worsen in this case (Fig. \ref{F_ns_EDU} and Table \ref{T_sn_EDU}).   This shows that the effect of unitarity corrections alone are not sufficient and there seems to be the effect of mixing that should be taken into account.

\begin{figure}[H]
\begin{center}
\vskip 1cm
\epsfxsize = 7.5cm
 \includegraphics[width=8cm,height=5.8cm]{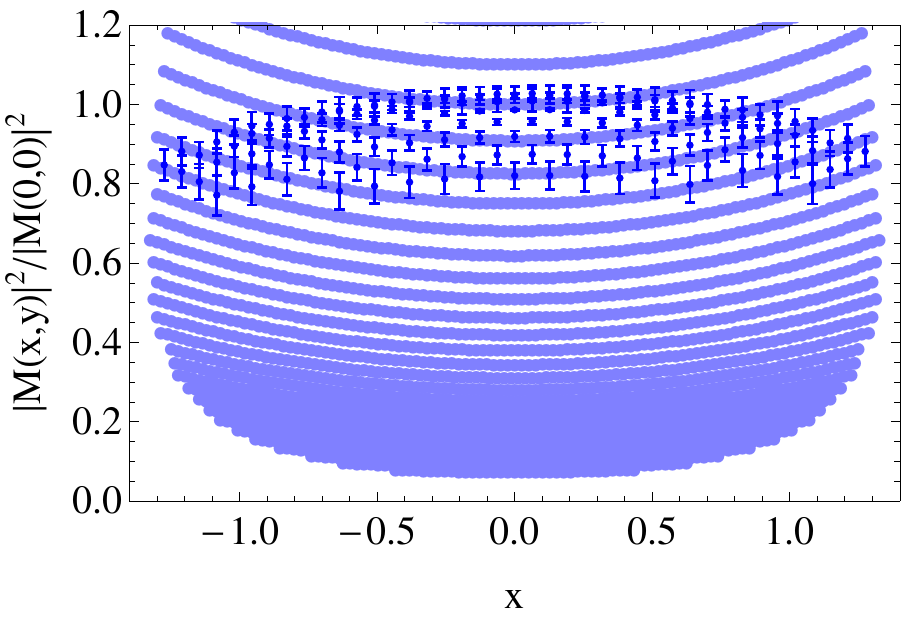}
\hskip 1cm
\epsfxsize = 7.5cm
 \includegraphics[width=8cm,height=6cm]{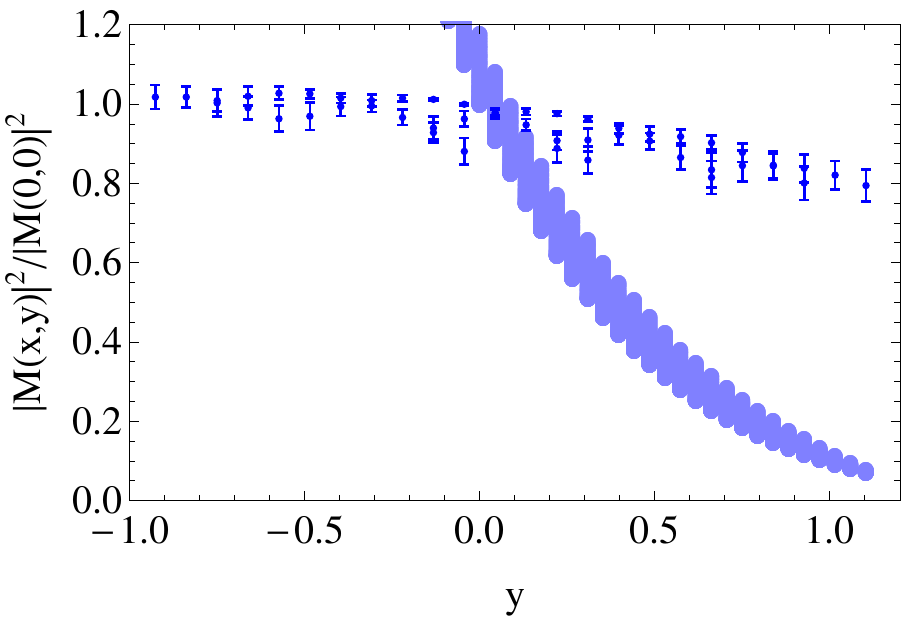}
 \caption{ Projections of $ |\hat{M}|^2=|M(x,y)|^2/|M(0,0)|^2$ onto the $ y - |\hat{M}|^2$ and
 $x - |\hat{M}|^2  $ planes (unitarized single nonet model).   While the effect of final state interactions improves the partial decay width predicted by  the single nonet model,   the energy dependencies worsen.   This shows that there is more into this decay that just the effect of final-state interactions.}
\end{center}
\label{F_ns_EDU}
\end{figure}

\begin{table}[H]
\centering
\caption{Predicted decay
 parameters in the unitarized single nonet approach of ref. \cite{LsM}.}
\renewcommand{\tabcolsep}{0.4pc} % enlarge column spacing
\renewcommand{\arraystretch}{2.5} % enlarge line spacing
\begin{tabular}{lcP{1.0cm}P{1.0cm}}
\hline
 Parameter       &  single nonet model  \\ \hline
a &  $-2.17\pm 0.01$  \\
b& $2.37 \pm 0.01$\\
d & $0.11 \pm 0.01$ \\ \hline
\end{tabular}\\[2pt]
\label{T_sn_EDU}
\end{table}

\begin{figure}[H]
\begin{center}
\vskip 1cm
\epsfxsize = 7.5cm
 \includegraphics[width=7cm,height=5cm]{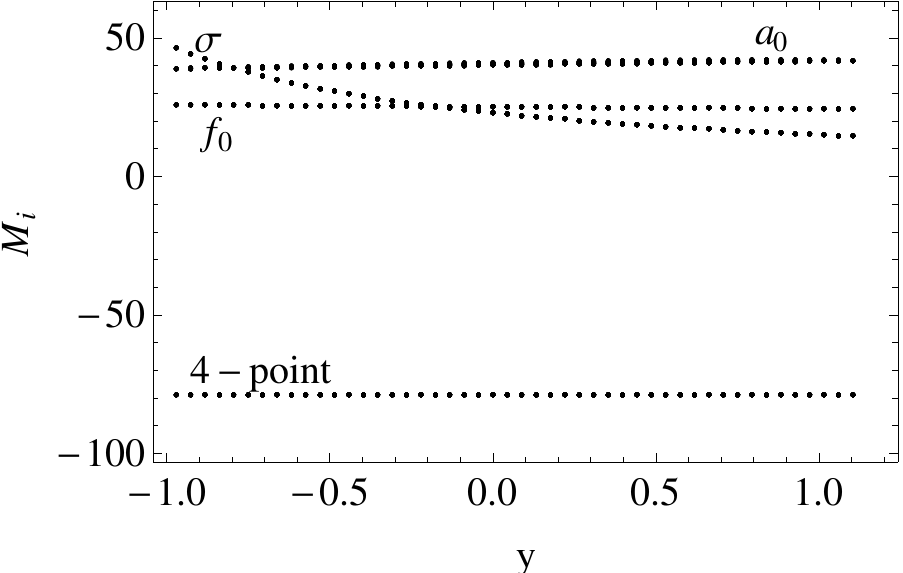}
 \caption{Individual contributions to the $\eta'\rightarrow\eta\pi\pi$ decay amplitude in unitarized single nonet model.  The large contribution of contact term $M_{4p}$ is balanced with the contributions of scalars.  Unitarity corrections are taken into account.}
\end{center}
\end{figure}

%%%%%%%%%%%%%%%%%%%%%%%%%%%%%%%%%%%%%%%%%%%%%%%%

%%%%%%%%%%%%%%%%%%%%%%%%%%%%%%%%%%%%%%%%%%%%%%%%

 \section{Concluding discussion}
 In this work, we examined the $\eta'\rightarrow\eta\pi\pi$ decay as a probe of scalar mesons substructure and mixing patterns within a generalized linear sigma model of low-energy QCD that is formulated in terms of two scalar meson nonets  and two pseudoscalar meson nonets (a two- and a four-quark nonet for each spin).   We first showed that the single nonet model of ref. \cite{LsM}, despite its considerable success in describing $\pi\pi$, $\pi K$ and $\pi \eta$ low-energy scatterings,  gives inaccurate predictions for the partial decay width of $\eta'\rightarrow\eta\pi\pi$ as well as the energy dependencies of its normalized decay amplitude squared.   Since this decay involves $\eta$ and $\eta'$ as well as intermediate scalar mesons and that these states are known to have nontrivial  mixings with states with the same quantum numbers above 1 GeV,    and since such mixings have been previously \cite{global} given important insights into the physical properties of both scalar as well as pseudoscalar mesons,  in this work we explored the effect of these mixings on this decay.    We investigated whether the inclusion of mixing can have a tangible effect and whether such effects improve the predictions of the single nonet linear sigma model for this decay.
 We showed that inclusion of the underlying mixings (even without unitarity corrections) considerably improves the partial decay width prediction as well as the energy dependencies of the normalized decay amplitude squared.   We then showed that inclusion of the final state interaction of pions further improves the predictions and brings the partial decay width to within 1.2\% of its experimental value, and considerably improves the predictions for the Dalitz parameters.
Our findings are summarized in several tables in this final section.    Table \ref{T_res_overall} gives our results for the partial decay width and Dalitz parameters in single nonet linear sigma model as well as its generalized version,  both with and without accounting for the final state interaction of pions.

We note that while the predictions of Dalitz parameters are improved in the fourth column of  Table \ref{T_res_overall}, they are still far from their experimental values.   However, we further note that since the Dalitz variables $X$ and $Y$ are relatively small over much of their domain,  the difference in the normalized decay amplitude itself is not that large for most of the domain.  To illustrate this,   Fig. \ref{F_XY_domain}  zooms in on the $X$, $Y$ domain in four steps.   Inside each ``loop'' the closeness of the model prediction for the
energy dependence of the normalized decay amplitude squared is measured with the quantity
\begin{equation}
{\bar \chi}_{{\cal M}^2} = {1\over N}\, \sum_i^N
{
{ \left|  \left( {\cal M}^2\right)^{\rm exp.}(X_i,Y_i) - \left( {\cal M}^2\right)^{\rm theo.}(X_i,Y_i) \right|}
 \over {\left({\cal M}^2\right)^{\rm exp.}(X_i,Y_i)}
 },
 \label{chi_ave}
\end{equation}
where the normalized decay matrix element is defined in Eq.(\ref{M_ren}) and the averaged experimental data in Table \ref{T_abd_exp}.   The results are presented in Table \ref{T_chi_ave_zoom} and clearly show an averaged agreement with experiment (for the two cases that the best energy dependencies are obtained is around 6\%),  despite the much less agreement on Dalitz coefficients displayed in Table \ref{T_res_overall}.

The dependence of the results on the choice of points in the two dimensional parameter space  $m[\pi(1300)]$ and $A_3/A_1$  are summarized in Tables \ref{T_sum_mmp_bare} and \ref{T_sum_mmp_unitarized}. The fact that the best points for the partial decay width and energy dependencies of the normalized decay amplitude squared do not occur at the same point, can be interpreted as an estimate of our theoretical uncertainty.  At the present order of accuracy of this model, we have ignored effects such as terms in the potential with higher than eight quark and antiquark lines as well as the  scalar and pseudoscalar glueballs.  Both of these are expected to have some effects on the results.   Since the U(1)$_{\rm A}$ anomaly plays an important role in the eta sector, we have made an initial investigation of the effect of the higher order U(1)$_{\rm A}$ breaking term (which are related to higher order instanton contributions at the quark level) and have observed that this term  improves the
picture by bringing the two points in the parameter space closer together.   This is quite encouraging and will be presented in detail in a separate work \cite{N10U1A}.   It is also interesting to further apply the present model to study the isospin violating $\eta\rightarrow 3 \pi$ decay \cite{SU,e3p}, and to examine the effect of various unitarization methods \cite{unitarization_methods}.

\begin{table}[H]
 \begin{center}
\caption{Comparing with experiment the predictions by the single nonet linear sigma model (first two columns) and those by the generalized linear sigma model (the last two columns) for the decay width and the Dalitz parameters of $\eta'\rightarrow \eta\pi\pi$ decay.  The goodness of the predictions are measured by the smallness of the parameter $\chi$ defined for a generic quantity $q$ as $\chi_q =  \left| (q^{\rm exp.} - q^{\rm theo.}) / q^{\rm exp.} \right|$ (i.e. $\chi_q\times 100$ gives the percent difference between theory and experiment).
The predictions of the generalized linear sigma model depend on the choice of points in its two dimensional parameter space ($m[\pi(1300)]$, $A_3/A_1$):  In the third column, the minimum of  $\chi_{\Gamma}$ and of $\chi_{\rm Dalitz} = \chi_a +\chi_b + \chi_d$
occur at point  (1.22 GeV, 30.00) and  at point (1.38 GeV, 28.75), respectively,  whereas in the fourth column, the minimum  of $\chi_{\Gamma}$ and of $\chi_{\rm Dalitz}$ occur at (1.29 GeV, 29.75) and at (1.38 GeV, 29.75), respectively.   Clearly, the shortcomings of the single nonet linear sigma model of ref. \cite{LsM} can be seen in the first two columns:  the decay width is several times larger than the experimental value and the unitarity corrections do not improve the situation and in fact worsen the Dalitz parameter predictions.  On the other hand,  the generalized linear sigma model significantly improves the predictions and gives the decay width in the unitarized version to 1.2\% of the experimental value and also improves the Dalitz parameter predictions.}
$%
\begin{tabular}
[c]{||c|c|c|c|c||}\hline\hline
& single nonet  & single nonet  & MM'& MM'   \\
& (Bare)  &(Unitarized)& (Bare)& (Unitarized)\\ \hline
$\chi_{\Gamma}$ & $6.09$   &  $3.07$  & $0.74$   & $0.012$    \\
$\chi_a$ & $0.21$   &  $22.08$  & $0.74$   & $0.16$    \\
$\chi_b$ & $0.99$   &  $30.02$  & $1.0$   & $1.29$    \\
$\chi_d$ & $0.16$   &  $2.4$  & $0.61$   & $0.63$    \\ \hline
$\chi_{\rm{total}}$ & $7.45$   &  $57.57$  & $3.10$   & $2.09$    \\ \hline\hline
\end{tabular}
$
\end{center}
\label{T_res_overall}
\end{table}

\begin{figure}[H]
\begin{center}
\vskip 1cm
\epsfxsize = 7.5cm
 \includegraphics[width=7cm]{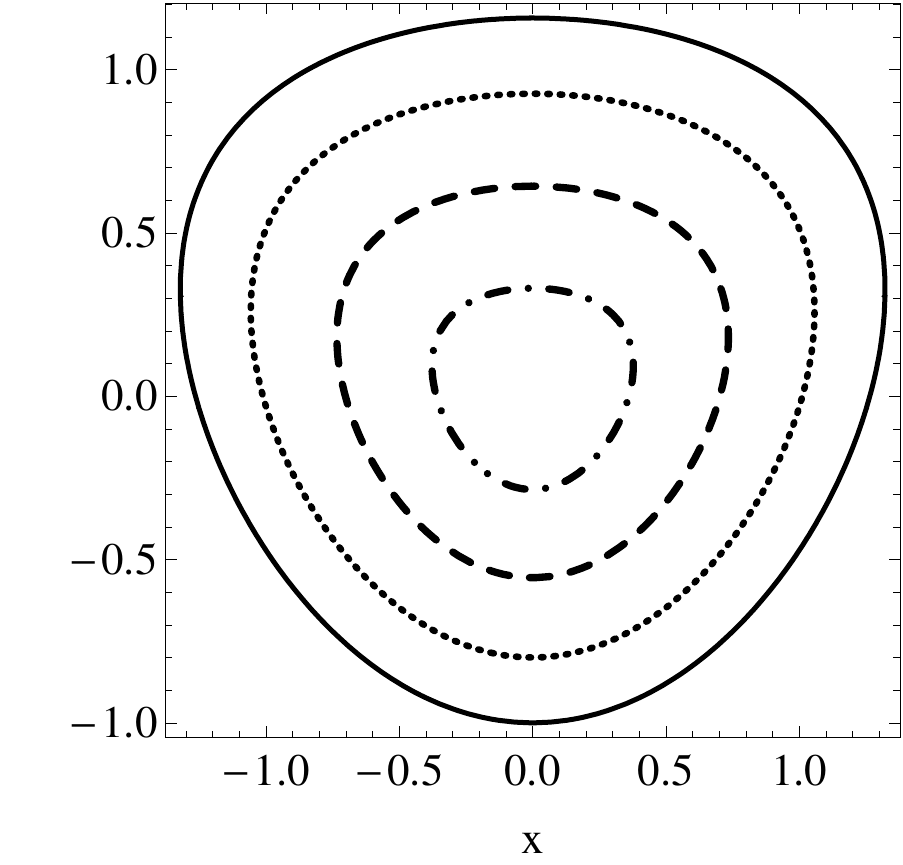}
 \caption{The breakdown of $XY$ domain into four subregions (``loops'').
         }
\label{F_XY_domain}
\end{center}
\end{figure}

\begin{table}[H]
\centering
\caption{Displayed numbers in the second to last columns  are  ${\bar \chi}_{{\cal M}^2}$ [defined in Eq. (\ref{chi_ave})] over the four ``loops''  of Fig. \ref{F_XY_domain} [see Eq.(\ref{M_ren})].
The predictions of the generalized linear sigma model depend on the choice of points in its two dimensional parameter space ($m[\pi(1300)]$, $A_3/A_1$).  The displayed values  of  $m[\pi(1300)]$ and $A_3/A_1$   give the best result  for partial decay width without/with
the final state interactions (first/third rows); and the best result for the energy dependencies of the normalized decay amplitude squared without/with the final-state interactions (second/fourth row).}
\renewcommand{\tabcolsep}{0.4pc} % enlarge column spacing
\renewcommand{\arraystretch}{1.5} % enlarge line spacing
\begin{tabular}{cP{1.4cm}||P{1.4cm}P{1.4cm}P{1.4cm}P{1.4cm}P{1.4cm}P{1.4cm}|}
\hline
$m[\pi(1300)]\rm{(GeV)}$& $A_3/A_1$ & Dotted-Dashed  & Dashed & Dotted & Solid     \\ \hline
1.22  &  30.00   & $0.14$       &  $0.27$          & $0.41$           & $0.54$        \\
1.38  &   28.75  & $0.01$       &  $0.03$          & $0.04$           & $0.07$         \\
1.29  &   29.75  & $1.0$        &  $2.2$           & $4.7$            & $7.6$        \\
1.38  &  29.75   & $0.005$      &  $0.02$          & $0.04$           & $0.06$        \\ \hline
\end{tabular}\\[2pt]
\label{T_chi_ave_zoom}
\end{table}

 \begin{table}[H]
 \begin{center}
\caption{Dependency on the choices of  $m[\pi(1300)]$, $A_3/A_1$ of the  ``bare'' model predictions (without the effect of unitarity corrections due to the final state interaction of pions).   In the first to last columns,  respectively,  the values of these two parameters are 1.30 GeV, 29.40  (best model prediction for the eta masses);   1.22 GeV, 30.00  (best prediction for the decay width)  and  1.38 GeV, 28.75 (best prediction for the energy dependencies).   In each column the targeted quantities are highlighted in bold and their closeness to experimental data  is measured with their corresponding $\chi$.
}
$%
\begin{tabular}
[c]{||c|c|c|c||}\hline\hline
MM'    &  $(\chi_{\rm{min}})_{\rm{mass}}$ & $(\chi_{\rm{min}})_{\Gamma}$ & $(\chi_{\rm{min}})_{\rm{E.D.}}$  \\
(Bare)&    $=0.14\% $       & $=74\%$            & $=235\%$\\ \hline

$m[\pi(1300)]$              & $1300$              &  $1220$             & $1380$      \\
$A_3/A_1$                   & $29.40$              &  $30.00$              & $28.75$      \\
$m_{\eta_{1}}(\rm{MeV})$    & \boldmath{$547$ }   &  $554$             & $539$    \\
$m_{\eta_{2}}(\rm{MeV})$    & \boldmath{$959$}    &  $979$             & $947$       \\
$m_{\eta_{3}}(\rm{MeV})$    & \boldmath{$1294$}   &  $1229$            & $1364$    \\
$m_{\eta_{4}}(\rm{MeV})$    & \boldmath{$1756$}   &  $1788$            & $1710$     \\
$\Gamma(\rm{MeV})$          & $0.42$              &  \boldmath{$0.15$}  & $0.97$     \\
$a$                         & $0.24$              &  $0.88$             & \boldmath{$-0.024$}    \\
$b$                         & $-0.026$            &  $0.07$             & \boldmath{$0.0001$ }    \\
$d$                         & $-0.037$            &  $-0.07$            & \boldmath{$-0.029$ }   \\\hline\hline
\end{tabular}
$
\label{T_sum_mmp_bare}
\end{center}

\end{table}

 \begin{table}[H]
 \begin{center}
\caption{Dependency on the choices of $m[\pi(1300)]$, $A_3/A_1$  of the  ``unitarized'' model predictions (with  the effect of the final state interaction of pions).   In the first to last columns,  respectively,  the values of these two parameters are 1.3 GeV, 29.40 (best model prediction for the eta masses);    1.29 GeV, 29.75 (best prediction for the decay width)  and 1.38 GeV, 29.75(best prediction for the energy dependencies).   In each column the targeted quantities are highlighted in bold and their closeness to experimental data  is measured with their corresponding $\chi$.  }

$%
\begin{tabular}
[c]{||c|c|c|c||}\hline\hline
MM' &  $(\chi_{\rm{min}})_{\rm{mass}}$ & $(\chi_{\rm{min}})_{\Gamma}$ & $(\chi_{\rm{min}})_{\rm{E.D.}}$  \\
(Unitarized)                &$=0.14\% $           & $=1.2\%$              & $=207\%$\\ \hline

$m[\pi(1300)]$              & $1300$              &  $1290$             & $1380$      \\
$A_3/A_1$                   & $29.40$              &  $29.75$              & $29.75$      \\
$m_{\eta_{1}}(\rm{MeV})$    & \boldmath{$547$}    &  $550$                & $544$        \\
$m_{\eta_{2}}(\rm{MeV})$    & \boldmath{$959$}    &  $956$                & $936$        \\
$m_{\eta_{3}}(\rm{MeV})$    & \boldmath{$1294$}   &  $1285$               & $1364$       \\
$m_{\eta_{4}}(\rm{MeV})$    & \boldmath{$1756$}   &  $1762$               & $1715$       \\
$\Gamma (\rm{MeV})$         & $0.072$             &  \boldmath{$0.085$}   & $0.62$        \\
$a$                         & $10.84$             &  $-9.48$              & \boldmath{$-0.079$}    \\
$b$                         & $24.72$             &  $26.2$               & \boldmath{$0.024$}     \\
$d$                         & $-0.29$             &  $0.22$               & \boldmath{$-0.028$}    \\\hline\hline
\end{tabular}
$
\label{T_sum_mmp_unitarized}
\end{center}

\end{table}

\newpage
\appendix

\section{Coupling constants in the single-nonet model}
The rotation matrices are
\begin{equation}
\left[
\begin{array}{c}
\pi^0 \\
\eta\\
\eta'
\end{array}
\right] =
R_\phi(\theta_p)
\left[
\begin{array}{c}
\phi_1^1 \\
\phi_2^2\\
\phi_3^3
\end{array}
\right]
=
\left[
\begin{array}{ccc}
{1\over \sqrt{2}} & -{1\over \sqrt{2}}      &   0\\
{a_p\over \sqrt{2}} & {a_p\over \sqrt{2}}  &  -b_p \\
{b_p\over \sqrt{2}} & {b_p\over \sqrt{2}}  &  a_p
\end{array}
\right]
\left[
\begin{array}{c}
\phi_1^1 \\
\phi_2^2\\
\phi_3^3
\end{array}
\right],
\label{Rp}
\end{equation}
with $a_p = ({{\rm cos} \theta_p - \sqrt{2} {\rm sin} \theta_p})/ {\sqrt{3}}$,
$b_p = ({\rm sin} \theta_p + \sqrt{2} {\rm cos} \theta_p) / {\sqrt{3}}$, where  $\theta_p$ is the
pseudoscalar (octet-singlet) mixing angle.  Similarly,
 \begin{equation}
\left[
\begin{array}{c}
a_0^0 \\
\sigma\\
f_0
\end{array}
\right] =
R_s(\theta_s)
\left[
\begin{array}{c}
S_1^1 \\
S_2^2\\
S_3^3
\end{array}
\right]
=
\left[
\begin{array}{ccc}
{1\over \sqrt{2}} & -{1\over \sqrt{2}}      &   0\\
{a_s\over \sqrt{2}} & {a_s\over \sqrt{2}}  &  -b_s \\
{b_s\over \sqrt{2}} & {b_s\over \sqrt{2}}  &  a_s
\end{array}
\right]
\left[
\begin{array}{c}
S_1^1 \\
S_2^2\\
S_3^3
\end{array}
\right],
\label{Rs}
\end{equation}
with $a_s = ({\rm cos} \theta_s - \sqrt{2} {\rm sin} \theta_s)/\sqrt{3}$, $b_s = ({\rm sin} \theta_s + \sqrt{2}{\rm cos} \theta_s
  / \sqrt{3}$ where $\theta_s$ is the
scalar (octet-singlet) mixing
angle.

The coupling constants are:

\begin{eqnarray}
\gamma^{(4)}=&&-\frac{1}{(2 F_K-F_{\pi }) F_{\pi }^3}a_p b_p \Big[-4 F_K^2 \Big(5 \sqrt{2} a_p b_p (m_{\eta }^2-m_{\eta '}^2)-84
F_{\pi } V_4\Big)\nonumber \\
&&+F_{\pi }^2 \Big(4 m^2_{\rm \tiny BARE}(a_0)+2 m^2_{\rm \tiny BARE}(\sigma)  a_s^2+2 \sqrt{2} (m^2_{\rm \tiny BARE}(f_0)-m_{\rm \tiny BARE}^2(\sigma) a_s b_s\nonumber \\
&&+2 m^2_{\rm \tiny BARE}(f_0) b_s^2-6 a_p^2 m_{\eta }^2-7 \sqrt{2} a_p b_p m_{\eta }^2+7 \sqrt{2} a_p b_p m_{\eta '}^2-6 b_p^2
m_{\eta '}^2+84 F_{\pi } V_4\Big)\nonumber \\
&&-4 F_K F_{\pi } \Big(2 m^2_{\rm \tiny BARE}(a_0)+m^2_{\rm \tiny BARE}(\sigma)  a_s^2+m^2_{\rm \tiny BARE}(f_0)
b_s^2-3 a_p^2 m_{\eta }^2-5 \sqrt{2} a_p b_p m_{\eta }^2\nonumber\\
&&+5 \sqrt{2} a_p b_p m_{\eta '}^2-3 b_p^2 m_{\eta '}^2+84 F_{\pi } V_4\Big)\Big],
\end{eqnarray}

\begin{eqnarray}
\gamma_{a\pi\eta} = \frac{\sqrt{2}}{F_\pi}a_p \left( m^2_{\rm \tiny BARE} (a_0) - m_\eta^2 \right),
\quad \gamma_{a\pi\eta^\prime} = \frac{\sqrt{2}}{F_\pi}b_p \left( m^2_{\rm
\tiny BARE} (a_0)  -
m_{\eta^\prime}^2 \right), \\ \nonumber
\gamma_{\sigma\pi\pi} = \frac{1}{F_\pi}a_s \left( m^2_{\rm \tiny BARE} (\sigma) - m_\pi^2
\right), \quad
\gamma_{f_0\pi\pi} = \frac{1 }{F_\pi}b_s\left( m^2_{\rm \tiny BARE} (f_0) - m_\pi^2
\right). \\
\label{trilinear-couplings}
\nonumber
\end{eqnarray}
For $\eta '$ decay we will also need:

\begin{eqnarray}
\gamma_{\sigma \eta \eta'}=&&-\frac{1}{(2 F_K-F_{\pi }) F_{\pi }^2}a_p b_p \Bigg[-2 m^2_{\rm \tiny BARE}(\sigma) a_s^2 b_s F_{\pi }^2+\sqrt{2}
m^2_{\rm \tiny BARE}(\sigma) a_s^3 F_{\pi } (-2 F_K+F_{\pi })\nonumber\\
&&+2 b_s F_{\pi } \Big(-m^2_{\rm \tiny BARE}(\sigma) b_s^2
F_{\pi }+b_p^2 F_{\pi } m_{\eta }^2+a_p^2 F_{\pi } m_{\eta '}^2-\sqrt{2} a_p b_p (2 F_K-F_{\pi }) (m_{\eta }^2-m_{\eta '}^2)\Big)\nonumber\\
&&+a_s
\bigg(\sqrt{2} m^2_{\rm \tiny BARE}(\sigma) b_s^2 F_{\pi } (-2 F_K+F_{\pi })-a_p b_p (4 F_K^2-4 F_K F_{\pi }+3 F_{\pi }^2)
(m_{\eta }^2-m_{\eta '}^2)\nonumber\\
&&+\sqrt{2} a_p^2 (2 F_K-F_{\pi }) F_{\pi } (2 m_{\eta }^2-m_{\eta '}^2)+\sqrt{2} (2
F_K-F_{\pi }) F_{\pi } \Big(-b_p^2 (m_{\eta }^2-2 m_{\eta '}^2)+18 (2 F_K-F_{\pi }) V_4\Big)\bigg)\Bigg],
\end{eqnarray}

\begin{eqnarray}
\gamma_{f_0 \eta \eta'}=&&-\frac{1}{(2 F_K-F_{\pi }) F_{\pi }^2}a_p b_p \Big[2 m^2_{\rm \tiny BARE}(f_0) a_s^3 F_{\pi }^2+\sqrt{2} m^2_{\rm \tiny BARE}(f_0)
a_s^2 b_s F_{\pi } (-2 F_K+F_{\pi })\nonumber\\
&&+2 a_s F_{\pi } \Big(m^2_{\rm \tiny BARE}(f_0) b_s^2 F_{\pi }-b_p^2 F_{\pi } m_{\eta }^2-a_p^2
F_{\pi } m_{\eta '}^2+\sqrt{2} a_p b_p (2 F_K-F_{\pi }) (m_{\eta }^2-m_{\eta '}^2)\Big)\nonumber\\
&&+b_s \Big(\sqrt{2} m^2_{\rm \tiny BARE}(f_0)
b_s^2 F_{\pi } (-2 F_K+F_{\pi })-a_p b_p (4 F_K^2-4 F_K F_{\pi }+3 F_{\pi }^2) (m_{\eta }^2-m_{\eta '}^2)\nonumber\\
&&+\sqrt{2}
a_p^2 (2 F_K-F_{\pi }) F_{\pi } (2 m_{\eta }^2-m_{\eta '}^2)+\sqrt{2} (2 F_K-F_{\pi }) F_{\pi } (-b_p^2 (m_{\eta
}^2-2 m_{\eta '}^2)\nonumber\\
&&+18 (2 F_K-F_{\pi }) V_4)\Big)\Big].
\end{eqnarray}

With five inputs of $m_{\pi}=137$ MeV, $m_{K}=493.677 \pm 0.016$ MeV, $m_{\eta}=547.853 \pm 0.024$ MeV, $m_{\eta'}=957.78 \pm 0.06$ MeV, and $F_{\pi}=131$ MeV, we find the five Lagrangian parameters: $\alpha_1=0.065$ GeV, $\alpha_3=0.13$ GeV,  $A_1=0.00061$ GeV$^3$, $A_3=0.024$ GeV$^3$ and $V_4=-0.23$ (in addition,  these inputs result in $\theta_p = 6.64^\circ$, and $F_K/F_\pi = 1.53$).      Together with the ``bare'' scalar masses found from fit to pion-pion $I=J=0$ scattering amplitude \cite{LsM}:  $m_{\rm \tiny BARE} (\sigma)= 0.847\,\rm GeV$,  $ m_{\rm \tiny BARE} (f_0)= 1.3\,\rm GeV$, $ m_{\rm \tiny BARE} (a_0)= 1.1\,\rm GeV$ and $\theta_s = -6.1^\circ$, we find the numerical values of the coupling constants:

 \begin{eqnarray}
\gamma_{\sigma\pi\pi}&=&3.53\,{\rm GeV},\nonumber\\
\gamma_{f_0\pi\pi}&=&9.57 \,{\rm GeV}, \nonumber \\
\gamma_{a_0\pi\eta}&=&4.71 \,{\rm GeV}, \nonumber \\
\gamma_{a_0\pi\eta'}&=&2.77 \,{\rm GeV}, \nonumber \\
\gamma_{\sigma\eta\eta'}&=&-0.56 \,{\rm GeV}, \nonumber \\
\gamma_{f_0\eta\eta'}&=&2.94 \,{\rm GeV}, \nonumber \\
\gamma^{(4)}&=&78.69 \,{\rm GeV}.
\end{eqnarray}

\section{three- and four-point bare couplings}

\begin{eqnarray}
\left\langle\frac{\partial^{3}V}{\partial (S_{1}^{2})_{1}\partial(\phi_{2}^{1})_{1}\partial \eta_{a}}\right\rangle &=&
\frac{4 \sqrt{2}\Big (2 c_4^a \alpha _1^5 \beta _1+c_4^a \alpha _1^4 \alpha _3 \beta _3
+2 c_3 \alpha _3 \beta _3 \gamma
_1^2+2 c_3 \alpha _1 \beta _1 \gamma _1 (1+\gamma _1)\Big)}{\alpha _1^3 (2 \alpha _1 \beta _1+\alpha _3 \beta_3)},\\ \nonumber\\
\left\langle\frac{\partial^{3}V}{\partial (S_{1}^{2})_{1}\partial(\phi_{2}^{1})_{2}\partial \eta_{a}}\right\rangle&=&
-\frac{8 \sqrt{2} c_3 \left(-1+\gamma _1\right) \Big (\alpha _3 \beta _3 \gamma _1+\alpha _1 \beta _1 \left(1+\gamma _1\right)\Big )}{\alpha
_1 \left(2 \alpha _1 \beta _1+\alpha _3 \beta _3\right){}^2},\\ \nonumber\\
\left\langle\frac{\partial^{3}V}{\partial (S_{1}^{2})_{2}\partial(\phi_{2}^{1})_{1}\partial \eta_{a}}\right\rangle&=&
\frac{8 \sqrt{2} c_3\left(-1+\gamma _1\right) \Big (\alpha _3 \beta _3 \gamma _1+\alpha _1 \beta _1 \left(1+\gamma _1\right)\Big )}{\alpha
_1 \left(2 \alpha _1 \beta _1+\alpha _3 \beta _3\right){}^2},\\ \nonumber\\
\left\langle\frac{\partial^{3}V}{\partial (S_{1}^{2})_{1}\partial(\phi_{2}^{1})_{1}\partial \eta_{b}}\right\rangle&=&
\frac{8 c_3 \gamma _1 \left(\alpha _3 \beta _3+2 \alpha _1 \beta _1 \gamma _1\right)}{\alpha _1^2 \alpha _3 \left(2 \alpha _1
\beta _1+\alpha _3 \beta _3\right)},\\ \nonumber\\
\left\langle\frac{\partial^{3}V}{\partial (S_{1}^{2})_{1}\partial(\phi_{2}^{1})_{2}\partial \eta_{b}}\right\rangle&=&
4 e_3^a-\frac{8 c_3 \left(-1+\gamma _1\right) \left(\alpha _3 \beta _3+2 \alpha _1 \beta _1 \gamma _1\right)}{\alpha _3 \left(2
\alpha _1 \beta _1+\alpha _3 \beta _3\right){}^2},\\ \nonumber\\
\left\langle\frac{\partial^{3}V}{\partial (S_{1}^{2})_{2}\partial(\phi_{2}^{1})_{1}\partial \eta_{b}}\right\rangle&=&
4 e_3^a+\frac{8 c_3 \left(-1+\gamma _1\right) \left(\alpha _3 \beta _3+2 \alpha _1 \beta _1 \gamma _1\right)}{\alpha _3 \left(2
\alpha _1 \beta _1+\alpha _3 \beta _3\right){}^2},\\ \nonumber\\
\left\langle\frac{\partial^{3}V}{\partial (S_{1}^{2})_{1}\partial(\phi_{2}^{1})_{1}\partial \eta_{c}}\right\rangle&=&
\frac{8 \sqrt{2} c_3 \left(-1+\gamma _1\right) \gamma _1}{\alpha _1 \left(2 \alpha _1 \beta _1+\alpha _3 \beta _3\right)},\\ \nonumber\\
\left\langle\frac{\partial^{3}V}{\partial (S_{1}^{2})_{1}\partial(\phi_{2}^{1})_{2}\partial \eta_{c}}\right\rangle&=&
-\frac{8 \sqrt{2} c_3 \alpha _1 \left(-1+\gamma _1\right){}^2}{\left(2 \alpha _1 \beta _1+\alpha _3 \beta _3\right){}^2},\\ \nonumber\\
\left\langle\frac{\partial^{3}V}{\partial (S_{1}^{2})_{2}\partial(\phi_{2}^{1})_{1}\partial \eta_{c}}\right\rangle&=&
\frac{8 \sqrt{2} c_3 \alpha _1 \left(-1+\gamma _1\right){}^2}{\left(2 \alpha _1 \beta _1+\alpha _3 \beta _3\right){}^2},\\ \nonumber\\
\left\langle\frac{\partial^{3}V}{\partial (S_{1}^{2})_{1}\partial(\phi_{2}^{1})_{1}\partial \eta_{d}}\right\rangle&=&
\frac{8 e_3^a \alpha _1^3 \beta _1+4 e_3^a \alpha _1^2 \alpha _3 \beta _3+8 c_3 \alpha _3 \left(-1+\gamma _1\right)
\gamma _1}{\alpha _1^2 \left(2 \alpha _1 \beta _1+\alpha _3 \beta _3\right)},\\ \nonumber\\
\left\langle\frac{\partial^{3}V}{\partial (S_{1}^{2})_{1}\partial(\phi_{2}^{1})_{2}\partial \eta_{d}}\right\rangle&=&
-\frac{8 c_3 \alpha _3 \left(-1+\gamma _1\right){}^2}{\left(2 \alpha _1 \beta _1+\alpha _3 \beta _3\right){}^2},\\ \nonumber\\
\left\langle\frac{\partial^{3}V}{\partial (S_{1}^{2})_{2}\partial(\phi_{2}^{1})_{1}\partial \eta_{d}}\right\rangle&=&
\frac{8 c_3 \alpha _3 \left(-1+\gamma _1\right){}^2}{\left(2 \alpha _1 \beta _1+\alpha _3 \beta _3\right){}^2},\\ \nonumber\\
\left\langle\frac{\partial^{3}V}{\partial (S_{2}^{3})_{1}\partial(\phi_{1}^{2})_{1}\partial(\phi_{3}^{1})_{1}}\right\rangle&=&
4 \alpha_{3} c_4^a,\\ \nonumber\\
\left\langle\frac{\partial^{3}V}{\partial (S_{2}^{3})_{2}\partial(\phi_{1}^{2})_{1}\partial(\phi_{3}^{1})_{1}}\right\rangle&=&
\left\langle\frac{\partial^{3}V}{\partial (S_{2}^{3})_{1}\partial(\phi_{1}^{2})_{1}\partial(\phi_{3}^{1})_{2}}\right\rangle=
\left\langle\frac{\partial^{3}V}{\partial (S_{2}^{3})_{1}\partial(\phi_{1}^{2})_{2}\partial(\phi_{3}^{1})_{1}}\right\rangle=
-4 e_3^a,\\ \nonumber\\
\left\langle\frac{\partial^{3}V}{\partial f_{a}\partial  \eta_ a \partial\eta_ a}\right\rangle &=&
\frac{1}{\alpha _1^3 \left(2 \alpha _1 \beta _1+\alpha _3 \beta _3\right){}^3}4 \sqrt{2} \Big(8 c_4^a \alpha _1^7 \beta _1^3+12
c_4^a \alpha _1^6 \alpha _3 \beta _1^2 \beta _3+6 c_4^a \alpha _1^5 \alpha _3^2 \beta _1 \beta _3^2\nonumber\\
&&+ c_4^a \alpha _1^4 \alpha _3^3\beta _3^3+4 c_3 \alpha _3^3 \beta _3^3 \gamma _1^2+24 c_3 \alpha _1^2 \alpha _3 \beta _1^2 \beta _3 \gamma _1 \left(1+\gamma _1\right)\nonumber\\
&&+8
c_3 \alpha _1^3 \beta _1^3 \left(1+\gamma _1\right){}^2+4 c_3 \alpha _1 \alpha _3^2 \beta _1 \beta _3^2 \gamma _1 \left(1+5 \gamma _1\right)\Big),\\ \nonumber\\
\left\langle\frac{\partial^{3}V}{\partial f_{a}\partial  \eta_ a \partial\eta_ b}\right\rangle &=&
\frac{8 c_3 \Big(6 \alpha _1 \alpha _3^2 \beta _1 \beta _3^2 \gamma _1+\alpha _3^3 \beta _3^3 \gamma _1+4 \alpha _1^3 \beta _1^3
\gamma _1 \left(1+\gamma _1\right)
+2 \alpha _1^2 \alpha _3 \beta _1^2 \beta _3 \left(2+\gamma _1+3 \gamma _1^2\right)\Big)}{\alpha
_1^2 \alpha _3 \left(2 \alpha _1 \beta _1+\alpha _3 \beta _3\right){}^3},\\ \nonumber\\
\left\langle\frac{\partial^{3}V}{\partial f_{a}\partial  \eta_ a \partial\eta_ c}\right\rangle &=&
\frac{8 \sqrt{2} c_3 \beta _1 \left(-1+\gamma _1\right) \Big(2 \alpha _1 \beta _1 \left(1+\gamma _1\right)+\alpha _3 \beta _3
\left(-1+3 \gamma _1\right)\Big)}{\left(2 \alpha _1 \beta _1+\alpha _3 \beta _3\right){}^3},\\ \nonumber\\
\left\langle\frac{\partial^{3}V}{\partial f_{a}\partial  \eta_ a \partial\eta_ d}\right\rangle &=&\frac{-4}{\alpha _1^2 \left(2 \alpha _1 \beta _1+\alpha _3 \beta _3\right){}^3}
 \bigg[8 e_3^a \alpha _1^5 \beta _1^3+12 e_3^a \alpha _1^4 \alpha _3 \beta _1^2 \beta _3+6 e_3^a \alpha _1^3
\alpha _3^2 \beta _1 \beta _3^2\nonumber\\
&&-12 c_3 \alpha _1 \alpha _3^2 \beta _1 \beta _3 \left(-1+\gamma _1\right) \gamma _1
-2 c_3 \alpha _3^3
\beta _3^2 \left(-1+\gamma _1\right) \gamma _1\nonumber\\
&&+\alpha _1^2 \alpha _3 \Big(e_3^a \alpha _3^2 \beta _3^3-8 c_3 \beta _1^2 (-1+\gamma
_1^2)\Big)\bigg] ,\\ \nonumber\\
\left\langle\frac{\partial^{3}V}{\partial f_{a}\partial  \eta_ b \partial\eta_ b}\right\rangle &=&
-\frac{16 \sqrt{2} c_3 \beta _1 \beta _3 \left(-1+\gamma _1\right) \left(\alpha _3 \beta _3+2 \alpha _1 \beta _1 \gamma _1\right)}{\alpha
_3 \left(2 \alpha _1 \beta _1+\alpha _3 \beta _3\right){}^3},\\ \nonumber\\
\left\langle\frac{\partial^{3}V}{\partial f_{a}\partial  \eta_ b \partial\eta_ c}\right\rangle &=&\frac{- 4}{\left(2 \alpha _1 \beta _1+\alpha _3 \beta _3\right){}^3}
\bigg[  8 e_3^a\alpha _1^3 \beta _1^3+12 e_3^a \alpha _1^2 \alpha _3 \beta _1^2 \beta _3+\alpha _3 \beta _3^2 \Big(e_3^a
\alpha _3^2 \beta _3+2 c_3 \left(-1+\gamma _1\right)\Big)\nonumber\\
&&+\,2 \alpha _1 \beta _1 \beta _3 \Big(3 e_3^a \alpha _3^2 \beta _3+2 c_3
(1-3 \gamma _1+2 \gamma _1^2)\Big)\bigg] ,\\\nonumber\\
\left\langle\frac{\partial^{3}V}{\partial f_{a}\partial  \eta_ b \partial\eta_ d}\right\rangle &=&
\frac{8 \sqrt{2} c_3 \beta _1 \left(-1+\gamma _1\right) \Big(-\alpha _3 \beta _3 \left(-2+\gamma _1\right)+2 \alpha _1 \beta
_1 \gamma _1\Big)}{\left(2 \alpha _1 \beta _1+\alpha _3 \beta _3\right){}^3},\\ \nonumber\\
\left\langle\frac{\partial^{3}V}{\partial f_{a}\partial  \eta_ c \partial\eta_ c}\right\rangle &=&
-\frac{16 \sqrt{2} c_3 \alpha _1 \alpha _3 \beta _3 \left(-1+\gamma _1\right){}^2}{\left(2 \alpha _1 \beta _1+\alpha _3 \beta
_3\right){}^3},\\\nonumber\\
\left\langle\frac{\partial^{3}V}{\partial f_{a}\partial  \eta_ c \partial\eta_ d}\right\rangle &=&
-\frac{8 c_3 \alpha _3 \left(-2 \alpha _1 \beta _1+\alpha _3 \beta _3\right) \left(-1+\gamma _1\right){}^2}{\left(2 \alpha _1
\beta _1+\alpha _3 \beta _3\right){}^3},\\ \nonumber\\
\left\langle\frac{\partial^{3}V}{\partial f_{a}\partial  \eta_ d \partial\eta_ d}\right\rangle &=&
\frac{16 \sqrt{2} c_3 \alpha _3^2 \beta _1 \left(-1+\gamma _1\right){}^2}{\left(2 \alpha _1 \beta _1+\alpha _3 \beta _3\right){}^3},\\ \nonumber\\
\left\langle\frac{\partial^{3}V}{\partial f_{b}\partial  \eta_ a \partial\eta_ a}\right\rangle &=&
-\frac{32 c_3 \beta _1 \beta _3 \left(-1+\gamma _1\right) \Big(\alpha _3 \beta _3 \gamma _1+\alpha _1 \beta _1 \left(1+\gamma
_1\right)\Big)}{\alpha _1 \left(2 \alpha _1 \beta _1+\alpha _3 \beta _3\right){}^3},\\ \nonumber\\
\left\langle\frac{\partial^{3}V}{\partial f_{b}\partial  \eta_ a \partial\eta_ b}\right\rangle &=&
\frac{1}{\alpha_1 \alpha _3^2 \left(2 \alpha _1 \beta _1+\alpha _3 \beta _3\right){}^3}
\bigg[8 \sqrt{2} c_3 \Big(\alpha _3^3 \beta _3^3 \gamma _1+4 \alpha _1^3 \beta _1^3 \gamma _1 \left(1+\gamma _1\right)\nonumber\\
&&+\,6 \alpha
_1^2 \alpha _3 \beta _1^2 \beta _3 \gamma _1 \left(1+\gamma _1\right)+2 \alpha _1 \alpha _3^2 \beta _1 \beta _3^2 (1+2 \gamma _1^2)\Big)\bigg],\\ \nonumber\\
\left\langle\frac{\partial^{3}V}{\partial f_{b}\partial  \eta_ a \partial\eta_ c}\right\rangle &=&\frac{-4}{\left(2 \alpha _1 \beta _1+\alpha _3 \beta _3\right){}^3}
 \Bigg[8 e_3^a \alpha _1^3 \beta _1^3+12 e_3^a \alpha _1^2 \alpha _3 \beta _1^2 \beta _3
+2 \alpha _1 \beta _1 \beta
_3 \Big(3 e_3^a  \alpha _3^2 \beta _3-4 c_3 \left(-1+\gamma _1\right)\Big)\nonumber\\
&&+\,\alpha _3 \beta _3^2 \Big(e_3^a  \alpha _3^2 \beta _3-4
c_3 \left(-1+\gamma _1\right) \gamma _1\Big)\Bigg],\\ \nonumber\\
\left\langle\frac{\partial^{3}V}{\partial f_{b}\partial  \eta_ a \partial\eta_ d}\right\rangle &=&
-\frac{8 \sqrt{2} c_3 \beta _1 \left(-1+\gamma _1\right) \Big(2 \alpha _1 \beta _1 \left(1+\gamma _1\right)+\alpha _3 \beta _3
\left(-1+3 \gamma _1\right)\Big)}{\left(2 \alpha _1 \beta _1+\alpha _3 \beta _3\right){}^3},\\ \nonumber\\
\left\langle\frac{\partial^{3}V}{\partial f_{b}\partial  \eta_ b \partial\eta_ b}\right\rangle &=&\frac{8}{\alpha _3^3 \left(2\alpha _1 \beta _1+\alpha _3 \beta _3\right){}^3}
 \bigg[\alpha _3^3 \left(2 c_3+c_4^a \alpha _3^4\right) \beta _3^3+6 \alpha _1 \alpha _3^2 \beta _1 \beta _3^2 \left(c_4^a
\alpha _3^4+2 c_3 \gamma _1\right)\nonumber\\
&&+\,8 \alpha _1^3 \beta _1^3 \left(c_4^a \alpha _3^4+2 c_3 \gamma _1^2\right)
+4 \alpha _1^2 \alpha _3 \beta _1^2 \beta _3 \Big(3 c_4^a \alpha _3^4
+\,2 c_3 \gamma _1 (1+2 \gamma _1)\Big)\bigg],\\ \nonumber\\
\left\langle\frac{\partial^{3}V}{\partial f_{b}\partial  \eta_ b \partial\eta_ c}\right\rangle &=&
\frac{16 \sqrt{2} c_3 \alpha _1 \left(-1+\gamma _1\right) \left(\alpha _3^2 \beta _3^2+2 \alpha _1^2 \beta _1^2 \gamma _1+3 \alpha
_1 \alpha _3 \beta _1 \beta _3 \gamma _1\right)}{\alpha _3^2 \left(2 \alpha _1 \beta _1+\alpha _3 \beta _3\right){}^3},\\ \nonumber\\
\left\langle\frac{\partial^{3}V}{\partial f_{b}\partial  \eta_ b \partial\eta_ d}\right\rangle &=&
\frac{8 c_3 \beta _3 \left(-1+\gamma _1\right) \Big(\alpha _3 \beta _3+2 \alpha _1 \beta _1 \left(-1+2 \gamma _1\right)\Big)}{\left(2
\alpha _1 \beta _1+\alpha _3 \beta _3\right){}^3},\\ \nonumber\\
\left\langle\frac{\partial^{3}V}{\partial f_{b}\partial  \eta_ c \partial\eta_ c}\right\rangle &=&
\frac{32 c_3 \alpha _1^2 \beta _3 \left(-1+\gamma _1\right){}^2}{\left(2 \alpha _1 \beta _1+\alpha _3 \beta _3\right){}^3},\\ \nonumber\\
\left\langle\frac{\partial^{3}V}{\partial f_{b}\partial  \eta_ c \partial\eta_ d}\right\rangle &=&
-\frac{8 \sqrt{2}c_3\alpha _1 \left(2 \alpha _1 \beta _1-\alpha _3 \beta _3\right) \left(-1+\gamma _1\right){}^2}{\left(2 \alpha
_1 \beta _1+\alpha _3 \beta _3\right){}^3},\\ \nonumber\\
\left\langle\frac{\partial^{3}V}{\partial f_{b}\partial  \eta_ d \partial\eta_ d}\right\rangle &=&
-\frac{32 c_3 \alpha _1 \alpha _3 \beta _1 \left(-1+\gamma _1\right){}^2}{\left(2 \alpha _1 \beta _1+\alpha _3 \beta _3\right){}^3},\\ \nonumber\\
\left\langle\frac{\partial^{3}V}{\partial f_{c}\partial  \eta_ a \partial\eta_ a}\right\rangle &=&
\frac{16 \sqrt{2} c_3 \alpha _3 \beta _3 \left(-1+\gamma _1\right) \Big(\alpha _3 \beta _3 \gamma _1+\alpha _1 \beta _1 \left(1+\gamma
_1\right)\Big)}{\alpha _1 \left(2 \alpha _1 \beta _1+\alpha _3 \beta _3\right){}^3},\\ \nonumber\\
\left\langle\frac{\partial^{3}V}{\partial f_{c}\partial  \eta_ a \partial\eta_ b}\right\rangle &=&
-\frac{ 4 \Big(8 e_3^a \alpha _1^3 \beta c_3 \left(-1+\gamma _1\right)\Big)+\alpha _3 \beta _3^2 \Big(e_3^a \alpha _3^2 \beta _3+2
c_3 (1-3 \gamma _1+2 \gamma _1^2)\Big)}
{\left(2 \alpha _1 \beta _1+\alpha _3 \beta _3\right){}^3}
,\\ \nonumber\\
\left\langle\frac{\partial^{3}V}{\partial f_{c}\partial  \eta_ a \partial\eta_ c}\right\rangle &=&
\frac{8 \sqrt{2} c_3 \alpha _1 \left(-1+\gamma _1\right) \Big(2 \alpha _1 \beta _1 (1+\gamma _1)+\alpha _3 \beta _3
\left(-1+3 \gamma _1\right)\Big)}{\left(2 \alpha _1 \beta _1+\alpha _3 \beta _3\right){}^3},\\ \nonumber\\
\left\langle\frac{\partial^{3}V}{\partial f_{c}\partial  \eta_ a \partial\eta_ d}\right\rangle &=&
\frac{8 c_3 \alpha _3 \left(-1+\gamma _1\right) \Big(2 \alpha _1 \beta _1 \left(1+\gamma _1\right)+\alpha _3 \beta _3 \left(-1+3
\gamma _1\right)\Big)}{\left(2 \alpha _1 \beta _1+\alpha _3 \beta _3\right){}^3},\\ \nonumber\\
\left\langle\frac{\partial^{3}V}{\partial f_{c}\partial  \eta_ b \partial\eta_ b}\right\rangle &=&
-\frac{16 \sqrt{2} c_3 \alpha _1 \beta _3 \left(-1+\gamma _1\right) \left(\alpha _3 \beta _3+2 \alpha _1 \beta _1 \gamma _1\right)}{\alpha
_3 \left(2 \alpha _1 \beta _1+\alpha _3 \beta _3\right){}^3},\\ \nonumber\\
\left\langle\frac{\partial^{3}V}{\partial f_{c}\partial  \eta_ b \partial\eta_ c}\right\rangle &=&
\frac{16 c_3 \alpha _1^2 \left(-1+\gamma _1\right) \Big(-\alpha _3 \beta _3 \left(-2+\gamma _1\right)+2 \alpha _1 \beta _1 \gamma
_1\Big)}{\alpha _3 \left(2 \alpha _1 \beta _1+\alpha _3 \beta _3\right){}^3},\\ \nonumber\\
\left\langle\frac{\partial^{3}V}{\partial f_{c}\partial  \eta_ b \partial\eta_ d}\right\rangle &=&
\frac{8 \sqrt{2} c_3 \alpha _1 \left(-1+\gamma _1\right) \Big(-\alpha _3 \beta _3 \left(-2+\gamma _1\right)+2 \alpha _1 \beta
_1 \gamma _1\Big)}{\left(2 \alpha _1 \beta _1+\alpha _3 \beta _3\right){}^3},\\ \nonumber\\
\left\langle\frac{\partial^{3}V}{\partial f_{c}\partial  \eta_ c \partial\eta_ c}\right\rangle &=&
\frac{32 \sqrt{2} c_3 \alpha _1^3 \left(-1+\gamma _1\right){}^2}{\left(2 \alpha _1 \beta _1+\alpha _3 \beta _3\right){}^3},\\ \nonumber\\
\left\langle\frac{\partial^{3}V}{\partial f_{c}\partial  \eta_ c \partial\eta_ d}\right\rangle &=&
\frac{32 c_3 \alpha _1^2 \alpha _3 \left(-1+\gamma _1\right){}^2}{\left(2 \alpha _1 \beta _1+\alpha _3 \beta _3\right){}^3},\\ \nonumber\\
\left\langle\frac{\partial^{3}V}{\partial f_{c}\partial  \eta_ d \partial\eta_ d}\right\rangle &=&
\frac{16 \sqrt{2} c_3 \alpha _1 \alpha _3^2 \left(-1+\gamma _1\right){}^2}{\left(2 \alpha _1 \beta _1+\alpha _3 \beta _3\right){}^3},\\ \nonumber\\
\left\langle\frac{\partial^{3}V}{\partial f_{d}\partial  \eta_ a \partial\eta_ a}\right\rangle &=&\frac{-4}{\alpha _1 \left(2 \alpha _1 \beta _1+\alpha_3 \beta _3\right){}^3}
\bigg[  \bigg(8 e_3^a \alpha _1^4 \beta _1^3+12 e_3^a \alpha _1^3 \alpha _3 \beta _1^2 \beta _3+6 e_3^a \alpha _1^2
\alpha _3^2 \beta _1 \beta _3^2\nonumber \\
&&+8 c_3 \alpha _3^2 \beta _1 \beta _3 \left(-1+\gamma _1\right) \gamma _1
+\alpha _1 \alpha _3 \Big(e_3^a
\alpha _3^2 \beta _3^3+8 c_3 \beta _1^2 (-1+\gamma _1^2)\Big)\bigg)\bigg],\\ \nonumber\\
\left\langle\frac{\partial^{3}V}{\partial f_{d}\partial  \eta_ a \partial\eta_ b}\right\rangle &=&
\frac{8 \sqrt{2} c_3 \beta _1 \left(-1+\gamma _1\right) \Big(2 \alpha _1 \beta _1+\alpha _3 \beta _3 \left(-1+2 \gamma _1\right)\Big)}{\left(2
\alpha _1 \beta _1+\alpha _3 \beta _3\right){}^3},\\ \nonumber\\
\left\langle\frac{\partial^{3}V}{\partial f_{d}\partial  \eta_ a \partial\eta_ c}\right\rangle &=&
\frac{16 c_3 \alpha _3 \left(-1+\gamma _1\right) \left(2 \alpha _1 \beta _1+\alpha _3 \beta _3 \gamma _1\right)}{\left(2 \alpha
_1 \beta _1+\alpha _3 \beta _3\right){}^3},\\ \nonumber\\
\left\langle\frac{\partial^{3}V}{\partial f_{d}\partial  \eta_ a \partial\eta_ d}\right\rangle &=&
\frac{8 \sqrt{2} c_3 \alpha _3^2 \left(-1+\gamma _1\right) \left(2 \alpha _1 \beta _1+\alpha _3 \beta _3 \gamma _1\right)}{\alpha
_1 \left(2 \alpha _1 \beta _1+\alpha _3 \beta _3\right){}^3},\\ \nonumber\\
\left\langle\frac{\partial^{3}V}{\partial f_{d}\partial  \eta_ b \partial\eta_ b}\right\rangle &=&
\frac{32 c_3 \alpha _1 \beta _1 \left(-1+\gamma _1\right) \left(\alpha _3 \beta _3+2 \alpha _1 \beta _1 \gamma _1\right)}{\alpha
_3 \left(2 \alpha _1 \beta _1+\alpha _3 \beta _3\right){}^3},\\ \nonumber\\
\left\langle\frac{\partial^{3}V}{\partial f_{d}\partial  \eta_ b \partial\eta_ c}\right\rangle &=&
\frac{8 \sqrt{2} c_3 \alpha _1 \left(-1+\gamma _1\right) \Big(\alpha _3 \beta _3+2 \alpha _1 \beta _1 \left(-1+2 \gamma _1\right)\Big)}{\left(2
\alpha _1 \beta _1+\alpha _3 \beta _3\right){}^3},\\ \nonumber\\
\left\langle\frac{\partial^{3}V}{\partial f_{d}\partial  \eta_ b \partial\eta_ d}\right\rangle &=&
\frac{8 c_3 \alpha _3 \left(-1+\gamma _1\right) \Big(\alpha _3 \beta _3+2 \alpha _1 \beta _1 \left(-1+2 \gamma _1\right)\Big)}{\left(2
\alpha _1 \beta _1+\alpha _3 \beta _3\right){}^3},\\ \nonumber\\
\left\langle\frac{\partial^{3}V}{\partial f_{d}\partial  \eta_ c \partial\eta_ c}\right\rangle &=&
\frac{32 c_3 \alpha _1^2 \alpha _3 \left(-1+\gamma _1\right){}^2}{\left(2 \alpha _1 \beta _1+\alpha _3 \beta _3\right){}^3},\\ \nonumber\\
\left\langle\frac{\partial^{3}V}{\partial f_{d}\partial  \eta_ c \partial\eta_ d}\right\rangle &=&
\frac{16 \sqrt{2} c_3 \alpha _1 \alpha _3^2 \left(-1+\gamma _1\right){}^2}{\left(2 \alpha _1 \beta _1+\alpha _3 \beta _3\right){}^3},\\ \nonumber\\
\left\langle\frac{\partial^{3}V}{\partial f_{d}\partial  \eta_ d \partial\eta_ d}\right\rangle &=&
\frac{16 c_3 \alpha _3^3 \left(-1+\gamma _1\right){}^2}{\left(2 \alpha _1 \beta _1+\alpha _3 \beta _3\right){}^3},\\ \nonumber\\
\left\langle\frac{\partial^{4}V}{\partial\eta_{a}\partial\eta_{a}\partial  (\phi_{1}^{2})_{1}\partial(\phi_{2}^{1})_{1}}\right\rangle &=&\frac{4 \Big(6 c_4^a \alpha _1^5 \beta _1+3 c_4^a\alpha _1^4 \alpha _3 \beta _3+8 c_3\alpha _3 \beta _3 \gamma _1^2+8
c_3 \alpha _1 \beta _1 \gamma _1 \left(1+\gamma _1\right)\Big)}{\alpha _1^4 \left(2 \alpha _1 \beta _1+\alpha _3 \beta _3\right)},\\ \nonumber\\
\left\langle\frac{\partial^{4}V}{\partial\eta_{a}\partial\eta_{a}\partial  (\phi_{1}^{2})_{1}\partial(\phi_{2}^{1})_{2}}\right\rangle
&=&\left\langle\frac{\partial^{4}V}{\partial\eta_{a}\partial\eta_{a}\partial  (\phi_{1}^{2})_{2}\partial(\phi_{2}^{1})_{1}}\right\rangle
=-\frac{32 c_3 \beta _1 \left(-1+\gamma _1\right) \Big(\alpha _3 \beta _3 \gamma _1+\alpha _1 \beta _1 \left(1+\gamma _1\right)\Big)}{\alpha
_1 \left(2 \alpha _1 \beta _1+\alpha _3 \beta _3\right){}^3}
,\\ \nonumber \\
\left\langle\frac{\partial^{4}V}{\partial\eta_{a}\partial\eta_{b}\partial  (\phi_{1}^{2})_{1}\partial(\phi_{2}^{1})_{1}}\right\rangle &=&\frac{8 \sqrt{2} c_3 \gamma _1 \left(\alpha _3 \beta _3+2 \alpha _1 \beta _1 \gamma _1\right)}{\alpha _1^3 \alpha _3 \left(2 \alpha
_1 \beta _1+\alpha _3 \beta _3\right)},\\ \nonumber\\
\left\langle\frac{\partial^{4}V}{\partial\eta_{a}\partial\eta_{b}\partial  (\phi_{1}^{2})_{1}\partial(\phi_{2}^{1})_{2}}\right\rangle
&=&\left\langle\frac{\partial^{4}V}{\partial\eta_{a}\partial\eta_{b}\partial  (\phi_{1}^{2})_{2}\partial(\phi_{2}^{1})_{1}}\right\rangle\nonumber\\
&=&\frac{-8 \sqrt{2} c_3 \left(-1+\gamma _1\right) \Big(2 \alpha _1^2 \beta _1^2 \gamma _1+\alpha _3^2 \beta _3^2 \gamma _1+\alpha
_1 \alpha _3 \beta _1 \beta _3 \left(2+\gamma _1\right)\Big)}{\alpha _1 \alpha _3 \left(2 \alpha _1 \beta _1+\alpha _3 \beta _3\right){}^3},\nonumber\\\\
\left\langle\frac{\partial^{4}V}{\partial\eta_{a}\partial\eta_{c}\partial  (\phi_{1}^{2})_{1}\partial(\phi_{2}^{1})_{1}}\right\rangle &=&
\frac{16 c_3 \left(-1+\gamma _1\right) \gamma _1}{\alpha _1^2 \left(2 \alpha _1 \beta _1+\alpha _3 \beta _3\right)},\\ \nonumber\\
\left\langle\frac{\partial^{4}V}{\partial\eta_{a}\partial\eta_{c}\partial  (\phi_{1}^{2})_{1}\partial(\phi_{2}^{1})_{2}}\right\rangle &=&
\left\langle\frac{\partial^{4}V}{\partial\eta_{a}\partial\eta_{c}\partial  (\phi_{1}^{2})_{2}\partial(\phi_{2}^{1})_{1}}\right\rangle
=\frac{16 c_3 \left(-1+\gamma _1\right) \left(2 \alpha _1 \beta _1+\alpha _3 \beta _3 \gamma _1\right)}{\left(2 \alpha _1 \beta
_1+\alpha _3 \beta _3\right){}^3},\\ \nonumber\\
\left\langle\frac{\partial^{4}V}{\partial\eta_{a}\partial\eta_{d}\partial  (\phi_{1}^{2})_{1}\partial(\phi_{2}^{1})_{1}}\right\rangle &=&
\frac{8 \sqrt{2} c_3 \alpha _3 \left(-1+\gamma _1\right) \gamma _1}{\alpha _1^3 \left(2 \alpha _1 \beta _1+\alpha _3 \beta _3\right)},\\ \nonumber\\
\left\langle\frac{\partial^{4}V}{\partial\eta_{a}\partial\eta_{d}\partial  (\phi_{1}^{2})_{1}\partial(\phi_{2}^{1})_{2}}\right\rangle &=&
\left\langle\frac{\partial^{4}V}{\partial\eta_{a}\partial\eta_{d}\partial  (\phi_{1}^{2})_{2}\partial(\phi_{2}^{1})_{1}}\right\rangle
=\frac{8 \sqrt{2} c_3 \alpha _3 \left(-1+\gamma _1\right) \left(2 \alpha _1 \beta _1+\alpha _3 \beta _3 \gamma _1\right)}{\alpha
_1 \left(2 \alpha _1 \beta _1+\alpha _3 \beta _3\right){}^3},\\ \nonumber\\
\left\langle\frac{\partial^{4}V}{\partial\eta_{b}\partial\eta_{b}\partial  (\phi_{1}^{2})_{1}\partial(\phi_{2}^{1})_{2}}\right\rangle &=&\left\langle\frac{\partial^{4}V}{\partial\eta_{b}\partial\eta_{b}\partial  (\phi_{1}^{2})_{2}\partial(\phi_{2}^{1})_{1}}\right\rangle
=-\frac{16 c_3 \beta _3 \left(-1+\gamma _1\right) \left(\alpha _3 \beta _3+2 \alpha _1 \beta _1 \gamma _1\right)}{\alpha _3 \left(2
\alpha _1 \beta _1+\alpha _3 \beta _3\right){}^3},\\ \nonumber\\
\left\langle\frac{\partial^{4}V}{\partial\eta_{b}\partial\eta_{c}\partial  (\phi_{1}^{2})_{1}\partial(\phi_{2}^{1})_{2}}\right\rangle &=&
\left\langle\frac{\partial^{4}V}{\partial\eta_{b}\partial\eta_{b}\partial  (\phi_{1}^{2})_{2}\partial(\phi_{2}^{1})_{1}}\right\rangle
=\frac{8 \sqrt{2} c_3 \alpha _1 \left(-1+\gamma _1\right) \Big(-\alpha _3 \beta _3 \left(-2+\gamma _1\right)+2 \alpha _1 \beta_1 \gamma _1\Big)}{\alpha _3 \left(2 \alpha _1 \beta _1+\alpha _3 \beta _3\right){}^3},\\ \nonumber\\
\left\langle\frac{\partial^{4}V}{\partial\eta_{b}\partial\eta_{d}\partial  (\phi_{1}^{2})_{1}\partial(\phi_{2}^{1})_{2}}\right\rangle &=&\left\langle\frac{\partial^{4}V}{\partial\eta_{b}\partial\eta_{d}\partial  (\phi_{1}^{2})_{2}\partial(\phi_{2}^{1})_{1}}\right\rangle
=\frac{8 c_3 \left(-1+\gamma _1\right) \Big(-\alpha _3 \beta _3 \left(-2+\gamma _1\right)+2 \alpha _1 \beta _1 \gamma _1\Big)}{\left(2
\alpha _1 \beta _1+\alpha _3 \beta _3\right){}^3},\\ \nonumber\\
\left\langle\frac{\partial^{4}V}{\partial\eta_{c}\partial\eta_{c}\partial  (\phi_{1}^{2})_{1}\partial(\phi_{2}^{1})_{2}}\right\rangle &=&\left\langle\frac{\partial^{4}V}{\partial\eta_{c}\partial\eta_{c}\partial  (\phi_{1}^{2})_{2}\partial(\phi_{2}^{1})_{1}}\right\rangle
=\frac{32 c_3 \alpha _1^2 \left(-1+\gamma _1\right){}^2}{\left(2 \alpha _1 \beta _1+\alpha _3 \beta _3\right){}^3},\\ \nonumber\\
\left\langle\frac{\partial^{4}V}{\partial\eta_{c}\partial\eta_{d}\partial  (\phi_{1}^{2})_{1}\partial(\phi_{2}^{1})_{2}}\right\rangle &=&
\left\langle\frac{\partial^{4}V}{\partial\eta_{c}\partial\eta_{d}\partial  (\phi_{1}^{2})_{2}\partial(\phi_{2}^{1})_{1}}\right\rangle
=\frac{16 \sqrt{2} c_3 \alpha _1 \alpha _3 \left(-1+\gamma _1\right){}^2}{\left(2 \alpha _1 \beta _1+\alpha _3 \beta _3\right){}^3},\\ \nonumber\\
\left\langle\frac{\partial^{4}V}{\partial\eta_{d}\partial\eta_{d}\partial  (\phi_{1}^{2})_{1}\partial(\phi_{2}^{1})_{2}}\right\rangle &=&\left\langle\frac{\partial^{4}V}{\partial\eta_{d}\partial\eta_{d}\partial  (\phi_{1}^{2})_{2}\partial(\phi_{2}^{1})_{1}}\right\rangle
=\frac{16 c_3\alpha _3^2 \left(-1+\gamma _1\right){}^2}{\left(2 \alpha _1 \beta _1+\alpha _3 \beta _3\right){}^3}.
\end{eqnarray}

\section{Recovering current algebra}
In this Appendix we show how the known current algebra result for this decay is obtained from the present model.   The four-quark fields are decoupled in the limit $d_2, e_{3}^{a}\rightarrow 0$ and $\gamma_1 \rightarrow 1$, in which:
\begin{eqnarray}
m_{\pi}^2&=&-2 c_2 +4 c_4^a \alpha_1^2,\nonumber \\
m_{f_1}^2&=&m_{a}^2=-2 c_ 2 +12 c_4^a \alpha_1^2,\nonumber \\
m_{f_2}^2&=&-2 c_ 2 +12 c_4^a \alpha_3^2,\nonumber \\
F_{\pi}&=& 2 \alpha _{1}\nonumber, \\
m_{\eta}^2+m_{\eta^{'}}^2&=&-4 c_2-\frac{16 c_3 }{\alpha_1^2}+4 c_4^a \alpha_1^2-\frac{8 c_3 }{\alpha_3^2}+4 c_4^a \alpha_3^2.
\end{eqnarray}

From the above equations we can solve for  the five model parameters:

\begin{eqnarray}\label{lpa}
\alpha_1&=&\frac{F_{\pi}}{2},\nonumber \\
\alpha_3&=&F_{\pi}\sqrt{\frac{2 m_{f_2}^2+ m_{f_1}^2-3 m_{\pi}^2 }{12(m_{f_1}^2-m_{\pi}^2)}},\nonumber \\
c_2&=&\frac{1}{4}(m_{f_1}^2-3m_{\pi}^2),\nonumber \\
 c_3 &=& -\frac{ F_{\pi}^2(m_{f_1}^2+ 2 m_{f_2}^2-3 m_{\pi}^2)\Big(m_{f_1}^2-m_{f_2}^2+3 (m_{\eta}^2+m_{\eta ' }^2-2 m_{\pi}^2 )\Big) }{96(5 m_{f_1}^2 +4 m_{f_1}^2-9 m_{\pi}^2)},\nonumber \\
 c_4^a &=& \frac{m_{f_1}^2-m_{\pi}^2}{2 F_{\pi}^2}.
\end{eqnarray}
We expect to recover the current algebra result when the scalars are decoupled as a result of becoming very heavy, i.e. in the limit $m_{f_1}=m_{f_2}=m_{f}\rightarrow \infty$.    In this limit,
\begin{eqnarray}\label{lpb}
\lim _{m_{f}\rightarrow \infty} \alpha_3 &=& \frac{ F_{\pi}}{2},\nonumber \\
\lim _{m_{f}\rightarrow \infty} c_2&=&{m_f^2\over {4}},\nonumber \\
\lim _{m_{f}\rightarrow \infty} c_3 &=& \frac{-1}{96}F_{\pi}^2(m_{\eta}^2+m_{\eta ' }^2-2 m_{\pi}^2 )\nonumber \\
\lim _{m_{f}\rightarrow \infty} c_4^a&=&{m_f^2\over {2 F_\pi^2}} \nonumber \\
\label{par_decoupling}
 \end{eqnarray}
The physical vertices (in the limit of $d_2,e_{3}^{a}\rightarrow 0$ and $\gamma_1 \rightarrow 1$) become:
\begin{eqnarray}\label{gf2pipi}
\gamma^{(4)}&=&6\, c_4^a \sin(2 \theta_p)+\frac{16\, c_3 \sin(2 \theta_p)}{\alpha_1^4}+\frac{8\,\sqrt{2}c_3 \cos(2 \theta_p)}{\alpha_1^3 \alpha_3},\nonumber \\
\gamma_{f_1 \pi \pi }&=&4 c_4^a \alpha_1,\nonumber \\
\gamma_{f_2 \pi \pi }&=&0,\nonumber \\
\gamma_{f_1 \eta \eta' }&=&2\sqrt{2}\, c_4^a \sin(2 \theta_p) \alpha_1 +\frac{8\sqrt{2}\, c_3 \sin(2 \theta_p)}{\alpha_1^3}+\frac{8\,c_3 \cos(2 \theta_p)}{\alpha_1^2 \alpha_3},\nonumber \\
\gamma_{f_2 \eta \eta' }&=&\frac{8\sqrt{2}\,c_3 \cos(2 \theta_p)\alpha_3-4\sin(2 \theta_p)\alpha_1(2\,c_3+c_4^a \alpha_3^4)}{\alpha_1 \alpha_3^3},\nonumber \\
\gamma_{a_0 \pi \eta }&=&\frac{8\sqrt{2}\,c_3 \cos(\theta_p)}{\alpha_{1}^3}+4\sqrt{2}\,c_4^a\cos(\theta_p)\alpha_1-\frac{8\,c_3\sin(\theta_p)}{\alpha_{1}^2\alpha_3},\nonumber \\
\gamma_{a_0 \pi \eta' }&=&\frac{8\sqrt{2}\,c_3 \sin(\theta_p)}{\alpha_{1}^3}+4\sqrt{2}\,c_4^a\sin(\theta_p)\alpha_1+\frac{8\,c_3\cos(\theta_p)}{\alpha_{1}^2\alpha_3},\nonumber \\
\end{eqnarray}
which together with (\ref{par_decoupling}),

\begin{eqnarray}
\gamma ^{(4)} &=&\frac{1}{3 F^2_\pi}\bigg[\left(m^2_\eta+m^2_{\eta'} -2 m^2_\pi \right)\Big(-4 \sqrt{2}  \cos(2 \theta _p)-8\sin(2 \theta _p)\Big)  +9 \left(m_{f}^2-m^2_\pi\right) \sin(2 \theta _p)\bigg] \nonumber\\
\gamma _{f_1\pi \pi }&=&\frac{m_{f}^2-m^2_\pi }{F_\pi}, \nonumber\\
\gamma _{f_1\eta \eta'}&=&\frac{1}{3 F_\pi}
\bigg[\left(m^2_\eta+m^2_{\eta'} -2 m^2_\pi \right) \Big(-2 \cos(2 \theta _p)-2 \sqrt{2} \sin(2 \theta _p)\Big) +\frac{3}{\sqrt{2}}\left(m_{f}^2-m^2_\pi\right) \sin(2 \theta _p)\bigg]
,\nonumber\\
%\gamma _{f_2\pi \pi }&=&0\nonumber\\
\gamma _{f_2 \eta \eta '}&=&\frac{2 }{3 F_\pi} \bigg[ \left(m^2_\eta+m^2_{\eta'}-2 m^2_\pi \right) \Big(-\sqrt{2}\cos(2 \theta _p)+ \sin(2 \theta _p) \Big) -\frac{3}{2} \left(m_{f}^2-m^2_\pi\right)\sin(2 \theta _p) \bigg]  , \nonumber\\
\gamma _{a_0\pi \eta }&=&\frac{1}{3 F_\pi}\bigg[  \left(m^2_\eta+m^2_{\eta'} -2 m^2_\pi \right)\Big(-2 \sqrt{2} \cos(\theta _p)+2\sin( \theta _p)\Big) +3 \sqrt{2} \left(m_{f}^2-m^2_\pi \right) \cos( \theta _p) \bigg] ,\nonumber\\
\gamma _{a_0\pi \eta' }&=&\frac{1}{3
F_\pi}\bigg[  \left(m^2_\eta+m^2_{\eta'} -2 m^2_\pi \right)\Big(-2 \cos( \theta _p)-2\sqrt{2}\sin( \theta _p)\Big) +3 \sqrt{2} \left(m_{f}^2-m^2_\pi \right) \sin( \theta _p)  \bigg].\nonumber\\
\end{eqnarray}

Each individual  decay amplitude inherits the scalar mass dependency via the physical vertices and propagators.  The four-point amplitude will have the scalar mass dependency

\begin{equation}
M_{4p}=\xi_0+\xi_1 m_{f}^2.
\end{equation}
The isosinglet scalar contribution  has the general structure
\begin{equation}
M_{f_i}=\sqrt{2}\gamma_{f_i\pi\pi}\gamma_{f_i \eta \eta'}\times (\rm{propagator}),
\end{equation}
with
\begin{eqnarray}
\sqrt{2}\gamma_{f_i\pi\pi}\gamma_{f_i \eta \eta'}&=&\rho_0+\rho_1 m_{f}^2+\rho_2 m_{f}^4,\nonumber\\
\rm{propagator}&=&\frac{1}{m_{f}^2+x}\simeq \frac{1}{m_{f}^2}-\frac{x}{m_{f}^4}+ {\cal O} (\frac{1}{m_{f}^6}).
\end{eqnarray}
Thus
\begin{equation}
\lim _{m_{f}\rightarrow \infty} M_{f_i}=\rho_1 - x \rho_2 +\rho_2 m_{f}^2.
\end{equation}
Similarly for the $a_0$ contribution
\begin{equation}
M_{a_0}=\gamma_{a_0\pi\eta}\gamma_{a_0 \pi \eta'}\big[\frac{1}{m_{f}^2+y_1}+\frac{1}{m_{f}^2+y_2}\big],
\end{equation}
with
\begin{eqnarray}
\gamma_{a_0\pi\eta}\gamma_{a_0 \pi \eta'}=\delta_0 +\delta_1 m_{f}^2 +\delta_2 m_{f}^4, \nonumber\\
\frac{1}{m_{f}^2+y_i}
\simeq\frac{1}{m_{f}^2}-\frac{y_i}{m_{f}^4}+{\cal O} (\frac{1}{m_{f}^6}).
\end{eqnarray}
Thus
\begin{equation}
\lim _{m_{f}\rightarrow \infty} M_{a_0}=2\, \delta_1-\sum_i y_i\delta_2 +2\, \delta_2 m_{f}^2.
\end{equation}
Now putting everything together, we expect:
\begin{equation}
\lim_{m_{f}\rightarrow\infty} M_{\rm{total}}=M_{\rm{C.A.}}
\end{equation}
which implies that the following two sum rules must be upheld
\begin{eqnarray}
\xi_0+\rho_1-x \rho_2+2\,\delta_1-\sum _i y_i \delta _2&=&M_{\rm{C.A.}},\nonumber\\
\xi_1 +\rho_2 +2\,\delta_2 &=&0.
\end{eqnarray}
We find that the second sum-rule is identically upheld, and the first one gives:
\begin{equation}
M_{\rm{C.A.}}=\frac{-1}{3 F_\pi^2}\Bigg(\sin (2 \theta _p) \left(m^2_{\eta}+m^2_{\eta'}-5 m^2_\pi\right)+2 \sqrt{2} \cos (2 \theta_p) \left(m^2_{\eta}+m^2_{\eta'}-2 m^2_\pi\right)\Bigg).
\end{equation}
Since in the decoupling limit $c_3=0$ and $m_{f}\rightarrow \infty$ we have
\begin{equation}
2m_{\pi}^{2} \rightarrow m_{\eta}^{2}+m_{\eta^{'}}^{2},
\end{equation}
which results in
\begin{equation}
M_{\rm{C.A.}}=\frac{m_{\pi}^2}{F_{\pi}^2}\sin (2\theta p),
\end{equation}
in agreement with Eq. (2.4) of ref. \cite{99FS}.

\newpage

\section*{Acknowledgments}
\vskip -.5cm
A.H.F. wishes to thank the Physics Dept. of Shiraz University for its hospitality in Summer of 2012 where this work was initiated. A.H.F. also wishes to thank Prof. M. Amaryan for many helpful discussions.  The work of J.S. is supported in part by the US DOE under the Contract No. DE-FG-02-85ER 40231.


\begin{thebibliography}{10}

\bibitem{PDG}
J. Beringer et al. (Particle Data Group), Phys. Rev. D {\bf 86}, 010001 (2012).


%  Begin General Refs. -----------------------------------------------------

\bibitem{Weinberg_13} S. Weinberg,   Phys. Rev. Lett. {\bf 110}, 261601 (2013).


\bibitem{Eli}
R.T. Kleiv, T.G. Steele, A. Zhang and I. Blokland, Phys. Rev. D {\bf 87}, 125018 (2013);
D. Harnett, R.T. Kleiv, K. Moats  and T.G. Steele,  Nucl. Phys. A {\bf 850}, 110 (2011);
J. Zhang, H.Y. Jin, Z.F. Zhang, T.G. Steele and D.H. Lu, Phys. Rev. D {\bf 79}, 114033 (2009);
Fang Shi, T.G. Steele, V. Elias, K.B. Sprague, Ying
Xue and A.H.  Fariborz, Nucl. Phys. A {\bf 671}, 416
(2000);
V. Elias, A.H. Fariborz, Fang Shi and
T.G. Steele, Nucl.  Phys.  A {\bf 633}, 279 (1998).

\bibitem{Lattice_scalars}
M. Wagner, et al., Acta Phys. Polon. Supp. {\bf 6} 847 (2013);
C. Alexandrou, et al,  JHEP {\bf 137}, 1304 (2013);
T. Kunihiro, S. Muroya, A. Nakamura, C. Nonaka, M. Sekiguchi, H. Wada, in proceedings of {\it International IUPAP Conference on Few-Body Problems in Physics (FB 19), Bonn, Germany, 31 Aug - 5 Sep 2009},  EPJ Web Conf.3:03010 (2010);
C. McNeile, in proceedings of {\it 11th Int. Conf. on Meson-Nucleon
Physics and
the Structure of the Nucleon}, 10-14 Sept. 2007, J\"ulich, Germany;
C. McNeile and C. Michael (UKQCD Collaboration), Phys. Rev. D {\bf 74},
014508 (2006);
N. Mathur et al, hep-ph/0607110;
A. Hart et al (UKQCD Collaboration),
Phys. Rev. D {\bf 74}, 114504 (2006);
H. Wada (SCALAR Collaboration), Nucl. Phys. Proc. Suppl. {\bf 129}, 432
(2004); T. Kunihiro et al (SCALAR Collaboration), Phys. Rev. D {\bf 70},
034504 (2003);
N. Ishii, H. Suganuma and H.
Matsufuru, Phys. Rev. D {\bf 66}, 014507 (2002);
Xi-Yan Fang, Ping Hui, Qi-Zhou Chen and D. Schutte,
Phys. Rev. D {\bf 65}, 114505 (2002);
M.G. Alford and R.L. Jaffe, Nucl. Phys. B {\bf
578}, 367 (2000);
C.J.
Morningstar and M. Peardon, Phys. Rev. D {\bf 60},
034509 (1999); J. Sexton, A. Vaccarino and D.
Weingarten, Phy. Rev. Lett. {\bf 75}, 4563 (1995);
G. Bali et al., Phys. Lett. B {\bf 309}, 378 (1993).



\bibitem{06_G}
 I. Eshraim, S. Janowski, F. Giacosa and  D.H. Rischke,
Phys. Rev. D {\bf 87}, 054036 (2013);
F. Giacosa, Phys. Rev. D {\bf 74}, 014028 (2006).



\bibitem{06_Pelaez}
J.R. Pelaez, PoS CD12, 047 (2013);
R. Garcia-Martin, R. Kaminski, J.R. Pelaez, J. Ruiz de Elvira
Phys. Rev. Lett. {\bf 107}, 072001 (2011);
J.R. Pelaez, Phys. Rev. Lett. {\bf 97}, 242002 (2006).


% 2012



% 2011


\bibitem{3flavor}
A.H. Fariborz,  R. Jora, J. Schechter and M.N. Shahid,
Phys. Rev. D {\bf 83}, 034018 (2011).

\bibitem{Ds_decays}
A.H. Fariborz, R. Jora, J. Schechter and M.N. Shahid, Phys. Rev. D {\bf 84}, 094024 (2011);
arXiv:1108.3581 [hep-ph].



% 2009


\bibitem{LsM_scatt_length}
D. Black, A.H. Fariborz, R. Jora, N.W. Park, J. Schechter and M.N. Shahid,  Mod. Phys. Lett. A {\bf 24}, 2285 (2009).


\bibitem{LsM_gauged}
A.H. Fariborz,  N.W. Park, J. Schechter and M.N. Shahid, Phys. Rev. D {\bf 80}, 113001 (2009).

\bibitem{bfjpss09}D. Black, A.H. Fariborz, R.
Jora, N.W. Park, J. Schechter and M.N. Shahid,
Mod. Phys. Lett. A {\bf 28}, 2285 (2009).





% 2008




\bibitem{08_tHooft} G. 't Hooft, G. Isidori, L.
Maiani, A.D. Polosa anf V. Riquer,
arXiv: 0801.2288 [hep-ph].



\bibitem{07_FJS2} A.H. Fariborz, R. Jora and J.
Schechter, Phys. Rev. D {\bf 77}, 034006
(2008).

\bibitem{07_FJS4} A.H. Fariborz, R. Jora and J.
Schechter, Phys. Rev. D {\bf 77}, 094004
(2008).



%  2007




\bibitem{06_MPPR}
L. Maiani, F. Piccinini, A.D. Polosa, V. Riquer, Eur. Phys. J. C {\bf 50},
609 (2007);  hep-ph/0604018.



\bibitem{07_FJS1} A.H. Fariborz, R. Jora and J.
Schechter, Phys. Rev. D {\bf 76}, 014011
(2007).

\bibitem{07_FJS3} A.H. Fariborz, R. Jora and J.
Schechter, Phys. Rev. D {\bf 76}, 114001
(2007).



% 2006


\bibitem{05_N}
S. Narison, Phys. Rev. D {\bf 73}, 114024 (2006).



\bibitem{06_CCY} H.Y. Cheng, C.K. Chua and K.C. Yang, Phys. Rev. D {\bf
73}, 014017 (2006).

\bibitem{06_KKNHH} Yu. Kalashnikova, A. Kudryavtsev, A.V. Nefediev, J.
Haidenbauer and C. Hanhart, Phys. Rev. C {\bf 73}, 045203 (2006).

\bibitem{06_BCKR} E. van Beveren, J. Costa, F. Kleefeld and G. Rupp,
Phys. Rev. D {\bf 74}, 037501 (2006).

\bibitem{06_Aetal} M. Ablikim et al, Phys. Lett. B {\bf 633}, 681 (2006).

\bibitem{06_T}N.A. T\"ornqvist, hep-ph/0606041.




\bibitem{06_CCL} I. Caprini, G. Colangelo and
H. Leutwyler, Phys. Rev. Lett. {\bf 96} (2006).







% 2005




\bibitem{04_Ynd}
F.J. Yndurain, Phys. Lett. B {\bf 578}, 99
(2004); Phys. Lett. B {\bf
612}, 245 (2005).

\bibitem{05_TKM} T. Teshima, I. Kitamura and N. Morisita,  Nucl. Phys. A
{\bf 759}, 131 (2005).

\bibitem{05_GGF}
F. Giacosa, T. Gutsche, A. Faessler, Phys. Rev. C {\bf 71}, 025202
(2005).

\bibitem{05_VVFS}
J. Vijande, A. Valcarce, F. Fernandez, B. Silvestre-Brac,
Phys. Rev. D {\bf 72}, 034025 (2005).



\bibitem{05_BNNB} T.V. Brito, F.S. Navarra, M. Nielsen,
M.E. Bracco, Phys. Lett. B {\bf 608}, 69 (2005).

\bibitem{05_GGLF}
F. Giacosa, Th. Gutsche, V.E. Lyubovitskij, A. Faessler,
Phys. Lett. B {\bf 622}, 277 (2005)




\bibitem{05_FJS} A.H. Fariborz, R. Jora and J.
Schechter, Phys. Rev. D {\bf 72}, 034001
(2005).

\bibitem{05_FJS2} A.H. Fariborz, R. Jora and J.
Schechter, Int. J. of Mod. Phys. A {\bf 20},
6178 (2005).






% 2004





\bibitem{04_KMNNSW} T. Kunihiro, S. Muroya, A. Nakamura,
C. Nonaka, M. Sekiguchi and H. Wada, Phys. Rev. D {\bf 70}, 034504 (2004).

\bibitem{04_UNOT} T. Umekawa, K. Naito, M. Oka and M. Takizawa, Phys. Rev.
C {\bf 70}, 055205 (2004).

\bibitem{04_MPPR} L. Maiani, F. Piccinini, A.D. Polosa and  V. Riquer,
Phys. Rev. Lett. {\bf 93}, 212002 (2004).

\bibitem{04_TKM} T. Teshima, I. Kitamura and N. Morisita, J. Phys. G.
{\bf 30}, 663 (2004).

\bibitem{04_NR}
M. Napsuciale and  S. Rodriguez, Phys. Rev. D {\bf 70}, 094043 (2004).

\bibitem{04_Pelaez}
J.R. Pelaez, Phys. Rev. Lett. {\bf 92}, 102001 (2004).

\bibitem{04_ACCGL}
A. Ananthanarayan, I. Caprini, G. Colangelo, J. Gasser and H. Leutwyler,
Phys. Lett. B {\bf 602}, 218 (2004).









% 2003










.






% 2004









% 2003








% 2002



\bibitem{E791} E.M. Aitala et al, Phys. Rev. Lett.
{\bf 89}, 121801 (2002).






% 2001




\bibitem{01_CGL} G. Colangelo, J. Gasser and H.
Leutwyler, Nucl. Phys. B {\bf 603}, 125 (2001).



% 2000



\bibitem{pieta}D. Black, A.H. Fariborz and J.
Schechter, Phys. Rev.  D {\bf 61}, 074030 (2000).


\bibitem{BFSS2}D. Black, A.H. Fariborz, F. Sannino
and J. Schechter, Phys.  Rev. D {\bf 59}, 074026
(1999).

\bibitem{BFSS1}D. Black, A.H. Fariborz, F. Sannino
and J. Schechter, Phys.  Rev. D {\bf 58}, 054012
(1998).


















\bibitem{Blk_rad} D. Black, M. Harada and J.
Shechter, Phys. Rev.  Lett. {\bf 88}, 181603 (2002).



\bibitem{CLEO} The CLEO collaboration, Phys. Rev. D {\bf 61}, 012002
(2000).




\bibitem{Teige} S. Teige et al, Phys. Rev. D
{\bf 59}, 012001 (1999).


\bibitem{Pelaez}
M. Albaladejo and J.A. Oller,  Phys. Rev. D {\bf 86}, 034003 (2012); J.A. Oller, E. Oset and J.R.
Pelaez, Phys. Rev. D {\bf 59}, 074001 (1999).


\bibitem{Ach} N.N. Achasov, Phys. Usp. {\bf 41}, 1149 (1999), hep-ph/9904223;
N.N. Achasov and G.N. Shestakov, hep-ph/9904254.


\bibitem{IH} K. Igi and K. Hikasa, Phys. Rev. {\bf
D59}, 034005 (1999).



\bibitem{OOP} J.A. Oller, E. Oset and J.R. Pelaez, Phys. Rev. Lett.
{\bf 80}, 3452 (1998).






\bibitem{Ishida_kappa} S.~Ishida, M.~Ishida, T.~Ishida,
K.~Takamatsu and T.~Tsuru, Prog. Theor. Phys. {\bf 98}, 621
(1997). See also M. Ishida and S. Ishida, Talk given at 7th
International Conference on Hadron Spectroscopy (Hadron
97), Upton, NY, 25-30 Aug. 1997, hep-ph/9712231.


\bibitem{AnSa}A.V. Anisovich and A.V. Sarantsev, Phys. Lett. {\bf B413},
137 (1997).













\bibitem{Ishida} S. Ishida, M.Y. Ishida, H. Takahashi, T. Ishida,
K. Takamatsu and T Tsuru, Prog. Theor. Phys. {\bf 95}, 745 (1996).


\bibitem{T}
{N.A.~T\"ornqvist} and M. Roos, Phys. Rev. Lett. {\bf 76}, 1575
(1996).

\bibitem{Sv} M. Svec, Phys. Rev. {\bf D53}, 2343 (1996).

\bibitem{HSS1}M. Harada, F. Sannino and J. Schechter, Phys. Rev. {\bf
D54}, 1991 (1996).






\bibitem{Tor} N.A. T\"ornqvist, Z. Phys. C {\bf 68},
647 (1995).


\bibitem{SS}F.~Sannino and J.~Schechter, Phys. Rev.  {\bf D52},  96
(1995).

\bibitem{JPHS}
{G.~Janssen, B.C.~Pearce, K.~Holinde and J.~Speth}, Phys. Rev. {\bf
D52},  2690  (1995).

\bibitem{DS} R. Delbourgo and M.D. Scadron, Mod. Phys. Lett. {\bf
A10}, 251 (1995).








\bibitem{AS94} N.N. Achasov and G.N. Shestakov, Phys. Rev. {\bf
D49}, 5779 (1994). A summary of the recent work of the Novosibirsk
group is given in N.N. Achasov, arXiv:0810.2601[hep-ph].

\bibitem{Kam94}{R. Kam\'inski}, {L. Le\'sniak} and J. P. Maillet,
Phys. Rev. {\bf D50}, 3145 (1994).


\bibitem{AS} N.N. Achasov and G.N. Shestakov,
Phys. Rev. D {\bf 49}, 5779 (1994).







\bibitem{MP} D. Morgan and M. Pennington, Phys.
Rev. {\bf D48}, 1185 (1993).

\bibitem{BMPV} A.A. Bolokhov, A.N. Manashov, M.V. Polyakov and
V.V. Vereshagin, Phys. Rev. {\bf D48}, 3090 (1993).




\bibitem{Isg} J. Weinstein and N. Isgur, Phys. Rev.
D {\bf 41}, 2236 (1990).


\bibitem{Aston} D. Aston et al., Nucl. Phys. B
{\bf 296}, 493 (1988).


\bibitem{vanBeveren} E. van Beveren, T.A. Rijken, K.
Metzger, C. Dullemond, G. Rupp and J.E. Ribeiro, Z. Phys.
C {\bf 30}, 615 (1986).

\bibitem{vanBev} E. van Beveren, T.A. Rijken,
K. Metzger, C. Dullemond, G. Rupp and J.E.
Ribeiro, Z. Phys. {\bf C30}, 615 (1986).






% End General refs. -------------------------------------------------







% MIT bag model --------------------------------------

\bibitem{Jaf} R.L. Jaffe, Phys. Rev. D {\bf 15}, 267
(1977).


% END:   MIT bag mode -------------------------














% ---- Begin mixing



\bibitem{06_F}
A.H. Fariborz,  Phys. Rev. D {\bf 74},
054030 (2006).




\bibitem{NR04}M. Napsuciale and S. Rodriguez, Phys. Rev. D {\bf 70},
094043 (2004).


\bibitem{Far_IJMPA} A.H. Fariborz, Int. J. of
Mod. Phys. A {\bf 19}, 2095 (2004).

\bibitem{04_F} A.H. Fariborz, Int. J. of
Mod. Phys. A {\bf 19}, 5417 (2004).

\bibitem{mixing}T. Teshima, I. Kitamura and N. Morisita,
J. Phys. G
{\bf 28}, 1391 (2002); {\it ibid} {\bf 30}, 663 (2004).


\bibitem{close}F. Close and N.
Tornqvist, {\it ibid.}
{\bf 28}, R249 (2002).


\bibitem{Mec}D. Black, A.H. Fariborz and J.
Schechter, Phys. Rev. D {\bf 61}, 074001 (2000).



% ---- end mixing





\bibitem{global}
A.H. Fariborz, R. Jora and J. Schechter,
Phys. Rev. D {\bf 79}, 074014 (2009).






\bibitem{07_KZ}
E. Klempt and A. Zaitsev, Phys. Rept. {\bf 454},1 (2007); arXiv:0708.4016v1.


\bibitem{10_AOR}  M. Albaladejo, J.A. Oller and  L. Roca,
Phys. Rev. {\bf D} 82, 094019 (2010).




% --- Begin CHPT

\bibitem{ChPT} S. Weinberg, Physica A {\bf 96},
327 (1979); J. Gasser and H. Leutwyler, Annalas
Phys. {\bf 158}, 142 (1984); Nucl. Phys. B {\bf
250}, 465 (1985).


\bibitem{bijnens11}
J. Bijnens, in proceedings of the 2nd International PrimeNet Workshop, Forschungszentrum Juelich, Germany,  26-28 (2011).
CNUM: C11-09-26.8,  arXiv:1110.6004 [hep-ph].



\bibitem{RCPT}
R. Escribano, P. Masjuan, J.J. Sanz-Cillero, JHEP 1105, 094 (2011).



% --- end ChPT






%  ----  Begin Exp

\bibitem{VES} V. Dorofeev et al,  Phys. Lett. B {\bf 651}, 22 (2007).

\bibitem{GAM4} A.M. Blik et al, Phys. Atom Nucl.  {\bf 72}, 231 (2009).

\bibitem{Moskov} M. Amaryan et al,  ``Decays of Light Mesons in CLAS,'' in
proceedings of the second International PrimeNet Workshop, September 26-28, 2011,
J\"ulich, Germany, p. 80-82; arXiv: 1204.5509 [nucl-ex].


\bibitem{07_BJ}
B.R. Jany, in proceedings of
Symposium on Meson Physics at COSY-11 and
WASA-at-COSY, Cracow, Poland, June 17-22, 2007
[AIP Conf. Proc. {\bf 950}, 209 (2007)].\\
WASA-at-COSY Collaboration, B.R. Jany et al,
%----------------------------------------------
    in proceedings of MENU 2007, the 11th
International Conference on Meson-Nucleon
Physics and the Structure of the Nucleon,
September 10-14, 2007, J\"ulich, Germany [SLAC
eConf {\bf C070910}, 169 (2007)].\\
%----------------------------------------------
M. B\"uscher,
        in proceedings of {\it Workshop on
Scalar Mesons and Related Topics}, February
11-16, 2008, Lisbon, Portugal [AIP Conf. Proc.
{\bf 1030}, 40 (2008)].


% --- End Exp





\bibitem{LsM}D. Black, A.H. Fariborz, S. Moussa, S.
Nasri and J.  Schechter, Phys. Rev. D {\bf 64},
014031 (2001).




\bibitem{LsM_Maple}
A.H. Fariborz, Int. J. Mod. Phys. A {\bf 26}, 2327 (2011).







\bibitem{99FS}A.H. Fariborz and J. Schechter, Phys.
Rev. D {\bf 60}, 034002 (1999).





\bibitem{mixing_pipi}
A.H. Fariborz, R. Jora, J. Schechter and M.N. Shahid, Phys. Rev. D {\bf 84}, 113004 (2011); arXiv:1106.4538 [hep-ph].


\bibitem{mixing_piK}
A.H. Fariborz,  E. Pourjafarabadi, J. Schechter and M. Zebarjad, ``Chiral nonet mixing in $\pi K$ scattering,''  in preperation.


\bibitem{mixing_pieta}
A.H. Fariborz,  E. Pourjafarabadi, J. Schechter, S. Zarepour and M. Zebarjad, ``Chiral nonet mixing in $\pi\eta$ scattering,''  in preperation.



\bibitem{N10U1A}
A.H. Fariborz,  E. Pourjafarabadi, J. Schechter, S. Zarepour and M. Zebarjad, ``Effect of higher order U(1)$_{\rm A}$ breaking on eta systems,''
in preperation.




\bibitem{SU} The isospin violation case for the
single-M
linear sigma model was treated in J. Schechter and
Y. Ueda, Phys. Rev. D {\bf 4}, 733 (1971).


\bibitem{e3p} A. Abdel-Rehim, D. Black, A.H.
Fariborz and J. Schechter,  Phys. Rev. D
{\bf 67}, 054001 (2003).





\bibitem{unitarization_methods}
Various unitarization schemes are contrasted in Zhi-Hui Guo, L.Y. Xiao and H.Q. Zheng, Int. J .Mod. Phys. A {\bf 22}, 4603 (2007).







\end{thebibliography}
\end{document}